\documentclass[aps,prb,a4paper,twocolumn,superscriptaddress,showkeys,longbibliography,floatfix]{revtex4-2}
\usepackage{comment} 
\usepackage[normalem]{ulem}
\usepackage{xcolor}
\usepackage{cancel}
\usepackage[dvipsnames]{xcolor}
\usepackage{graphicx}
\usepackage{subfigure}
\usepackage{amsmath}
\usepackage{amsmath}
\usepackage{amssymb}
\usepackage{physics}
\usepackage{subcaption}
\usepackage{bm}% bold math
\usepackage[hidelinks,colorlinks=true,linkcolor=blue,citecolor=blue]{hyperref}
\usepackage{blindtext}
\hypersetup{colorlinks=true, linkcolor=blue, citecolor=magenta, filecolor=magenta, urlcolor=magenta }

\begin{document}

\preprint{APS/123-QED}

\title{Emergent Nonperturbative Universal Floquet Localization}
%%%%%% AUTHORS %%%%%%%%%%%%
\author{Soumadip Pakrashi}
%\email{soumadippakrashi@students.iisertirupati.ac.in}
\affiliation{Department of Physics, Indian Institute of Science Education and Research (IISER)
Tirupati, Tirupati - 517507, Andhra Pradesh, India}
\author{Atanu Rajak}
%\email{atanu@phy.iith.ac.in}
\affiliation{Department of Physics, Indian Institute of Technology, Hyderabad 502284, India}
\author{Sambuddha Sanyal}
%\email{sambuddha.sanyal@iisertirupati.ac.in}
\affiliation{Department of Physics, Indian Institute of Science Education and Research (IISER)
Tirupati, Tirupati - 517507, Andhra Pradesh, India}

\date{\today}
%%%%%%%%%%%%%%%%%%%%%%%%%%%

\date{\today}
\begin{abstract}
We show that a robust, nonperturbative localization plateau emerges in periodically driven quasiperiodic lattices, independent of the static localization properties and drive protocol. Using exact Floquet dynamics, Floquet perturbation theory, and optimal-order van Vleck analysis, we identify a fine-tuned amplitude-to-frequency ratio where all Floquet states become localized despite dense resonances. The van Vleck expansion achieves superasymptotic accuracy up to an optimal orde; it ultimately breaks down due to resonant hybridization at a weak quasiperiodic potential, revealing that the observed localization is nonperturbative.
\end{abstract}
\maketitle

%\tableofcontents

%%%%%%%%%%%%%%%%%%%%%%%%%%%%%%%%%%%%%%%%%%%%%%%%%%%%%%%%%%%%%%%%%%%%%%%%%%%%%%%%%%%%%%%%%%%%%%%%%%%%%%%%%%%%%%%%%%%%%%%%%%%%%%%%%%
%%%%%%%%%%%%%%%%%%%%%%%%%%%%%%%%%%%%%%%%%%%%%%%%%%%%%%%%%%%%%%%%%%%%%%%%%%%%%%%%%%%%%%%%%%%%%%%%%%%%%%%%%%%%%%%%%%%%%%%%%%%%%%%%%%
\paragraph*{\textbf{Introduction}}

The localization phenomenon \cite{Andersonloc} of matter waves due to broken crystalline translation symmetry is one of the most intriguing phenomena in modern physics. In a tight-binding model, localization arises from suppression of the coherent superposition of localized Wannier orbital wavefunctions caused by sufficiently large differences in on-site potentials relative to the hopping amplitude. This insulating regime can be "melted" by a time-periodic drive that can restore the Wannier wave function hybridization \cite{AL_drive1,AL_drive2}. Conversely, localization can also be dynamically engineered in a clean tight-binding model \cite{DL_Kenkre} by special time-periodic drive protocols at resonant frequencies, via tuning a destructive interference of the extended wave functions. 

Over the past three decades, the idea of engineering localization by tuning the hybridization of Wannier orbitals using time-periodic drives has attracted significant interest \cite{AL_ac_1,DL_expt1,DL_expt2}. In low-dimensional systems, even infinitesimal deviations from crystalline symmetry can lead to complete localization. However, in many practical settings-such as mesoscopic systems \cite{AL_meso1}, engineered optical lattices \cite{Al_expt1,AL_expt2}, or materials with broken crystalline order—wavefunctions may instead exhibit partial localization, multifractality, or even coexist as a mixture of localized and delocalized states, separated by a mobility edge. This motivates a central question: \emph{is it possible to engineer complete localization using an external time-periodic drive, regardless of a system’s static properties?}

In this Letter, we identify a universal dynamical localization regime that emerges under staggered, time-periodic modulation of the hopping amplitudes in tight-binding models with crystalline symmetry-breaking onsite potential, where the universal localization regime doesn't depend on the properties of the static Hamiltonian or the nature of the drive. Using Van Vleck perturbation theory (VVPT) \cite{VV_Eckardt} up to optimal order \cite{Abanin_optimal}, we characterize the perturbative signatures of this universal localization transition. We find that the effective Floquet Hamiltonian contains algebraically decaying long-range hopping terms whose amplitudes inherit spatial modulations from the underlying onsite potential. The competition between these drive induced long-range modulated photon-assisted tunneling, and the photon-dressed onsite potential gives rise to a novel, robust localized phase. We further find that due to resonant hybridization, the perturbation theory doesn't capture the universal localization observed in the exact calculation at resonant energies. Those resonant energies are rare when the unperturbed Hamiltonian has a point spectrum or even a multifractal spectrum, but when the unperturbed Hamiltonian has a continuous spectrum, the resonances proliferate across the entire spectrum and the perturbation theory fails to converge. We demonstrate that universal localization is a novel non-perturbative phenomenon that occurs through an interplay of drive and disorder, leaving an impression on a superasymptotic perturbation theory. 

\paragraph*{\textbf{Model}}We consider a tight-binding model with staggered, time-periodic modulation of the hopping amplitudes, described by the Hamiltonian
\begin{eqnarray}
H(t) &=& \sum_{\langle i j \rangle} (\gamma + a \mathcal{F}(t))\, c_{i,A}^\dagger c_{i,B} + (\gamma - a \mathcal{F}(t))\, c_{j,A}^\dagger c_{i,B} \nonumber \\
&&  + \sum_{i} \Big(V_{i,A} n_{i,A}+ V_{i,B} n_{i,B}\Big)\ +\text{h.c.} 
\label{eqn1}
\end{eqnarray}
Due to the staggered nature of the drive, it is convenient to write it in a unit cell with a two-site basis, the Hamiltonian in Eq.~(\ref{eqn1}) represents a lattice of $\mathcal{N}$ unit cells consisting of two sublattices $(A,B)$, where $i,j$ represent the unit cell. Alternatively one can also consider uniform drive and staggered hopping\cite{Aditya,SS_AA}. Here \(c_{i,A}^\dagger\)  creates a particle on sublattice site \(A\) in unit cell \(i\), and \(n_{i,A}\) is the corresponding number operator. The hopping modulation is governed by a time-periodic function \(\mathcal{F}(t)\), with \(\mathcal{F}(t + T) = \mathcal{F}(t)\). We consider two representative drive protocols: (i) a sinusoidal drive, \(\mathcal{F}(t) = \sin(\omega t)\), and (ii) a square wave pulse alternating between \(\pm 1\) over half-periods. 

The onsite potential $V_{i ,A/B}$ is a crystalline symmetry-breaking potential that varies from site to site. In this letter, we consider the one-dimensional Aubry-Andre model \cite{aa1} with quasiperiodic potential and its generalizations\cite{aa_expt_spme1,aa_expt_me} as primary examples. The Aubry-Andre model is defined as $V^{AA}_{i,A}=\lambda cos(2\pi \alpha (2i)), V^{AA}_{i,B}=\lambda cos(2\pi \alpha (2i+1))$ in Eq.~(\ref{eqn1}). For generalized Aubry-Andre model $V^{GAA}_{i,A}=2\lambda \frac{1-(cos(2\pi \alpha (2i))}{1+\beta(cos(2\pi \alpha (2i))}, V^{GAA}_{i,B}=2\lambda \frac{1-(cos(2\pi \alpha (2i+1))}{1+\beta(cos(2\pi \alpha (2i+1))}$. In the Aubry-André model, all eigenstates undergo a transition from delocalization to localization as $\lambda$ increases, passing through a critical point at $\lambda_c=2$, where all eigenstates are critical. On the other hand, in the generalized Aubry-Andre (GAA) model, the localization transition occurs through a mobility edge for a range of values of $\lambda$ and $\beta$. The Aubry-Andre(AA) model and the generalized Aubry-Andre model are considered as archetypal models of localization transitions in one dimension, which allow computational tractability. However, we argue that the drive-induced universal localization phenomenon is not limited to quasiperiodic or lower-dimensional systems.

\begin{figure}[!h]
\centering
 {
\subfigure[]{
\includegraphics[width=0.49\linewidth]{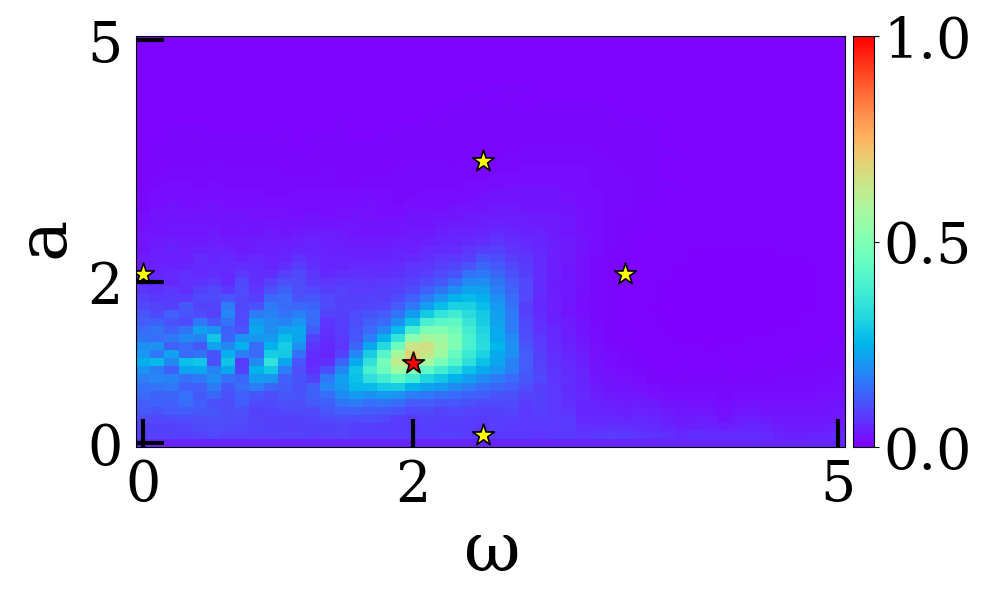}\label{exact_pd}
}
\hspace{-0.42cm}
\subfigure[]{
\includegraphics[width=0.49\linewidth]{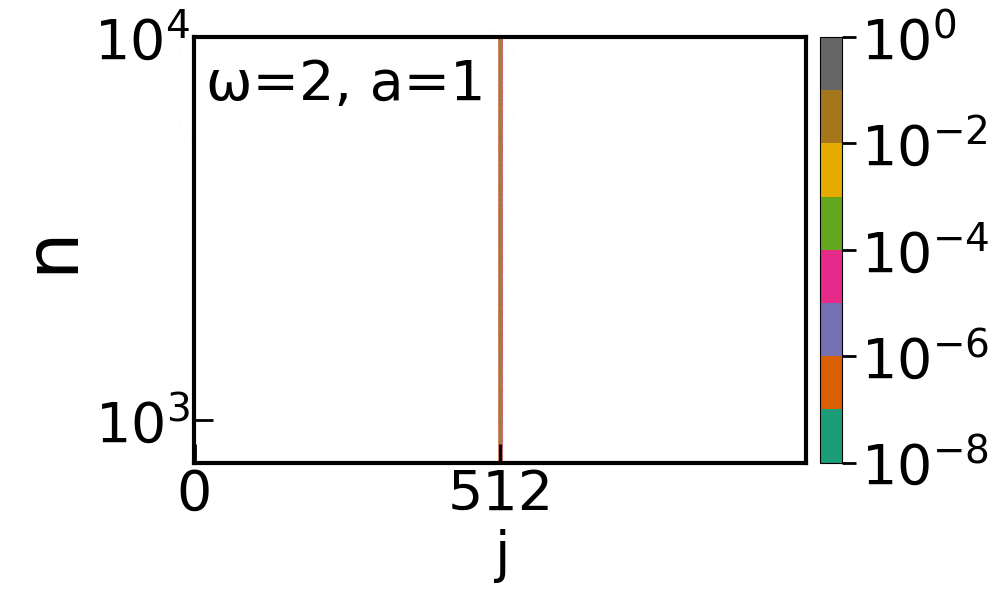}\label{exact_dynamics}}
\subfigure[]{
\includegraphics[width=0.48\linewidth]{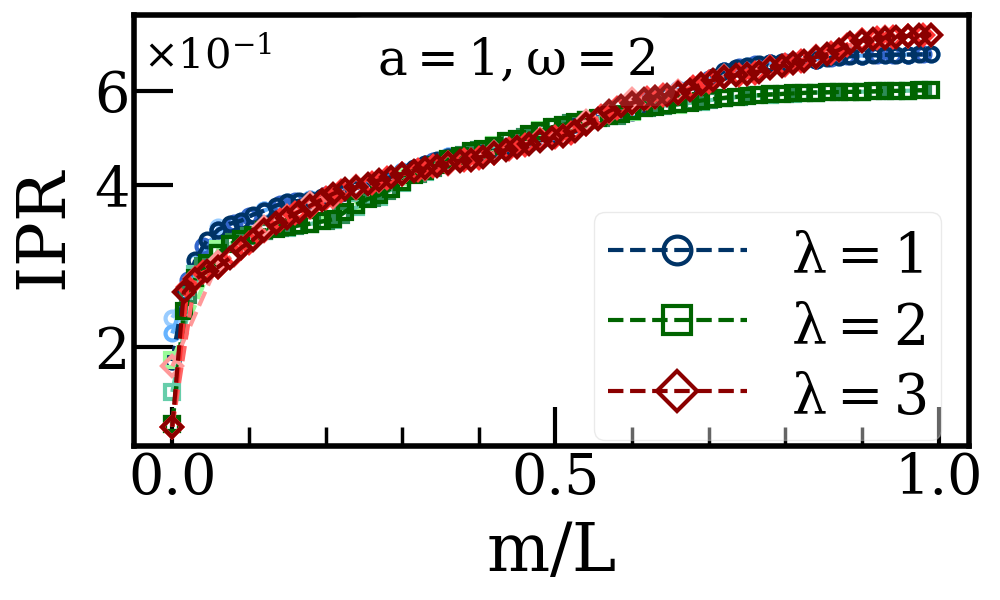}
\label{exact_scaling_aa}}
\hspace{-0.42cm}
\subfigure[]{
\includegraphics[width=0.48\linewidth]{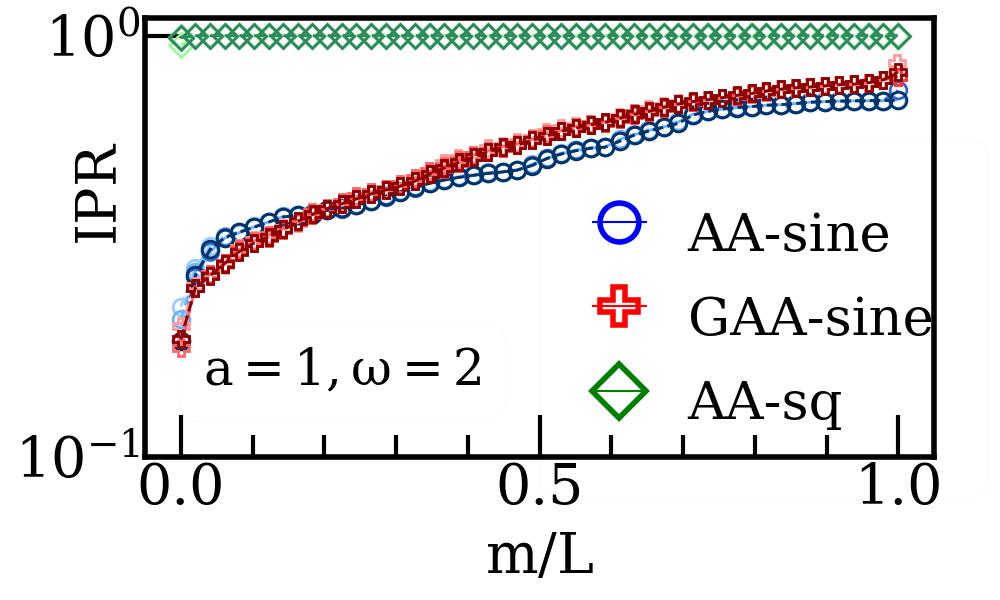}
\label{exact_scaling_gaa}}
\subfigure[]{
\includegraphics[width=0.48\linewidth]{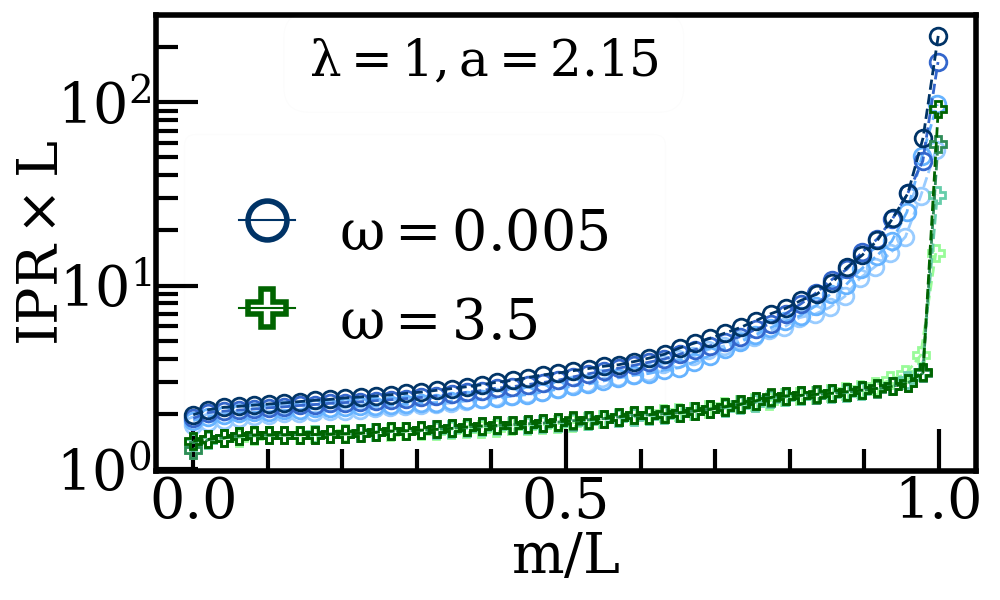}
\label{exact_scaling_om}}
\hspace{-0.42cm}
\subfigure[]{
\includegraphics[width=0.48\linewidth]{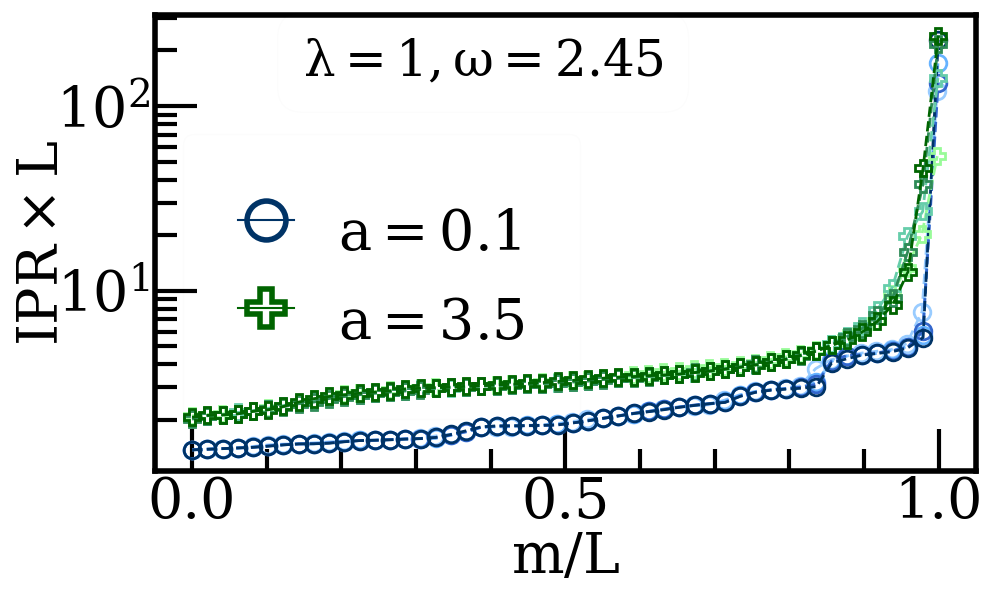}
\label{exact_scaling_a}}
\captionsetup{justification=centerlast, width=\linewidth} 
\caption{
{\bf{a)}} $\langle IPR\rangle_g$(in color code) is plotted for all Floquet eigenstates $\chi^m$ at $\lambda=1, 2, 3$ of AA model and $(\lambda,\beta)=(-0.8,0.5)$ of GAA model for both sinusoidal and square wave drive protocols, for $L=512, N=200$. A universal localization regime is found near $(a,\omega)=(1,2)$. {\bf{b)}} The density plot of the local probability $\langle p_{nT}(j) \rangle_g$ for same parameter values, models, and drive protocols. {\bf{c)}} $\mathcal{O}(1)$ IPR scaling is observed at $(a,\omega)=(1,2)$ for AA at $\lambda=1,2,3$, {\bf{d)}} The same $\mathcal{O}(1)$ scaling is seen across different driving protocols (square/sine pulses) and models (AA:$\lambda=1$; GAA: $(\lambda,\beta)=(-0.8,0.5)$). {\bf{e)}},{\bf{f)}} For AA: $\lambda=1$, IPR $\sim1/L$ for points lying outside the localized patch marked by yellow $\star$ in phase diagram 1(a) (Ballistic transport behavior of those regions shown in Fig.3 of supplementary section).  Here, $L=500, 1000, 1500$ and $2000$, the trotter steps $N=500$, and sinusoidal drive is used for scaling plots. In IPR scaling plots, states are organized in ascending order of IPR magnitude.
}
}
\end{figure}

%%%%%%%%%%%%%%%%%%%%%%%%%%%%%%%%%%%%%%%%%%%%%%%%%%%%%%%%%%%%%%%%%%%%%%%%%%%%%%%%%%%%%%%%%%%%%%%%%%%%%%%%%%%%%%%
\paragraph*{\textbf{Exact results:}}
We first study our model in Eq.~(\ref{eqn1}) using aymptotically exact numerical calculations of the Floquet operator $U(T,0)$ (details given in the End Matter). The Floquet eigenstates are simultaneous eigenkets of the Floquet operator and the Floquet Hamiltonian $H_F$, which are related by the equation $U(T,0)=e^{-iH_FT}$.

To measure localization of the individual Floquet eigenstate, we consider Inverse Participation Ratio (IPR), defined as ${\rm IPR}(m)=\frac{\sum_i (\chi_i^m)^4}{(\sum_i (\chi_i^m)^2)^2}$, where $\chi_i^m$ denotes the $m$-th Floquet eigenstate and $i$ indicates the lattice site. For a completely delocalized/extended state, IPR($m$) $\sim 1/L$, where $L$ is the size of the system, whereas, for a completely localized state, IPR is independent of system size. To demonstrate the universal dynamical localization in a compact form, we show a density plot(in Fig. \ref{exact_pd}) of geometric mean of the IPR ($\langle {\rm IPR}\rangle_g$) of all the Floquet eigenstates, for the parameters $\lambda=1,2,3$ for AA model and $(\lambda,\beta)= (-0.8,0.5)$ for GAA model, as a function of the driving parameters $a$ and $\omega$ considering sinusoidal, and square pulse driving protocols, and $\alpha=\frac{\sqrt{5}+1}{2}+\frac{\sqrt{13}+3}{2}$. We observe a region around $a=1$ and $\omega=2$, where $\langle {\rm IPR}\rangle_g$ has non-zero values, indicating a universal dynamical localization.

We further plot the scaling of IPR with system size for each individual eigenstate in the universal localization region, as well as at a few nearby points for different parameters of the AA and GAA models under different drive protocols. 
In Fig.~\ref{exact_scaling_aa}, we find that the IPRs of all the eigenstates at $(a,\omega)=(1,2)$ for different system sizes collapse for all the cases with $\lambda=1,2$ and $3$ of the AA model using a sinusoidal driving protocol, thus confirming complete localization.
Furthermore, we find that all eigenstates are localized for both square and sinusoidal driving protocols for AA model at $(a,\omega)=(1,2)$ and $\lambda=1$, as well as at $(\lambda,\beta)= (-0.8,0.5)$ for GAA (see Fig. \ref{exact_scaling_gaa}). In Figs.~\ref{exact_scaling_om} and \ref{exact_scaling_a}, we plot the system-size scaling of the IPR for all eigenstates at $\lambda=1$, around the universal localization regime (at the points marked in Fig.~\ref{exact_pd}), and we find that all the eigenstates are delocalized in this region.

Next we compute the spatiotemporal spread of a wave-packet at stroboscopic times ($\psi_{nT}(j)$) that was initially localized at the centre of the lattice, where $|\psi(nT)\rangle=U(T,0)^n|\psi_{\rm in}\rangle$, $\psi_{\rm in}$ referring to the initial wave packet. 
In Fig.~\ref{exact_dynamics}, we show the density plot of the local probability $p_{nT}(j)=|\psi_{nT}(j)|^2$ at the universal localization region, as indicated in Fig.~\ref{exact_pd}, obtained by taking the geometric mean($\langle p_{nT}(j) \rangle_g$) of the same parameters and drive protocols. We find that, in universal localization regime, the dynamics is completely frozen. 
%==========================================================
\begin{figure*}
\centering
\subfigure[]{
\includegraphics[width=0.24\linewidth]{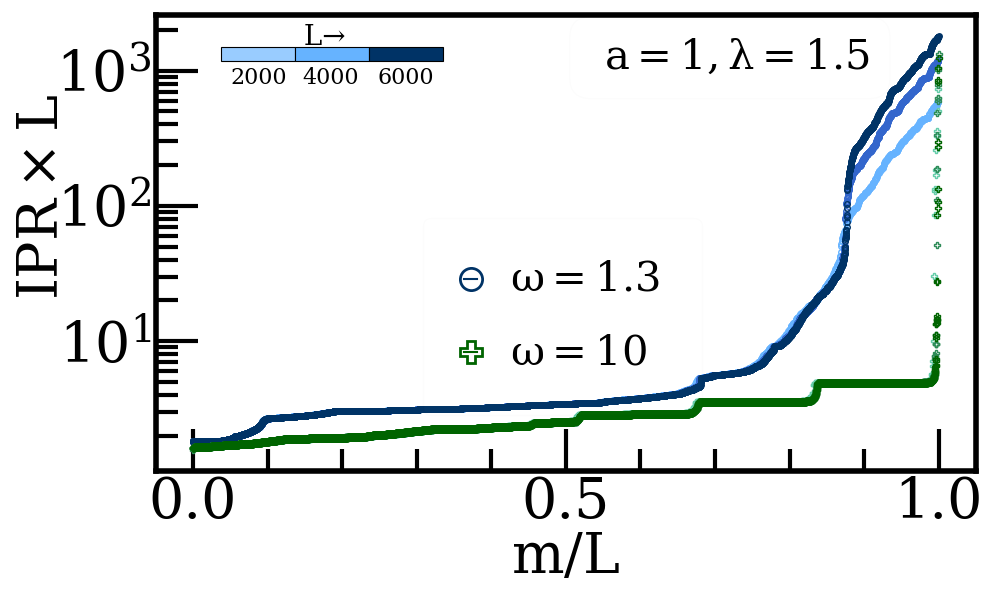}\label{VV_2_IPR_both}}
\hspace{-0.38cm}
\subfigure[]{
\includegraphics[width=0.24\linewidth]{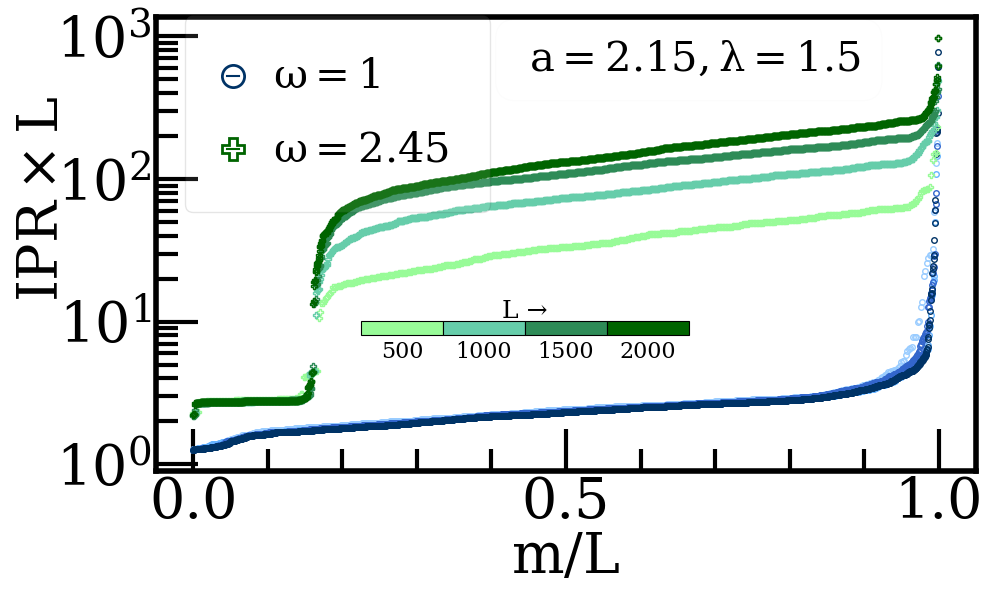}
\label{FPT_2_IPR_comb}}
\subfigure[]{
\includegraphics[width=0.24\linewidth]{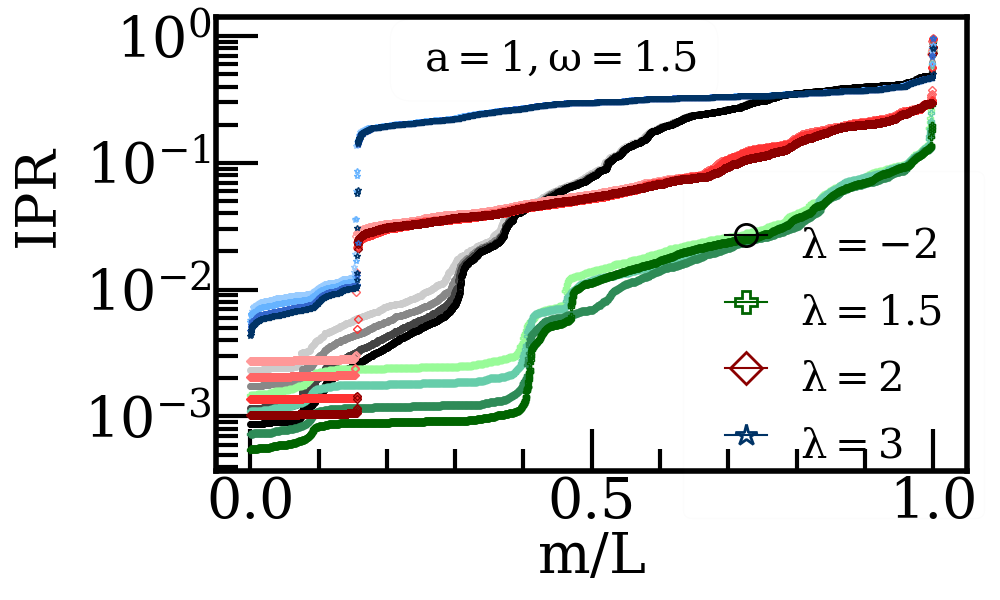}
\label{VV_IPR_sw}}
\hspace{-0.35cm}
\subfigure[]{
\includegraphics[width=0.24\linewidth]{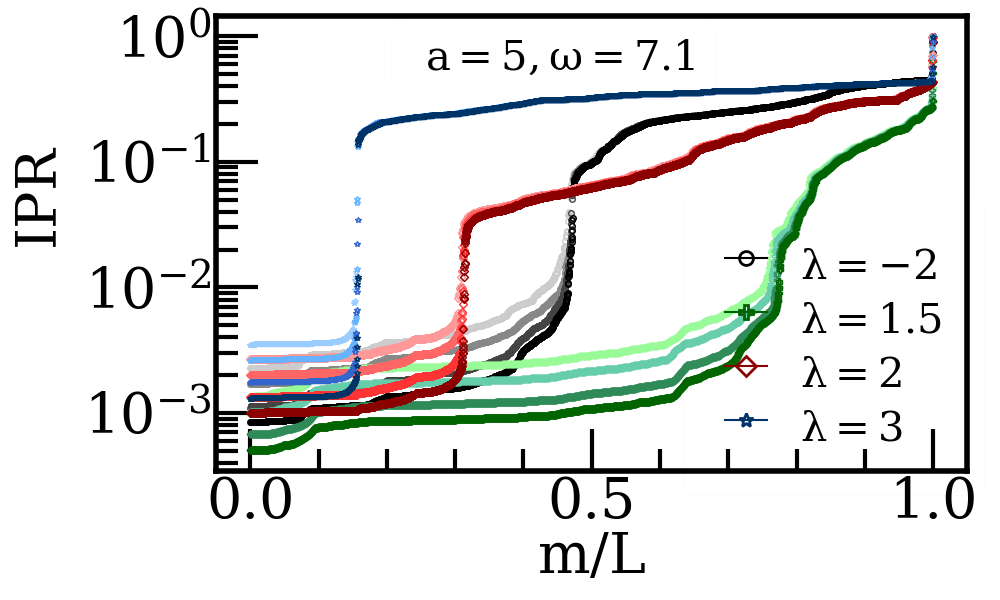}
\label{VV_IPR_osw}}
\captionsetup{justification=centerlast, width=\linewidth} 
\caption{{\bf{a)}} Using $2^{\rm nd}$ order VV, some eigenstates are found to be localized at $\omega=1.3$ for $a=1$, $\lambda=1.5$, while the revival of high frequency delocalized static limit is seen at $\omega=10$. \textbf{b)} Equivalent localized behavior can be depicted using FPT at $a=2.15,\omega=2.45$; while at low frequencies $\omega=1$ the FPT suggests retrieved delocalized phase. {\bf{c),d)}} Heuristic extrapolation of VVPT up to an optimal order reveals enhanced localization achieved at $ a=1,\omega=1.5$ universal localization region compared to $ a=5,\omega=7.1$ distant regions. Here, L=1500, 2000, 3000, 4000 are used for (c)-(d) IPR scaling and sine drive is considered throughout this analysis. In all plots, states are organized in ascending order of IPR magnitude.
}
\end{figure*}
%=========================================
The exact numerical results show that the universal localization regime in the amplitude-frequency plane of quasiperiodic chains with staggered driving is independent of both the localization properties of the static counterpart of the model and the nature of the drive. However, it is important to note that the universal localization arises from an interplay between disorder and periodic driving; even for $\lambda \rightarrow 0$ (shown in the Fig.3 of Supplementary Material), the universal localization persists, but it does not appear at $\lambda=0$. We study the detailed mechanism of this interplay using a superasymptotic perturbative approach.
\paragraph*{\textbf{Perturbative approaches}} We now turn to construct an effective time-independent Hamiltonian governing stroboscopic dynamics of the system using perturbative approaches.  We employ Van Vleck perturbation theory (VVPT), that systematically incorporates virtual processes arising from the time-periodic drive by expanding the full Hamiltonian in eqn. \ref{eqn1} in powers of $\frac{1}{\omega}$. VVPT is drive phase independent and gives a Floquet Hamiltonian in real space which can be directly used to study real space  dynamics. The effective Hamiltonian in VVPT is given by $H_{eff}=\sum_{n=0}^{\infty} \frac{H^{(n)}}{\omega^n}$, the
$H^{(n)}$ terms are given by $H^{(0)}=H_0, H^{(1)}=\sum_{p\neq0}\frac{[H_{-p},H_p]}{2p}$, and $H^{(2)}=\sum_{p\neq0}\frac{[[H_{-p},H_0],H_p]}{2p^2}+\sum_{p\neq0,q\neq0,p}\frac{[[H_{-p},H_{p-q}],H_q]}{3pq}$, where, $H_p=\frac{1}{T}\int_{0}^{T}e^{-ip\omega t}\hat{H}(t)dt$, and $H_0$ is the static part of the Hamiltonian in Eq.~\ref{eqn1}.
%%%%%%%%%%%%%%%%%%%%%%%%%%%%%%%%%%%%%%%%%%%%%%%%%%%%%%%%%%%%%%%%%%%%%%%%
At order \(n\), the Van Vleck effective Hamiltonian consists of a sum over all nested commutators involving \(n+1\) Fourier components of the full Hamiltonian in Eq.~\ref{eqn1}. These components are constrained by harmonic conservation (i.e., total harmonic index sums to zero), and their nested structure corresponds to distinct bracketings\cite{tree_magnus_1,tree_magnus_2}. 
%%%%%%%%%%%%%%%%%%%%%%%%%%%%%%%%%%%%%%%%%%%%%%%%%%%%%%%%%%%%%%%%%%%%%%%%
For drive protocols used in this letter, $H_p$ vanishes for $p \in 2\mathbb{Z} \setminus \{0\}$. From Eq.~\ref{eqn1}, we get $H_p=\sum_{j }\mathcal{A}(c_{j,A}^\dagger c_{j,B}-c_{j,A}^\dagger c_{j-1,B}+h.c)$. For sinusoidal and square drives, $\mathcal{A}=\frac{a}{2i}$ ($p\in1$) and $\frac{2ia}{p\pi}$ ($p\in 2\mathbb{Z}+1 $), respectively. For $p=0$, the time-dependent parts of full time-periodic Hamiltonian in Eq.~\ref{eqn1} gets integrated to zero and one gets back the static Hamiltonian $H_0$. The Fourier components of the time-dependent Hamiltonian (Eq.~\ref{eqn1}), denoted by $H_p$, satisfy the relation $H_{-p}=(-1)^p H_p$; consequently, $[H_{p}, H_{-p}]=[H_{-p}, H_{p-q}]=0$.

 From the properties of $H_p$, $\mathcal{O}(1/\omega)$ terms in $H_{eff}$ vanish, and at $\mathcal{O}(1/\omega^2)$ only the $[[H_{-p},H_0],H_p]$ term contributes, which, for the sinusoidal drive, is given by
\begin{equation}
\begin{split}
H_2 = -{\frac{a^2\gamma}{\omega^2}}\sum_j \Big[
c_{jA}^\dagger c_{j-1,B}
+ c_{jA}^\dagger c_{jB}
- c_{j+2,A}^\dagger c_{jB}
- c_{jA}^\dagger c_{j+1,B} \\\
 - \frac{\lambda}{4\gamma} \big( 
2\mathcal{C}\, n_{jA}
+ 2\mathcal{D}\, n_{jB}
+ \mathcal{D}\, c_{jA}^\dagger c_{j+1,A}
+ \mathcal{C}\, c_{jB}^\dagger c_{j-1,B}
\big) + \text{h.c.} \Big].
\end{split}
\label{vv_2_eqn}
\end{equation}

Here, $\mathcal{C}=\partial^{2}_j cos(2\pi \alpha(2j))$, and $\mathcal{D}=\partial^{2}_j cos(2\pi \alpha(2j+1))$, where the second-order lattice derivative is defined as $\partial^{2}_j f(j) \equiv f(j+1)+f(j-1)-2f(j)$.

Truncating the effective Hamiltonian $H^{(2)}_{VV}=H_0+H_2$ at $\mathcal{O}(\omega^{-2})$ is a valid approximation when $\omega \gg a$; however, as one approaches $\omega \rightarrow a$ limit, the nearest-neighbor hopping term of $H_2$ effectively cancels with the nearest-neighbor hopping term in $H_0$. At that point, the $H_{eff}$ has a dominant second-neighbor hopping, consisting a pure hopping, and a quasiperiodic hopping in which the quasiperiodic modulation is a linear combination of quasiperiodic potentials of neighboring sites ($\mathcal{C,D}$), with a prefactor $\lambda$. Moreover, same linear combination of quasiperiodic potential at neighboring sites ($\mathcal{C,D}$) also contributes in the onsite potential of the $H_{VV}^{(2)}$. The modulated hopping and onsite potential has the same pre-factor $\lambda$, the spectrum of the effective second-order Hamiltonian is expected to have both localized and delocalized states \cite{off_diag_qp}, where the localized states appear around $E\sim \omega$,  independent of $\lambda$. This description remains same for a square drive with appropriate modifications in the prefactors. 

%=========================================
We compute the eigenvalues and eigenvectors of effective Hamiltonian $H^{(2)}_{VV}$, and analyse the scaling properties of the IPR from the eigenstates in Fig. \ref{VV_2_IPR_both}. In Fig. \ref{VV_2_IPR_both} we set $\lambda=1.5$ where all eigenstates are delocalized in the absence of drive, when sinusoidal drive $\mathcal{F}(t)$ is applied with $a=1, \omega=10$ we find all the eigenstates are delocalized, as we decrease the frequency of drive to $\omega=1.3$ we find that few eigenstates are localized near $E\sim \omega$. 

We further probe this localization transition from the low-frequency side. In this limit, we employ the Floquet perturbation theory (FPT) \cite{FPT1} (details given in the End Matter) where the time-dependent part is considered as the unperturbed Hamiltonian and the time-independent part is considered as the perturbation. FPT is valid when $a\gg \lambda, \forall~\omega$. In Fig. \ref{FPT_2_IPR_comb}, we plot the system-size scaling of the IPR for $a=2.15, \lambda=1.5$ at driving frequencies $\omega=1$ and $2.45$. Both perturbative approaches suggest that, at intermediate amplitudes and frequencies ($a\sim 1-2.15$ and $ \omega\sim 1.3-2.5$), only a subset of eigenstates becomes localized, indicating a reentrant transition analogous to that found in the exact calculations. In contrast to the exact case, the localization region here contains only a few localized states rather than all eigenstates. However, using perturbative approaches, we find indications of localization of the Floquet states in the vicinity of the exact localization regime shown in Fig.~\ref{exact_pd}. Notably, in the universal localization region, the first-order FPT Hamiltonian ($H_F^{(1)}$) in position basis, shows a stronger weight in the diagonal terms compared to the off-diagonal term , which in turn promotes the signature of localization (details given in the Fig.4 of supplementary material).

Next we construct an ansatz for the time-independent Floquet Hamiltonian upto an arbitrary order using the recursive structure \cite{recursive_1} of Van-Vleck perturbation theory. The commutative structure of successive higher order terms ensures that if the $H_0$ is quadratic, at an arbitrary order the $H_{eff}$ remains quadratic.  The Van-Vleck perturbation expansion is a superasymptotic series that needs to be truncated at an optimal order $n_{*}(a, \omega, \lambda, \gamma)$ \cite{Abanin_optimal,tree_magnus_2,Bukov,goldman_dalibard,mori_trunc2,mori_trunc3}. The truncation criteria is determined by a ratio test of the appropriate norm of the $H_{eff}$ in successive order\cite{ratio_1}.

In this letter we will focus on $H_{eff}$ for sinusoidal drive only (without loss of any generalities), where $p \in 1$. For this discussion, we split the $H(t)$ into the kinetic term $H^0_h=\gamma\sum_{\langle i j \rangle}  c_{i,A}^\dagger c_{i,B} + h.c.$, static potential term $H^0_v= \sum_{i}V_{i,A} n_{i,A}+ \ \text{h.c.}$ and time-dependent drive $H_v=\mathcal{F}(t)\sum_{\langle i j \rangle} c_{i,A}^\dagger c_{i,B} + h.c.$ 
In order to construct the ansatz Hamiltonian, we use two important structures of the nested commutators as follows: 1) all odd order terms in the series vanishes due to the property $H_{-p}=(-1)^p H_p$. 2) From the harmonic sum conservation of a time-independent Floquet Hamiltonian, one can create the $2k+2$-th order term from the $2k$-th order term by adding two $H_0$ in the nested commutators, or by adding  $H_1$ and $H_{-1}$, and then permuting the combination. This property keeps the number of terms tractable as a branching tree \cite{tree_magnus_1,tree_magnus_2}. The final effective Hamiltonian will have three types of terms at $\mathcal{O}(1/\omega^{2k})$. First: pure hopping terms that connect maximum up to $2k+1$-th neighbors generated by nested commutators $H_h^0$ and $H_{\pm1}$ terms, due to the $H_{-1}$ term, the hopping amplitude between neighbors closer than $2k+1$ distance will be an oscillatory sum of the amplitudes of all the processes. 
Second: hopping terms that connect maximum up to $2k$-th neighbors generated by nested commutators of $H_h^0$, $H_v^0$ and $H_{\pm1}$ terms, since this is a hopping mediated by the $H_v^0$ term, this will contain the quasiperiodic potential. The $H_v^0$ mediated hopping terms will have the structure of a higher order lattice derivative and depending on the $i-j$-value it flips the sign of hopping. Third: Onsite potential terms generated by the combined action of $H_{\pm1},H_v^0$ and $H_h^0$, representing particle's return to origin. This combination of $H_{\pm1},H_v^0$ and $H_h^0$ proccesses also sets the total power associated with hopping amplitudes $\mathcal{Z}=\tilde{p}+\tilde{q}$, which further decides the order of lattice derivative, while remaining power $\tilde{r}=2\tilde{m}+1-\mathcal{Z}$ represents products of cosine harmonics.
Considering all these processes, we construct an effective Floquet Hamiltonian at an arbitrary order of perturbation theory as following,

\begin{equation}
      H_{eff}=\sum_{\substack{i,j}}\Big(\delta_{|i-j|,0}\lambda \cos(2\pi\beta j)+\delta_{|i-j|,1}\gamma+R_{i,j}^{2k}\Big)c^\dagger_ic_j+h.c..
      \label{VV_nth_order_1}
\end{equation}
where
\begin{eqnarray}
R^{2k}_{i,j} &=& \sum_{\substack{
    \tilde{m}=1,2,3 \\ 
    |i-j|\in I_{\tilde{m}}
}}^{k} \left(\frac{1}{\omega}\right)^{2\tilde{m}}
\Bigg[  \sum_{\tilde{p}=2,4,\ldots}^{2\tilde{m}} \chi\Big(\frac{a}{2}\Big)^{\tilde{p}} \\ \nonumber
& \times& \sum_{\substack{
    \tilde{q} = |i-j|-\tilde{p} + 2\tilde{l} \\ 
    \tilde{l}= 0,1,2,\ldots \\
    \tilde{q} \geq 0,\, \tilde{r} = 2\tilde{m}+1-\tilde{p}-\tilde{q}
}}^{2\tilde{m}+1-\tilde{p}} 
(\lambda f)^{\tilde{r}} \gamma^{\tilde{q}} \zeta_{\tilde{m},\tilde{p},\tilde{q},\tilde{r},|i-j|} \Bigg],
\label{R_term}
\end{eqnarray}
$I_1 = [0,3], I_2 = [0,5]$, $I_3 = [0,7]\dots$. Here $\zeta_{\tilde{m},\tilde{p},\tilde{q},|i-j|}=(-1)^{\frac{|i-j|+1}{2}}$ when  $(\tilde{p},\tilde{q})=(2\tilde{m},1)$, otherise $\zeta_{\tilde{m},\tilde{p},\tilde{q},|i-j|}\in\{\pm1\}$, taken randomly to mimic the sign flips for $H_{-}$ mediated processes, however we note that this sign is not very important given that it multiplies an oscillating trigonometric factor.  To match with Eq. \ref{vv_2_eqn}  the prefactor $\chi$ is chosen as $\chi$=4 for $|i-j|$=1, 3 at $(\tilde{p},\tilde{q})=(2\tilde{m},1)$, $\chi=2$ for $|i-j|$=0 and $\chi$=1 otherwise. $f=\partial^{\mathcal{Z}} f(i,j)=
\sum_{\tilde{r}=0}^{\mathcal{Z}}(-1)^{\tilde{r}} \binom{\mathcal{Z}}{\tilde{r}}
\cos\!\left(2\pi\beta\!\left(c + \tilde{r} - \tfrac{\mathcal{Z}}{2}\right)\right)$, where $c=\dfrac{i+j}{2}$. The closed-form analytical expression of $R^{2k}_{i,j}$ (Eq. \ref{R_term}) is obtained by a heuristic extrapolation of the lower order terms.

Here each real-space term $R_{i,j}^{(2k)}$ is a sum over virtual processes in the $H_0$ basis and hence equals a sum of products of matrix elements divided by detunings. Resonant divergences occur when one (or a product) of these denominators becomes anomalously small; thus the perturbation theory can only be used away from these resonances i.e. when the drive frequency matches the difference between unperturbed Hamiltonian eigenenergies. At lower values of $\lambda$, the spectrum becomes continuous, leading to a complete breakdown of the perturbation theory throughout the spectrum due to resonance proliferation. We determine the optimal order truncation of the superasymptotic VVPT by ratio test, $\overline{R}^{2k+2}/\overline{R}^{2k}<1$, where $\overline{R}^{2k}=\langle |R_{i,j}^{2k}| \rangle_{ij} $\cite{ratio_1}. Despite the resonances, the ratio test remains well-behaved; we also discuss the other norms \cite{ratio_1} in the Sec:3 of supplementary material. 

In Figs. \ref{VV_IPR_sw} and \ref{VV_IPR_osw}, we plot the IPR scaling at the optimal order of truncation of VVPT at $(a,\omega)=(1,1.5)$ and $(5,7.1)$, which is at the universal localization and away from it, respectively. The ratio test results for these two cases are shown in the supplementary section 3. From the IPR scaling, we find that more states are localized in the universal localization region than away from it. The difference is more clear from $\lambda=1.5$ and $\lambda=2$, in Figs.  \ref{VV_IPR_sw} and Fig. \ref{VV_IPR_osw}, where all eigenstates are delocalized and critical states without drive, respectively, and also in the case of GAA with mobility edge. Note that all the cases have resonant energy states in the spectrum, these are not the mobility edges, these resonances increase more as the spectrum gets continuous at lower $\lambda$ and the perturbation theory fails to explain the localization (Fig. \ref{exact_scaling_aa}). 

\paragraph*{\textbf{Conclusion}}

In summary, we have demonstrated the existence of a robust, non-perturbative localization phenomenon in periodically driven quasiperiodic lattices. Using exact Floquet dynamics, Floquet perturbation theory and Van-Vleck (VV) asymptotic expansions, we identified a region in the amplitude–frequency plane at which all Floquet eigenstates become localized irrespective of the properties of the undriven Hamiltonian. 
Although the Van-Vleck expansion is divergent, its optimally truncated form accurately captures the non-resonant part of the Floquet spectrum. The resulting agreement with exact results suggests that a fundamentally non-perturbative localization mechanism leaves an imprint on a divergent perturbative series. We further showed that the eventual breakdown of the superasymptotic VV series originates from rare, large resonant hybridizations that are invisible to global ratio tests but dominate the physical long-range couplings. This dichotomy between bulk asymptotic behavior and sparse resonances provides a single-particle analog of the rare-resonance–induced collapse of the Floquet-Magnus expansion known in many-body systems\cite{Bukov,mori_trunc3,tree_magnus_2,Abanin_optimal,Haga}.

The non-perturbative universality of the localization plateau persisting across quasiperiodic potentials, drive protocols, and parameter ranges suggests that this phenomenon is not restricted to only quasiperiodic model and is likely to extend to higher dimensions and even disordered lattices\cite{future}.

Since the mechanism of universal localization relies only on (i) a staggered or sublattice-asymmetric drive and (ii) a bounded on-site modulation, it is directly realizable in multiple experimental platforms. In ultracold atoms, driven Aubry–André and generalized quasiperiodic lattices have been implemented with high control \cite{aa_expt_bec,aa_expt_1}, while periodically modulated tunneling can be engineered via lattice shaking or Raman-assisted tunneling \cite{goldman_dalibard}. Photonic waveguide arrays provide an alternative platform where longitudinal modulation implements Floquet driving and quasiperiodic refractive-index patterns realize the AA potential \cite{aa_expt_photonic}. All these platforms can measure participation ratios, wave-packet spreading, and energy absorption, making the predicted localization plateau immediately testable.

Our results therefore establish emergent non-perturbative Floquet localization as a physically robust phenomenon, detectable in existing experiments with imprints captured by superasymptotically truncated perturbation theory. This interplay between rare resonances, asymptotic expansions, and exact localization opens a new direction for exploring controllable non-perturbative phases in driven quantum systems.

\paragraph*{\textbf{Acknowledgments}}
S.~S. and S.~P. acknowledges fruitful discussions with Krishnendu Sengupta and Arnab Das. All authors acknowledge early collaborations with Hrithik Nandy. S.~S. acknowledges support from the Anusandhan National Research Foundation (Department of Science and Technology) Govt. of India, under grant no. CRG/2023/007457 and an internal grant from the Indian Institute of Science Education and Research, Tirupati. A. R.  acknowledges support from IIT Hyderabad, India through the Seed Grant SG/IITH/F337/2023-24/SG-175, and  ANRF (DST), Govt. of India under the grant no. ANRF/IRG/2024/001653/PS.
\bibliography{master/FPT}

@article{off_diag_qp,
  title = {Mobility edges in off-diagonal disordered tight-binding models},
  author = {Liu, Tong and Guo, Hao},
  journal = {Phys. Rev. B},
  volume = {98},
  issue = {10},
  pages = {104201},
  numpages = {6},
  year = {2018},
  month = {Sep},
  publisher = {American Physical Society},
  doi = {10.1103/PhysRevB.98.104201},
  url = {https://link.aps.org/doi/10.1103/PhysRevB.98.104201}
}

@article{AL_ac_1,
  title = {ac-Field-Controlled Anderson Localization in Disordered Semiconductor Superlattices},
  author = {Holthaus, Martin and Ristow, Gerald H. and Hone, Daniel W.},
  journal = {Phys. Rev. Lett.},
  volume = {75},
  issue = {21},
  pages = {3914--3917},
  numpages = {0},
  year = {1995},
  month = {Nov},
  publisher = {American Physical Society},
  doi = {10.1103/PhysRevLett.75.3914},
  url = {https://link.aps.org/doi/10.1103/PhysRevLett.75.3914}
}

@article{Andersonloc,
  title = {Absence of Diffusion in Certain Random Lattices},
  author = {Anderson, P. W.},
  journal = {Phys. Rev.},
  volume = {109},
  issue = {5},
  pages = {1492--1505},
  numpages = {0},
  year = {1958},
  month = {Mar},
  publisher = {American Physical Society},
  doi = {10.1103/PhysRev.109.1492},
  url = {https://link.aps.org/doi/10.1103/PhysRev.109.1492}
}

@Article{AL_drive1,
author={Ducatez, Raphael
and Huveneers, Fran{\c{c}}ois},
title={Anderson Localization for Periodically Driven Systems},
journal={Annales Henri Poincar{\'e}},
year={2017},
month={Jul},
day={01},
volume={18},
number={7},
pages={2415-2446},
issn={1424-0661},
doi={10.1007/s00023-017-0574-1},
url={https://doi.org/10.1007/s00023-017-0574-1}
}

@article{AL_drive2,
  title = {Delocalization induced by low-frequency driving in disordered tight-binding lattices},
  author = {Martinez, Dario F. and Molina, Rafael A.},
  journal = {Phys. Rev. B},
  volume = {73},
  issue = {7},
  pages = {073104},
  numpages = {4},
  year = {2006},
  month = {Feb},
  publisher = {American Physical Society},
  doi = {10.1103/PhysRevB.73.073104},
  url = {https://link.aps.org/doi/10.1103/PhysRevB.73.073104}
}

@article{DL_Kenkre,
  title = {Dynamic localization of a charged particle moving under the influence of an electric field},
  author = {Dunlap, D. H. and Kenkre, V. M.},
  journal = {Phys. Rev. B},
  volume = {34},
  issue = {6},
  pages = {3625--3633},
  numpages = {0},
  year = {1986},
  month = {Sep},
  publisher = {American Physical Society},
  doi = {10.1103/PhysRevB.34.3625},
  url = {https://link.aps.org/doi/10.1103/PhysRevB.34.3625}
}

@article{DL_expt1,
  title = {Measuring a localization phase diagram controlled by the interplay of disorder and driving},
  author = {Dotti, Peter and Bai, Yifei and Shimasaki, Toshihiko and Dardia, Anna R. and Weld, David M.},
  journal = {Phys. Rev. Res.},
  volume = {7},
  issue = {2},
  pages = {L022026},
  numpages = {6},
  year = {2025},
  month = {Apr},
  publisher = {American Physical Society},
  doi = {10.1103/PhysRevResearch.7.L022026},
  url = {https://link.aps.org/doi/10.1103/PhysRevResearch.7.L022026}
}

@article{DL_expt2,
  title = {Reversible Phasonic Control of a Quantum Phase Transition in a Quasicrystal},
  author = {Shimasaki, Toshihiko and Bai, Yifei and Kondakci, H. Esat and Dotti, Peter and Pagett, Jared E. and Dardia, Anna R. and Prichard, Max and Eckardt, Andr\'e and Weld, David M.},
  journal = {Phys. Rev. Lett.},
  volume = {133},
  issue = {8},
  pages = {083405},
  numpages = {6},
  year = {2024},
  month = {Aug},
  publisher = {American Physical Society},
  doi = {10.1103/PhysRevLett.133.083405},
  url = {https://link.aps.org/doi/10.1103/PhysRevLett.133.083405}
}

@Article{Al_expt1,
author={Billy, Juliette
and Josse, Vincent
and Zuo, Zhanchun
and Bernard, Alain
and Hambrecht, Ben
and Lugan, Pierre
and Cl{\'e}ment, David
and Sanchez-Palencia, Laurent
and Bouyer, Philippe
and Aspect, Alain},
title={Direct observation of Anderson localization of matter waves in a controlled disorder},
journal={Nature},
year={2008},
month={Jun},
day={01},
volume={453},
number={7197},
pages={891-894},
issn={1476-4687},
doi={10.1038/nature07000},
url={https://doi.org/10.1038/nature07000}
}

@Article{AL_expt2,
author={Semeghini, G.
and Landini, M.
and Castilho, P.
and Roy, S.
and Spagnolli, G.
and Trenkwalder, A.
and Fattori, M.
and Inguscio, M.
and Modugno, G.},
title={Measurement of the mobility edge for 3D Anderson localization},
journal={Nature Physics},
year={2015},
month={Jul},
day={01},
volume={11},
number={7},
pages={554-559},
issn={1745-2481},
doi={10.1038/nphys3339},
url={https://doi.org/10.1038/nphys3339}
}

@article{SS_AA,
  title = {Emergent scale and anomalous dynamics in certain nearest-neighbor quasiperiodic tight binding models in one dimension},
  author = {Nair, Parvathy S. and Joy, Dintomon and Deb, Rohit and Pakrashi, Soumadip and Sanyal, Sambuddha},
  journal = {Phys. Rev. B},
  volume = {112},
  issue = {10},
  pages = {104203},
  numpages = {8},
  year = {2025},
  month = {Sep},
  publisher = {American Physical Society},
  doi = {10.1103/5t3z-f6bj},
  url = {https://link.aps.org/doi/10.1103/5t3z-f6bj}
}

@article{
AL_meso1,
author = {Anthony Richardella  and Pedram Roushan  and Shawn Mack  and Brian Zhou  and David A. Huse  and David D. Awschalom  and Ali Yazdani },
title = {Visualizing Critical Correlations Near the Metal-Insulator Transition in Ga$_{1-x}$ Mn$_x$As},
journal = {Science},
volume = {327},
number = {5966},
pages = {665-669},
year = {2010},
doi = {10.1126/science.1183640},
URL = {https://www.science.org/doi/abs/10.1126/science.1183640},
eprint = {https://www.science.org/doi/pdf/10.1126/science.1183640}}

@article{Aditya,
  title = {Periodically driven model with quasiperiodic potential and staggered hopping amplitudes: Engineering of mobility gaps and multifractal states},
  author = {Aditya, Sreemayee and Sengupta, K. and Sen, Diptiman},
  journal = {Phys. Rev. B},
  volume = {107},
  issue = {3},
  pages = {035402},
  numpages = {14},
  year = {2023},
  month = {Jan},
  publisher = {American Physical Society},
  doi = {10.1103/PhysRevB.107.035402},
  url = {https://link.aps.org/doi/10.1103/PhysRevB.107.035402}
}

@article{Haga,
  title = {Divergence of the Floquet-Magnus expansion in a periodically driven one-body system with energy localization},
  author = {Haga, Taiki},
  journal = {Phys. Rev. E},
  volume = {100},
  issue = {6},
  pages = {062138},
  numpages = {12},
  year = {2019},
  month = {Dec},
  publisher = {American Physical Society},
  doi = {10.1103/PhysRevE.100.062138},
  url = {https://link.aps.org/doi/10.1103/PhysRevE.100.062138}
}

@article{future,
  author = "Pakrashi, Soumadip and Rajak, Atanu and Sanyal, Sambuddha",
  title = "",
  journal = "",
  year = "in preparation",
  note = ""
}

@article{Abanin_optimal,
  title = {Effective Hamiltonians, prethermalization, and slow energy absorption in periodically driven many-body systems},
  author = {Abanin, Dmitry A. and De Roeck, Wojciech and Ho, Wen Wei and Huveneers, Fran\ifmmode \mbox{\c{c}}\else \c{c}\fi{}ois},
  journal = {Phys. Rev. B},
  volume = {95},
  issue = {1},
  pages = {014112},
  numpages = {8},
  year = {2017},
  month = {Jan},
  publisher = {American Physical Society},
  doi = {10.1103/PhysRevB.95.014112},
  url = {https://link.aps.org/doi/10.1103/PhysRevB.95.014112}
}

@article{aa1,
  title = {Analyticity breaking and Anderson localization in incommensurate lattices},
  author = {S. Aubry and G. Andre},
  journal = {Ann. Israel Phys. Soc},
  volume = {3},
  issue = {},
  pages = {18},
  numpages = {},
  year = {1980},
  month = {}
}

@article{aa_expt_spme1,
  title = {Observation of Many-Body Localization in a One-Dimensional System with a Single-Particle Mobility Edge},
  author = {Kohlert, Thomas and Scherg, Sebastian and Li, Xiao and L\"uschen, Henrik P. and Das Sarma, Sankar and Bloch, Immanuel and Aidelsburger, Monika},
  journal = {Phys. Rev. Lett.},
  volume = {122},
  issue = {17},
  pages = {170403},
  numpages = {7},
  year = {2019},
  month = {May},
  publisher = {American Physical Society},
  doi = {10.1103/PhysRevLett.122.170403},
  url = {https://link.aps.org/doi/10.1103/PhysRevLett.122.170403}
}

@article{aa_expt_1,
author = {Michael Schreiber  and Sean S. Hodgman  and Pranjal Bordia  and Henrik P. Lüschen  and Mark H. Fischer  and Ronen Vosk  and Ehud Altman  and Ulrich Schneider  and Immanuel Bloch },
title = {Observation of many-body localization of interacting fermions in a quasirandom optical lattice},
journal = {Science},
volume = {349},
number = {6250},
pages = {842-845},
year = {2015},
doi = {10.1126/science.aaa7432},
URL = {https://www.science.org/doi/abs/10.1126/science.aaa7432},
eprint = {https://www.science.org/doi/pdf/10.1126/science.aaa7432}}

@article{aa_expt_me,
  title = {Observation of Interaction-Induced Mobility Edge in an Atomic Aubry-Andr\'e Wire},
  author = {Wang, Yunfei and Zhang, Jia-Hui and Li, Yuqing and Wu, Jizhou and Liu, Wenliang and Mei, Feng and Hu, Ying and Xiao, Liantuan and Ma, Jie and Chin, Cheng and Jia, Suotang},
  journal = {Phys. Rev. Lett.},
  volume = {129},
  issue = {10},
  pages = {103401},
  numpages = {7},
  year = {2022},
  month = {Sep},
  publisher = {American Physical Society},
  doi = {10.1103/PhysRevLett.129.103401},
  url = {https://link.aps.org/doi/10.1103/PhysRevLett.129.103401}
}

@article{FPT1,
doi = {10.1088/1367-2630/aade37},
url = {https://doi.org/10.1088/1367-2630/aade37},
year = {2018},
month = {sep},
publisher = {IOP Publishing},
volume = {20},
number = {9},
pages = {093022},
author = {Rodriguez-Vega, M and Lentz, M and Seradjeh, B},
title = {Floquet perturbation theory: formalism and application to low-frequency limit},
journal = {New Journal of Physics}
}

@article{VV_Eckardt,
doi = {10.1088/1367-2630/17/9/093039},
url = {https://doi.org/10.1088/1367-2630/17/9/093039},
year = {2015},
month = {sep},
publisher = {IOP Publishing},
volume = {17},
number = {9},
pages = {093039},
author = {Eckardt, André and Anisimovas, Egidijus},
title = {High-frequency approximation for periodically driven quantum systems from a Floquet-space perspective},
journal = {New Journal of Physics}
}

@Article{aa_expt_bec,
author={Roati, Giacomo
and D'Errico, Chiara
and Fallani, Leonardo
and Fattori, Marco
and Fort, Chiara
and Zaccanti, Matteo
and Modugno, Giovanni
and Modugno, Michele
and Inguscio, Massimo},
title={Anderson localization of a non-interacting Bose--Einstein condensate},
journal={Nature},
year={2008},
month={Jun},
day={01},
volume={453},
number={7197},
pages={895-898},
issn={1476-4687},
doi={10.1038/nature07071},
url={https://doi.org/10.1038/nature07071}
}

@article{tree_magnus_1,
  title={On the solution of linear differential equations in Lie groups},
  author={Iserles, Arieh and N{\o}rsett, Syvert P},
  journal={Philosophical Transactions of the Royal Society of London. Series A: Mathematical, Physical and Engineering Sciences},
  volume={357},
  number={1754},
  pages={983--1019},
  year={1999},
  publisher={The Royal Society}
}

@article{tree_magnus_2,
  title={The Magnus expansion and some of its applications},
  author={Blanes, Sergio and Casas, Fernando and Oteo, Jose-Angel and Ros, Jos{\'e}},
  journal={Physics reports},
  volume={470},
  number={5-6},
  pages={151--238},
  year={2009},
  publisher={Elsevier}
}

@article{recursive_1,
  title = {Recursive generation of higher-order terms in the Magnus expansion},
  author = {Klarsfeld, S. and Oteo, J. A.},
  journal = {Phys. Rev. A},
  volume = {39},
  issue = {7},
  pages = {3270--3273},
  numpages = {0},
  year = {1989},
  month = {Apr},
  publisher = {American Physical Society},
  doi = {10.1103/PhysRevA.39.3270},
  url = {https://link.aps.org/doi/10.1103/PhysRevA.39.3270}
}

@article{Bukov,
author = {Marin Bukov and Luca D'Alessio and Anatoli Polkovnikov},
title = {Universal high-frequency behavior of periodically driven systems: from dynamical stabilization to Floquet engineering},
journal = {Advances in Physics},
volume = {64},
number = {2},
pages = {139--226},
year = {2015},
publisher = {Taylor \& Francis},
doi = {10.1080/00018732.2015.1055918},


URL = { 
    
        https://doi.org/10.1080/00018732.2015.1055918
    
    

},
eprint = { 
    
        https://doi.org/10.1080/00018732.2015.1055918
    
    

}

}

@article{goldman_dalibard,
  title = {Periodically Driven Quantum Systems: Effective Hamiltonians and Engineered Gauge Fields},
  author = {Goldman, N. and Dalibard, J.},
  journal = {Phys. Rev. X},
  volume = {4},
  issue = {3},
  pages = {031027},
  numpages = {29},
  year = {2014},
  month = {Aug},
  publisher = {American Physical Society},
  doi = {10.1103/PhysRevX.4.031027},
  url = {https://link.aps.org/doi/10.1103/PhysRevX.4.031027}
}

@article{aa_expt_photonic,
  title = {Observation of a Localization Transition in Quasiperiodic Photonic Lattices},
  author = {Lahini, Y. and Pugatch, R. and Pozzi, F. and Sorel, M. and Morandotti, R. and Davidson, N. and Silberberg, Y.},
  journal = {Phys. Rev. Lett.},
  volume = {103},
  issue = {1},
  pages = {013901},
  numpages = {4},
  year = {2009},
  month = {Jun},
  publisher = {American Physical Society},
  doi = {10.1103/PhysRevLett.103.013901},
  url = {https://link.aps.org/doi/10.1103/PhysRevLett.103.013901}
}

@article{mori_trunc2,
  title = {Rigorous Bound on Energy Absorption and Generic Relaxation in Periodically Driven Quantum Systems},
  author = {Mori, Takashi and Kuwahara, Tomotaka and Saito, Keiji},
  journal = {Phys. Rev. Lett.},
  volume = {116},
  issue = {12},
  pages = {120401},
  numpages = {5},
  year = {2016},
  month = {Mar},
  publisher = {American Physical Society},
  doi = {10.1103/PhysRevLett.116.120401},
  url = {https://link.aps.org/doi/10.1103/PhysRevLett.116.120401}
}

@article{mori_trunc3,
title = {Floquet–Magnus theory and generic transient dynamics in periodically driven many-body quantum systems},
journal = {Annals of Physics},
volume = {367},
pages = {96-124},
year = {2016},
issn = {0003-4916},
doi = {https://doi.org/10.1016/j.aop.2016.01.012},
url = {https://www.sciencedirect.com/science/article/pii/S0003491616000142},
author = {Tomotaka Kuwahara and Takashi Mori and Keiji Saito}
}

@article{ratio_1,
  title = {Application of average Hamiltonian theory to the NMR of solids},
  author = {Maricq, M. Matti},
  journal = {Phys. Rev. B},
  volume = {25},
  issue = {11},
  pages = {6622--6632},
  numpages = {0},
  year = {1982},
  month = {Jun},
  publisher = {American Physical Society},
  doi = {10.1103/PhysRevB.25.6622},
  url = {https://link.aps.org/doi/10.1103/PhysRevB.25.6622}
}

\section*{End Matter}
\section{Exact method}
We divide the time period $T$ into $N$ Trotter steps, where $N$ is chosen to be large enough such that $H(t)$ does not change significantly between $t_s$ and $t_{s + T/N}$ for any time instant $t_s$. Then we construct the stroboscopic time evolution operator  $U(T,0)=\prod_{s=1}^N\sum_ne^{-i\epsilon_n^s\frac{T}{N}}|\psi_n^s\rangle\langle \psi_n^s|$ using the Suzuki-Trotter decomposition, where $\epsilon_n^s$ and $|\psi_n^s\rangle$ are the eigenvalues and eigenvectors of the instantaneous Hamiltonian $H(t_s)$ on a lattice of size $L$, $H(t_s)|\psi_n^s\rangle=\epsilon_n^s|\psi_n^s\rangle$. Finally, we obtain the stroboscopic Floquet eigenstates, $|\chi_m\rangle$ and eigenvalues, $E^F_m$ by diagonalizing the unitary matrix $U(T,0)$, which satisfies $U(T,0)|\chi_m\rangle = e^{-i E^F_m t}|\chi_m\rangle$. The corresponding Floquet Hamiltonian is given by $H_F=\sum_m E_m^F |\chi_m\rangle \langle \chi_m|. $
\section{Floquet Perturbation Theory}
In FPT one considers the time-dependent part of the Hamiltonian as the unperturbed Hamiltonian and the static part of the Hamiltonian as the perturbation. The central idea is to construct the Floquet states in the presence of weak static perturbation from the known unperturbed floquet states by applying the time-dependent perturbation theory. The perturbation theory converges when the amplitude of the drive $a$ is larger than the amplitudes of all static terms $\gamma$, and $\lambda$.

In order to develop FPT for our model, we rewrite the Hamiltonian as follows: $H_{FPT}(t) = H_0(t) + V_1 + V_2$ and took Fourier transformation which takes $c_{i,A}\rightarrow a_k,c^\dagger_{i,A}\rightarrow a^\dagger_k$ and $c_{i,B}\rightarrow b_k,c^\dagger_{i,B}\rightarrow b^\dagger_k$. This yields zeroth order term in k -space as
\begin{equation}
H_0(t)=\sum_k[a \sin(\omega t)(1-e^{-2ik})a_k^\dagger b_k +  a \sin(\omega t)(1-e^{2ik})b_k^\dagger a_k],
\end{equation}
and the perturbation terms as 
\begin{equation}
V_1=\sum_k[\gamma(1+e^{-2ik})a_k^\dagger b_k \ + \ h.c],
\end{equation}
\vspace{-3mm}
and 
\vspace{-3mm}
\begin{equation}
V_2=\sum_k[f_1(k,k^\prime)a_k^\dagger a_{k^\prime} \ + \ f_1(k,k^\prime)b_k^\dagger b_{k^\prime}],
\end{equation}
where $f_1(k,k^\prime)=\sum_j\frac{2\lambda}{L}cos(2\pi\alpha(2j))e^{2i(k-k^\prime)j}$ \ and \ $f_2(k,k^\prime)=\sum_j\frac{2\lambda}{L}cos(2\pi\alpha(2j+1))e^{2i(k-k^\prime)j}$,   $k \in [\frac{-\pi}{2},\frac{\pi}{2}]$. Now, if we redefine our basis states as $|k\uparrow\rangle=a^\dagger_k|0\rangle$ and  $|k\downarrow\rangle=b^\dagger_k|0\rangle$, the
eigenstates of $H_0$ can be expressed as $|k+\rangle=\frac{1}{\sqrt{2}}(|k\uparrow\rangle - ie^{ik}|k\downarrow\rangle)$ and $|k-\rangle=\frac{1}{\sqrt{2}}(|k\uparrow\rangle + ie^{ik}|k\downarrow\rangle)$, in this basis the Hamiltonian can be written in terms of Pauli matrices. The first-order correction term  arising from$V_1$ in $|k \pm \rangle$ is given by:
\begin{equation}
H^{(1)}_{F1}=\frac{1}{2}\sum_k\psi_k^\dagger
\begin{pmatrix}
2\alpha_{2k} & \alpha_{1k}\\
\alpha^*_{1k} & -2\alpha_{2k}
\end{pmatrix}
\psi_k
\end{equation}
 Where, 
\begin{equation}
\alpha_{2k}=-2\gamma J_0(\mu_k)cos(k)sin(\mu_k)
\end{equation}
\begin{equation}
%\alpha_{1k}=\frac{1}{2}[(\gamma_1-\gamma_2)(1-e^{-2ik})+(\gamma_1+\gamma_2)J_0(\mu_k)cos(\mu_k)(1+e^{-2ik})
\alpha_{1k}=\gamma J_0(\mu_k)cos(\mu_k)(1+e^{-2ik})
\end{equation}

%\begin{equation}
%\beta_{1k}=\frac{1}{2}[(\gamma_1-\gamma_2)(1-e^{2ik})+(\gamma_1+\gamma_2)J_0(\mu_k)cos(\mu_k)(1+e^{2ik}).
%\beta_{1k}=\gamma J_0(\mu_k)cos(\mu_k)(1+e^{2ik}).    
%\end{equation}

We use a two-component operator $\psi_k=(a_k,b_k)$ and $\mu_k=\frac{4asin(k)}{\omega}$. 
%We can see from the expression itself when $\gamma_1=-\gamma_2$, $\alpha_{2k}=0$ and $H^{(1)}_{F1}$ will be independent of driving. 
Similarly, first order correction term $H_{F2}^{(1)}$ , arising from $V_2$.
%\begin{widetext}
\begin{eqnarray}
H_{F2}^{(1)}&=&\sum_k\frac{1}{2}[(X+\tilde{X}+Y+\tilde{Y})a_k^\dagger a_{k^\prime} \\ \nonumber
&+& e^{i(k-k^\prime)}(X+\tilde{X}-Y-\tilde{Y}))b_k^\dagger b_{k^\prime}\\ \nonumber
&+&ie^{-ik^\prime}(X-\tilde{X}-Y+\tilde{Y}))a_k^\dagger b_{k^\prime} 
\\ \nonumber
&-&ie^{ik}(X-\tilde{X}+Y-\tilde{Y}))b_k^\dagger a_{k^\prime}]
\end{eqnarray}
\\
%\end{widetext}
%\textcolor{RoyalBlue}{{\bf{Comment}}:- Considering Im[X] and Im[Y] contains the $i$ with it.}
Where,
$X$=$e^{i\mu^-_{kk^\prime}}J_0(\mu^-_{kk^\prime})P_+(k,k^\prime),  Y$=$e^{i\mu^+_{kk^\prime}}J_0(\mu^+_{kk^\prime})P_-(k,k^\prime),\\ \tilde{X}$=$e^{-i\mu^-_{kk^\prime}}J_0(\mu^-_{kk^\prime})P_+(k,k^\prime), \tilde{Y}$=$e^{-i\mu^+_{kk^\prime}}J_0(\mu^+_{kk^\prime})P_-(k,k^\prime)$. Where, $P_+(k,k^\prime)$=$\frac{1}{2}(f_1(k,k^\prime)+f_2(k,k^\prime)e^{i(k-k^\prime)})$  , $P_-(k,k^\prime)$=$\frac{1}{2}(f_1(k,k^\prime)-f_2(k,k^\prime)e^{i(k-k^\prime)})$ ,  $\mu^+_{kk^\prime}=\frac{\mu_k+\mu_{k^\prime}}{2}$ and  $\mu^-_{kk^\prime}=\frac{\mu_k-\mu_{k^\prime}}{2}$. Finally, we perform Inverse Fourier Transformation of $H_F=H_{F1}^{(1)}+H_{F2}^{(1)}$ to obtain approximated real space Hamiltonian.  
%It is important to note that eigenvalues of effective Hamiltonian may suffer from branch-cut ambiguity if driving frequencies are very low.

\onecolumngrid
\appendix
\newpage
\section*{Supplementary Material}

 \subsection{Analysis by Exact Method}
\subsubsection{Points outside the universal localization region}
    Using an exact numerical approach, we study few points away from the universal localization region for $\lambda=1$ of the AA model, where the static model has completely delocalized spectrum. We plot the probability profile (light cone) $p(nT,j)=|\psi_{nT}(j)|^2$ and the RMS spread of the wave packet $\sigma(n)=\sqrt{\sum_{j}(j-j_{0})^2|\psi_{nT}(j)|^2}$ in Figs. \ref{cone_a_0.1_om_2.45}, \ref{cone_a_3.5_om_2.45} and \ref{cone_a_2.15_om_3.5} for $(a,\omega)=(0.1,2.45), (3.5,2.45)$ and $(2.15,3.5)$ respectively. For each of these points we find signatures of ballistic transport, where $p(nT,j)$ rapidly spreads across the system and reaches the boundary within $\sim100n-200n$ cycles. This signature of early time boundary-hitting is consistent with $\sigma \ vs. \ n$ plots in \ref{sigma_vs_n}, showing $\eta\approx1$, where $\sigma \sim n^{\eta}$.
     
\subsubsection{ Weak disorder cases}
\begin{figure*}[b]
\centering
\subfigure[]{
\includegraphics[width=0.24\linewidth]{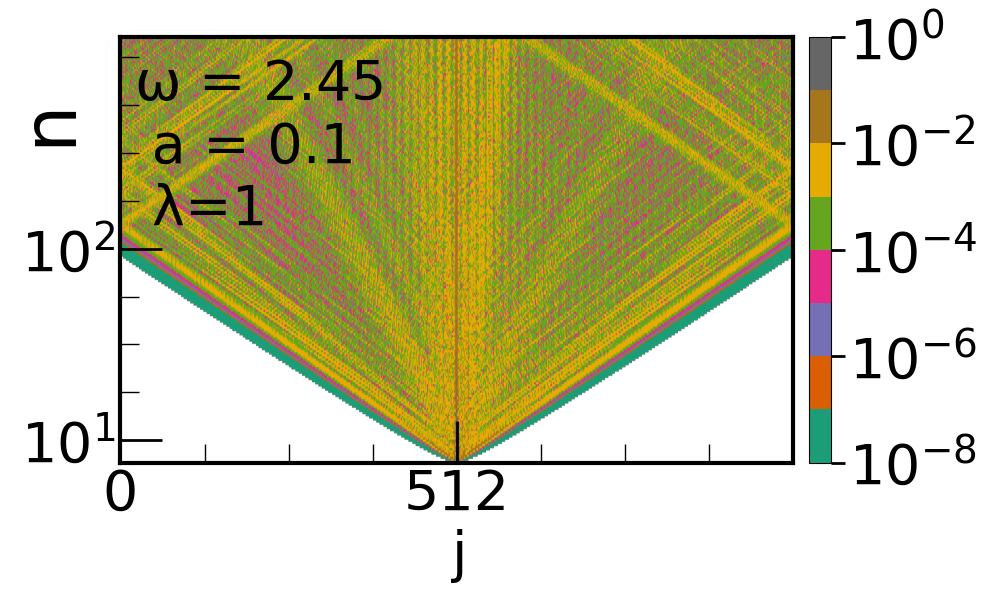}\label{cone_a_0.1_om_2.45}}
\hspace{-0.38cm}
\subfigure[]{
\includegraphics[width=0.24\linewidth]{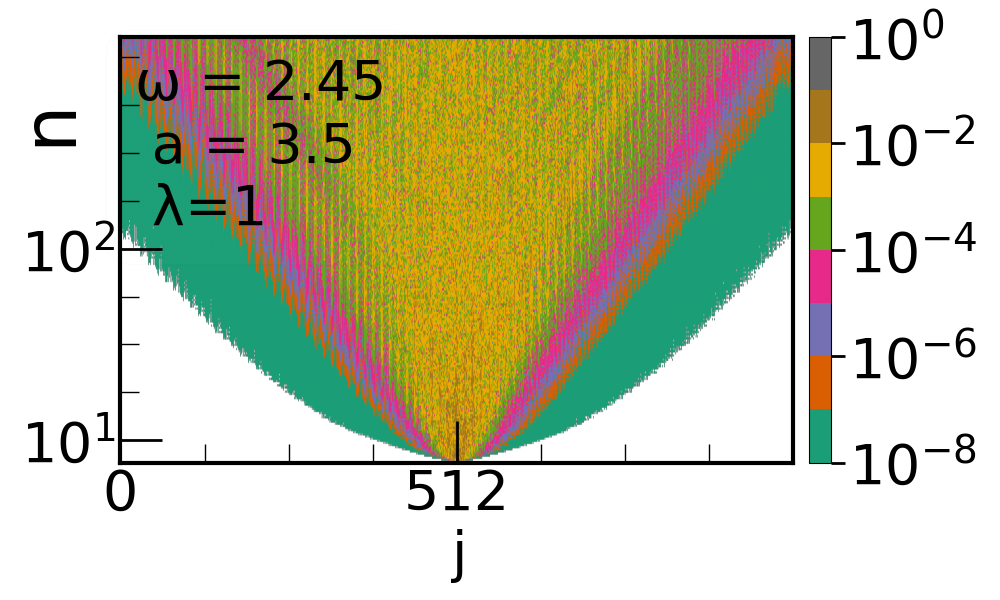}
\label{cone_a_3.5_om_2.45}}
\subfigure[]{
\includegraphics[width=0.24\linewidth]{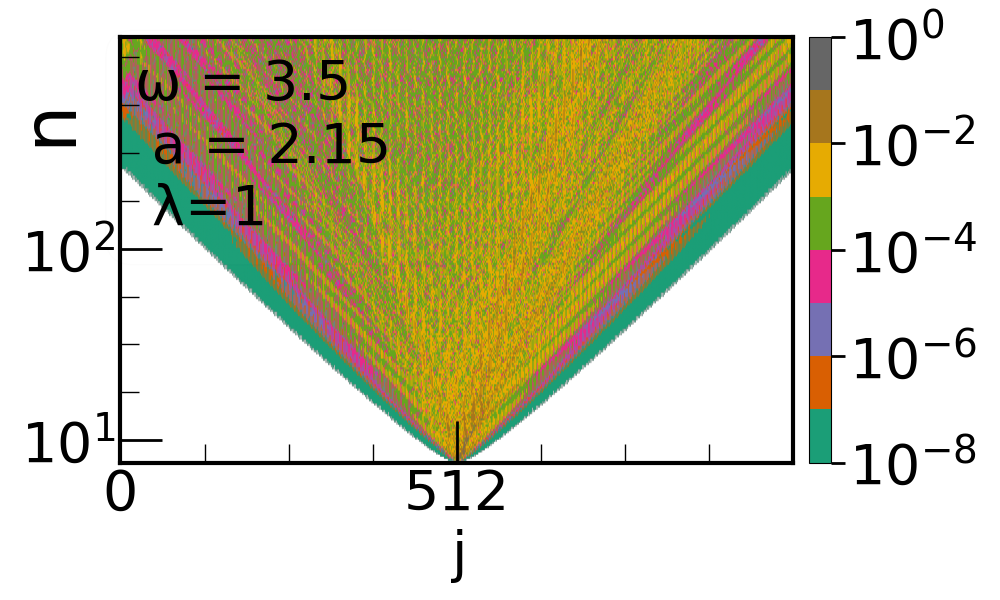}
\label{cone_a_2.15_om_3.5}}
\hspace{-0.35cm}
\subfigure[]{
\includegraphics[width=0.24\linewidth]{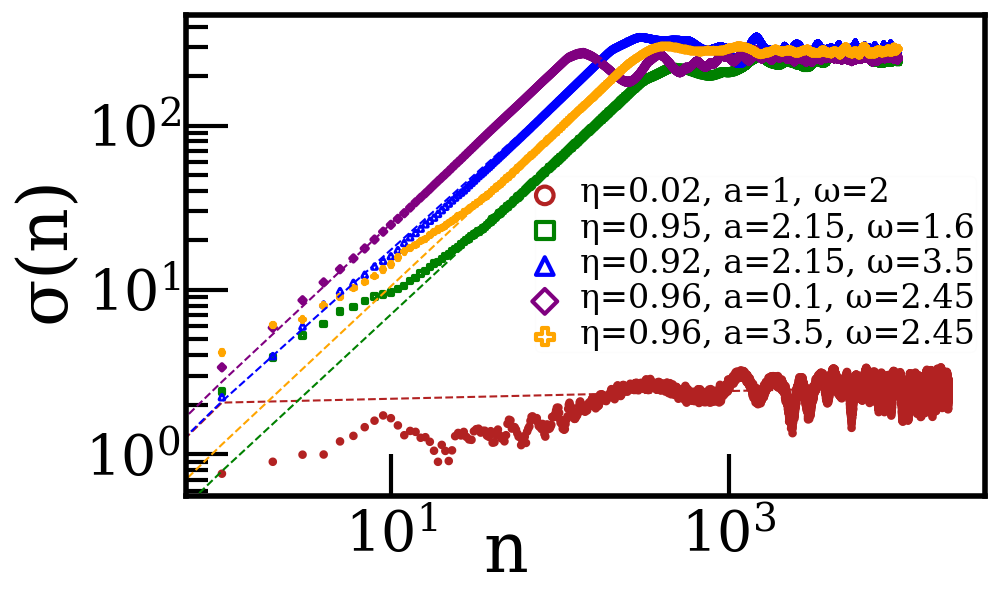}
\label{sigma_vs_n}}\\
%%%%%%%%%%%%%%%%%%%%%%%%%%%%%%%%%%%%%%%%%%%%%%%%%%%%%%%%%%%%%%%%%%%%%%
\subfigure[]{
\includegraphics[width=0.24\linewidth]{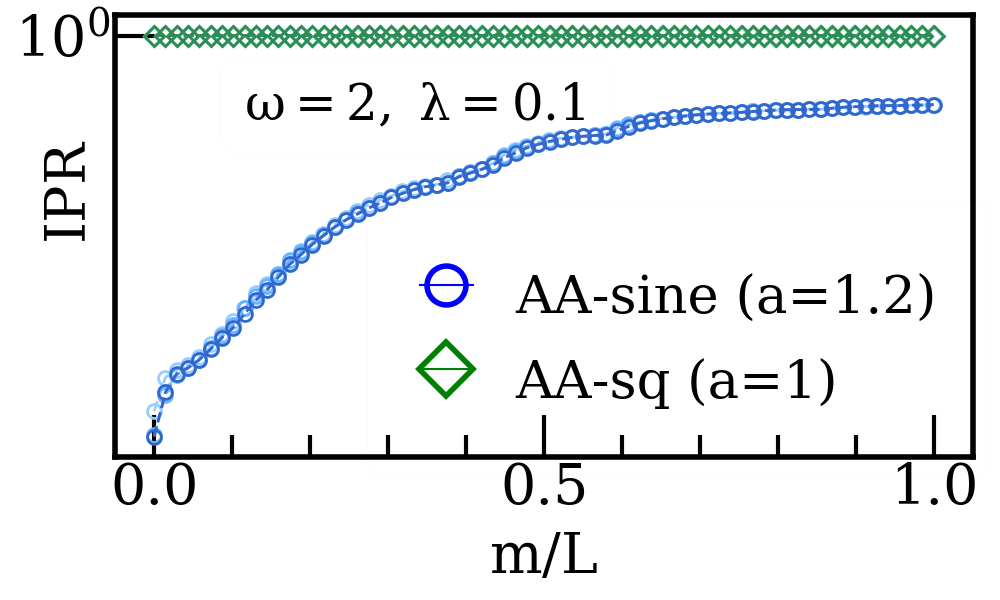}
\label{V_0.1_loc_scaling}}
\hspace{-0.38cm}
\subfigure[]{
\includegraphics[width=0.24\linewidth]{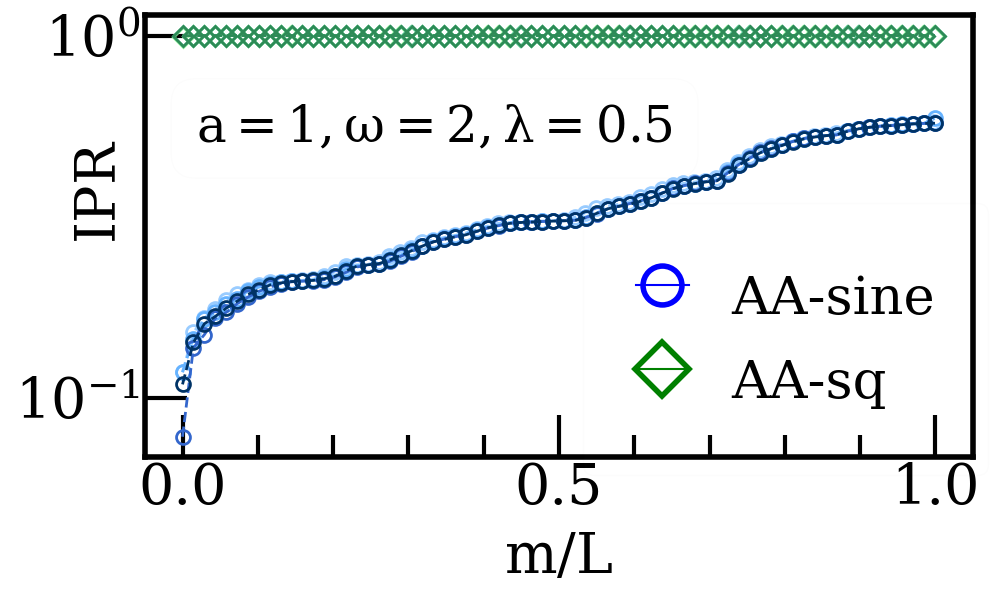}
\label{V_0.5_loc_scaling}}
\subfigure[]{
\includegraphics[width=0.24\linewidth]{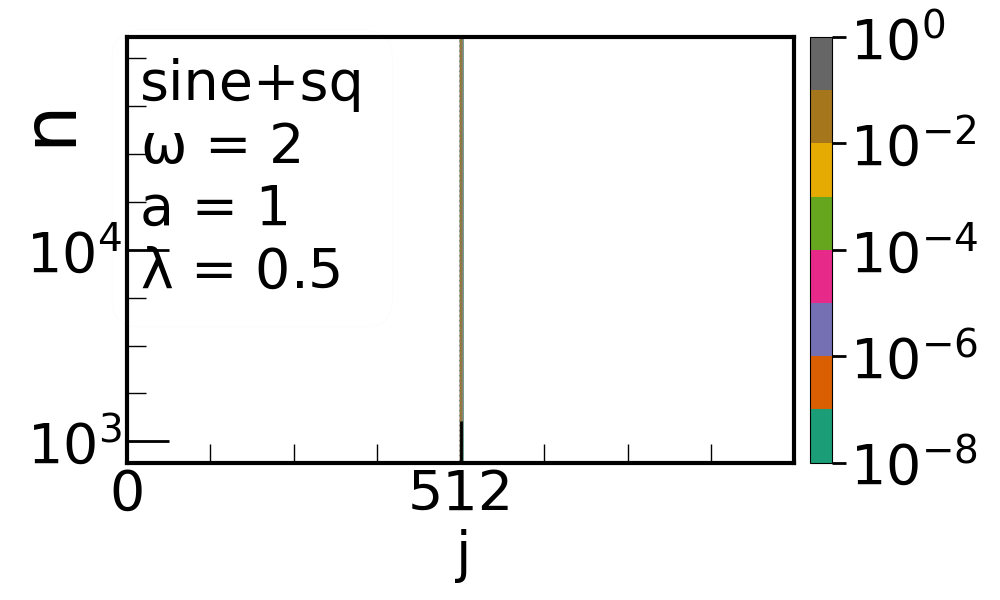}
\label{V_0.5_loc_light_cone_sine}}
\hspace{-0.35cm}
\subfigure[]{
\includegraphics[width=0.24\linewidth]{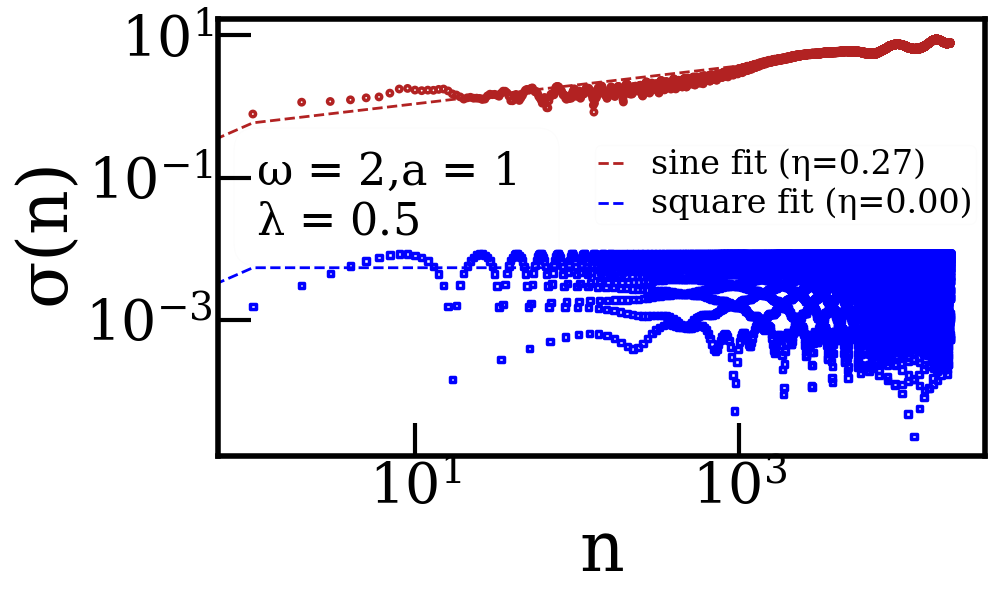}
\label{V_0.5_sigma_sine}}
\captionsetup{justification=centerlast, width=\linewidth} 
\caption{{\bf{a)-c)}} Rapid wave packet spreading observed for the points lying outside universal localization $(a,\omega)=(0.1,2.45),(3.5,2.45)$ and $(2.15,3.5)$ at $\lambda=1$, indicating ballistic transport with scaling exponent $\eta\approx1$ as quantified in {\bf{d)}}. Here, $L=1024$, $N=1000$ and sine drive used for the analysis.  {\bf{e)}}  IPR scaling plotted for $\lambda=0.1$, demonstrating drive-induced universal localization for weak disorder. Plots {\bf{f)-g)}} illustrate localized scaling in the form of IPR and slow wave packet evolution in the form of $\sigma(n)$ and wave packet profile using geometric mean of square and sine drive, proving the existence of universal localization. $L=500,1000,1500,2000$ is used for scaling and $L=1024$ for transport.}
\end{figure*}

     In Figs. \ref{V_0.1_loc_scaling} and \ref{V_0.5_loc_scaling} we study $\lambda \rightarrow 0$, here we plot the state resolved IPR for different drive protocol and show that the system remains localized, as $\mathcal{O}(1)$ IPR scaling persists even in weak disorder strength $\lambda=0.5,0.1$. This demonstrates how the universal localization arises as the interplay between disorder and drive, we also find the universal localization phenomena to complete disappear at $\lambda=0$. However, the universal localization region in Fig. \ref{exact_pd} effectively collapses to a point if $\lambda$ is decreased significantly. Nevertheless, the slow transport behavior of localized regime is clearly noticeable even at such weak disorder $\lambda=0.5$. Figs. \ref{V_0.5_loc_light_cone_sine} and \ref{V_0.5_sigma_sine} demonstrate the slow transport.   
 %======================================================
 %===================plots for exact numeics =======================================================
 %======================================================
 \subsection{Analysis by Floquet Perturbation Theory}
\begin{figure*}
\centering
\subfigure[]{
\includegraphics[width=0.24\linewidth]{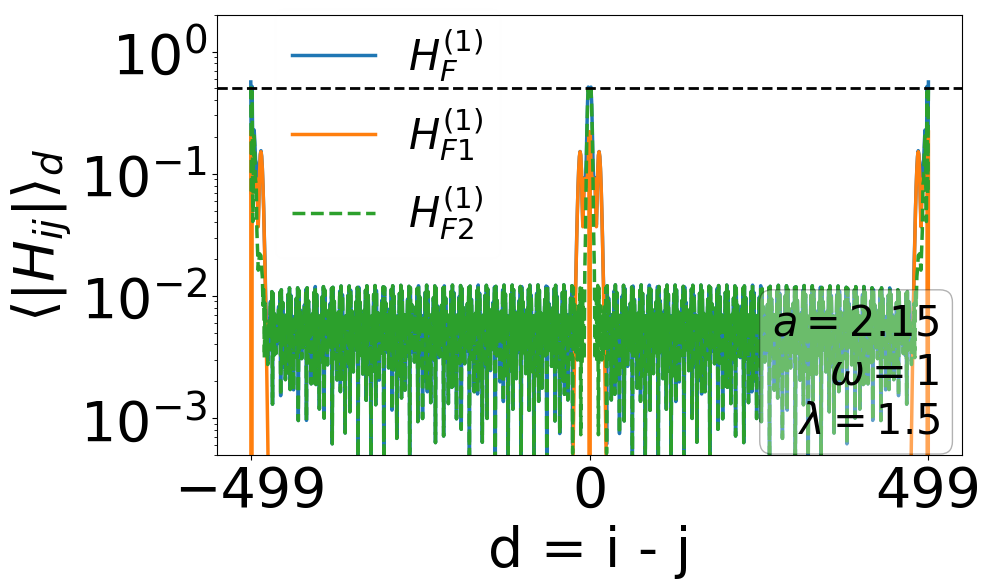}\label{fpt_HF_om_1_a_2.15}}
\hspace{-0.38cm}
\subfigure[]{
\includegraphics[width=0.24\linewidth]{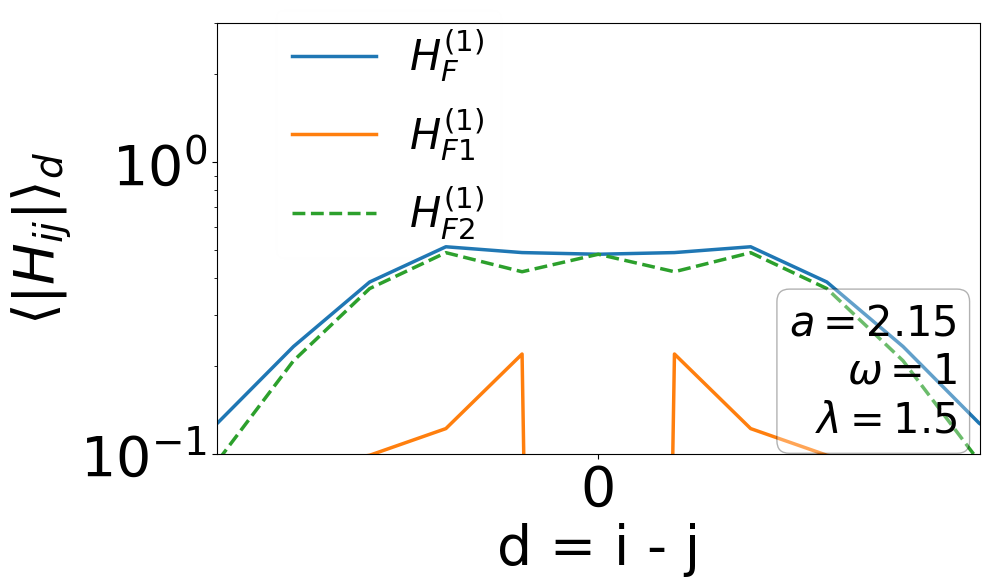}
\label{fpt_HF_om_1_a_2.15_zoom}}
\subfigure[]{
\includegraphics[width=0.24\linewidth]{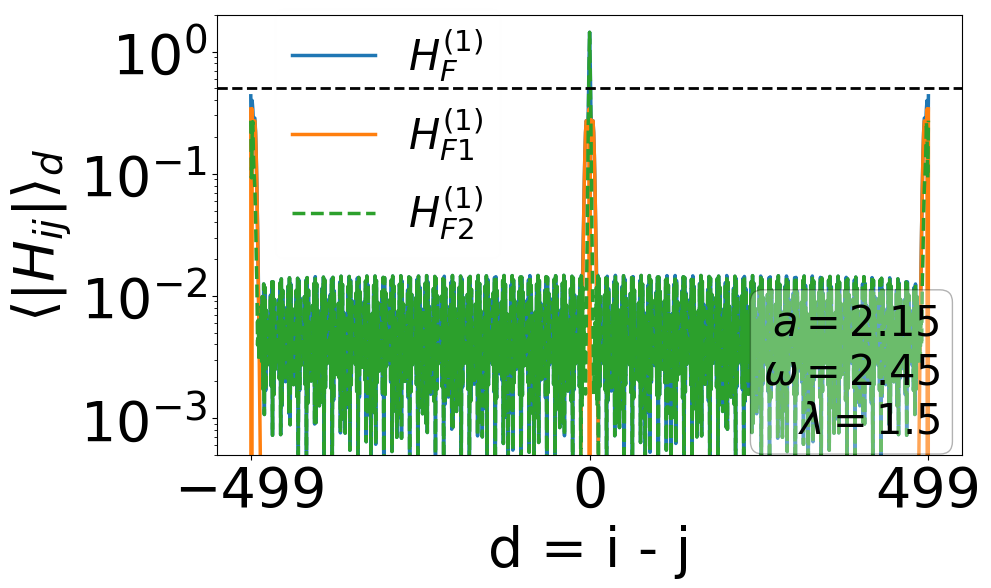}
\label{fpt_HF_om_2.45_a_2.15}}
\hspace{-0.35cm}
\subfigure[]{
\includegraphics[width=0.24\linewidth]{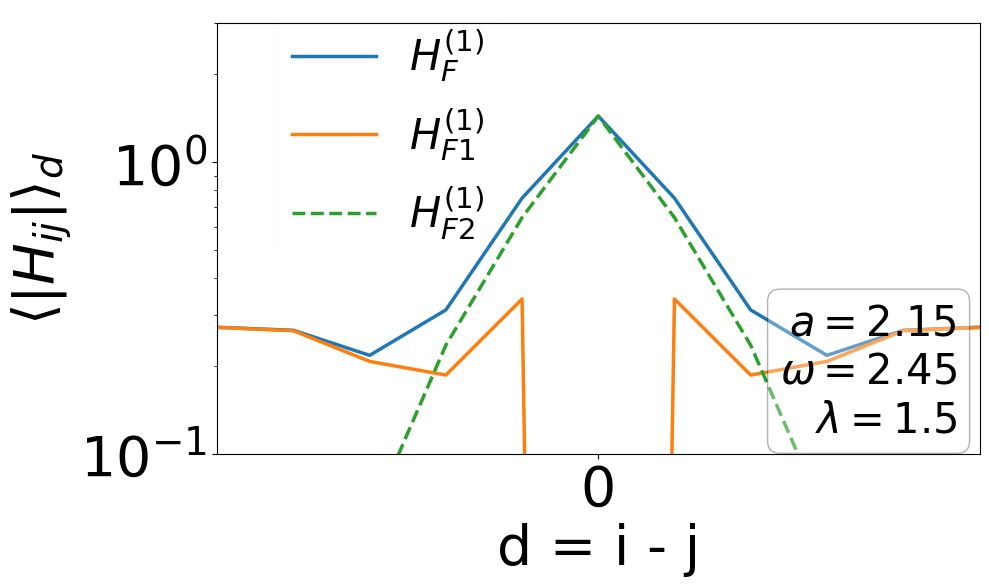}
\label{fpt_HF_om_2.45_a_2.15_zoom}}
\captionsetup{justification=centerlast, width=\linewidth} 
\caption{{\bf{a),b)}} The combined effect of $H_{F1}^{(1)}$ and $H_{F2}^{(1)}$ decreases the diagonal weight relative to off-diagonals at $(a,\omega)=(2.15,1)$, resulting in delocalization. Whereas, {\bf{{c)},d)}} $(a,\omega)=(2.15,{2.45})$ shows enhanced diagonal weight leading to localization. Here, $L=500$ and sine drive is implemented. 
}
\end{figure*}

    Performing a numerical Inverse Fourier Transformation (IFT) of the momentum space Hamiltonian reveals that $H^{(1)}_{F1}$ contributes predominantly to off-diagonal terms, while $H^{(1)}_{F2}$ generates both diagonal and off-diagonal components. The relative weight of diagonal to off-diagonal terms determines the nature of the phase at a given $(a,\omega)$. We show here that after IFT, the Floquet Hamiltonian at $(a,\omega)=(2.15,2.45)$ exhibits enhanced diagonal weight leading to localization. We plot the average matrix element value $\langle |H_{ij}|\rangle_d$ vs. distance from the diagonal $d=|i-j|$ in Fig. \ref{fpt_HF_om_2.45_a_2.15}, in Fig. \ref{fpt_HF_om_2.45_a_2.15_zoom} we zoom near the diagonal($d=0$). In contrast, for $(a,\omega)=(2.15,1)$, we plot the same average matrix element in Fig. \ref{fpt_HF_om_1_a_2.15} and zoomed near the diagonal in Fig. \ref{fpt_HF_om_1_a_2.15_zoom}, from these figures we see a dominant off-diagonal contribution leading to delocalization. 

%==========================================================
%=========================================
%=======================================================
\subsection{Analysis by Van Vleck perturbation theory}
 \subsubsection{Analysis of resonances in VVPT}
\begin{figure*}
\centering
\subfigure[]{
\includegraphics[width=0.3\linewidth]{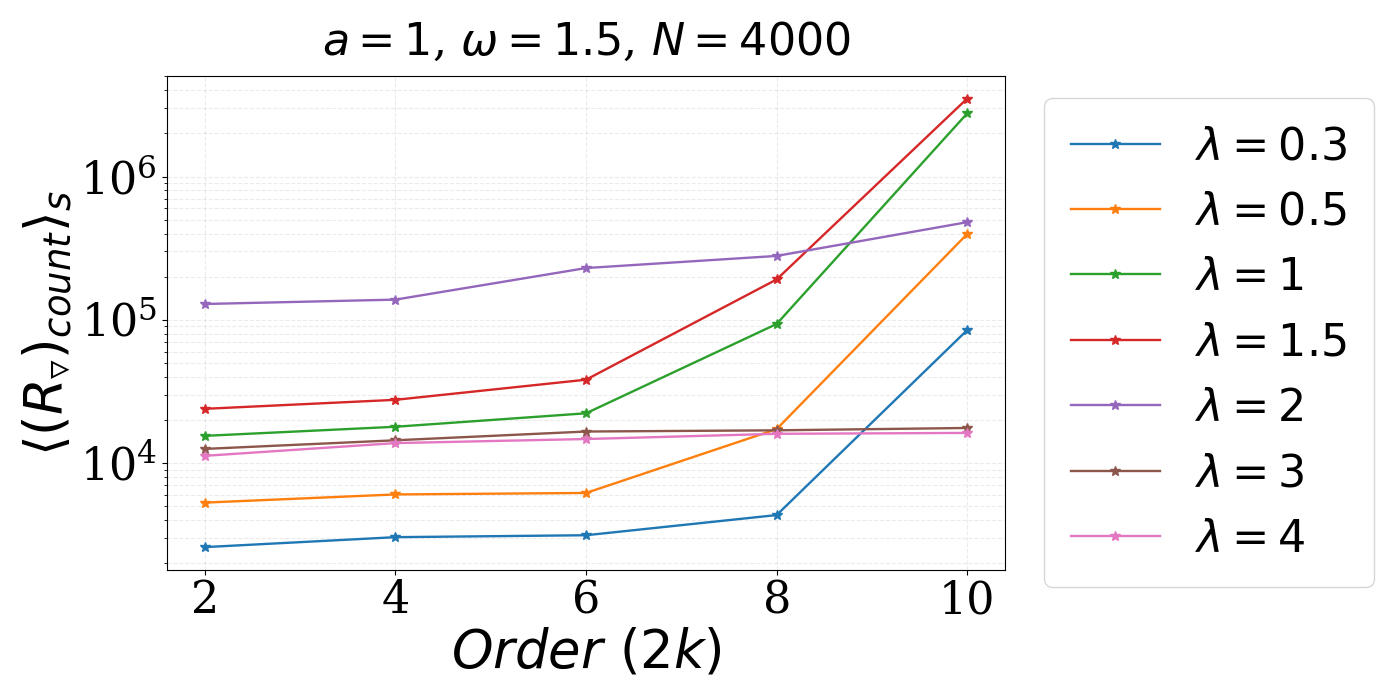}\label{res_count_all_mw}}
\hspace{0.2cm}
\subfigure[]{
\includegraphics[width=0.3\linewidth]{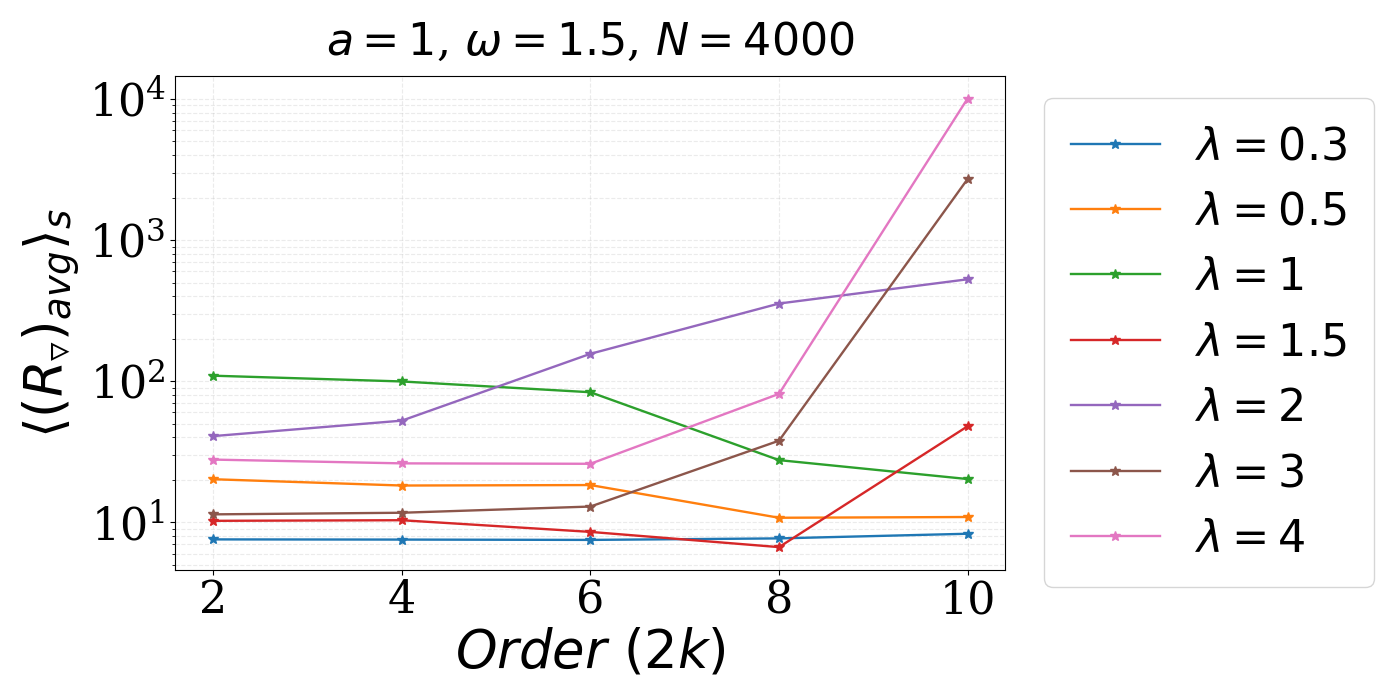}
\label{res_avg_all_mw}}
\hspace{0.2cm}
\subfigure[]{
\includegraphics[width=0.3\linewidth]{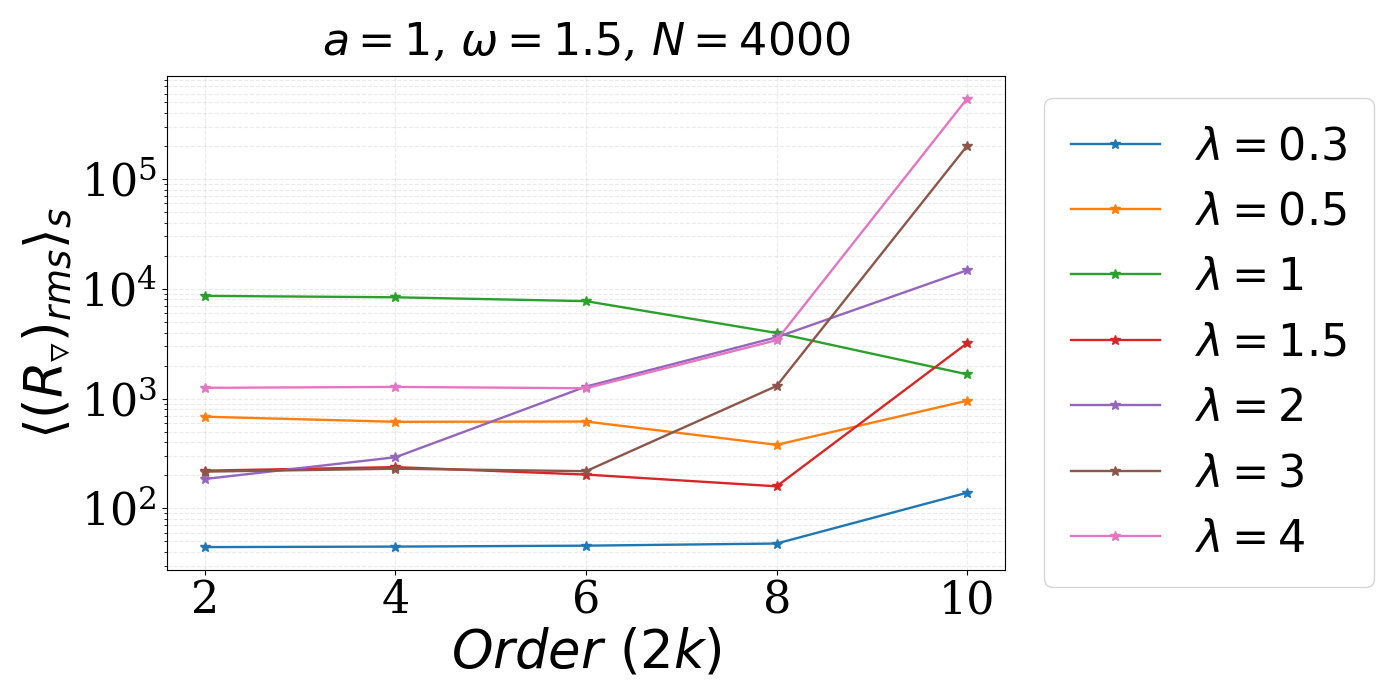}
\label{res_rms_all_mw}}
\captionsetup{justification=centerlast, width=\linewidth} 
\caption{{\bf{a)}} The count of resonance $\mathcal{R_r}$ is high and grows modearately with increasing order $2k$ at intermediate $\lambda=1,1.5$. {\bf{b),c)}} At higher $\lambda$, resonances are fewer but stronger, reflected in the form of average and rms $\mathcal{R_r}$ strength, while $\lambda=1$ contains a consistent high $\mathcal{R_r}$ average and rms in initial orders proving the breakdown of perturbation at this size $L=4000$. Here, $a=1$ and $\omega=1.5$; sine drive is used for calculation.  
}
\end{figure*}

We consider the Fourier representation $H(t)=\sum_{p}H_pe^{ip\omega t}$, where the $H_0$ represents the static part of the Hamiltonian and the remaining harmonics represent time dependent part of the drive. In the interaction picture we can define, $V_I(t)=e^{iH_0t}(H(t)-H_0)e^{-iH_0t}$ and $U_I(T)=I+S_I^{(1)}(T)+S_I^{(2)}(T)$. The first-order contribution in the Dyson series is $S_I^{(1)}(T)=\int_0^T dt\,V_I(t)$ in the eigenbasis of $H_0$ is,
$\langle w|S_1^{(I)}(T)|x\rangle
=\int_0^T dt\,
\langle w|e^{iH_0 t}H(t)e^{-iH_0 t}|x\rangle $.
Using the Fourier expansion of $H(t)$ and eigenvalue equation for $H_0|x\rangle=E_x|x\rangle$, the integrand becomes $\langle w|e^{iH_0 t}H(t)e^{-iH_0 t}|x\rangle=\langle w|H(t)e^{i(E_w-E_x+p\omega)}|x\rangle$. The time integral therefore generates denominators of the form $(E_w-E_x+p\omega)^{-1}$. However, the explicit prefactor $\frac{1}{T}$ appearing in the definition of $H_{eff}\ (H_{eff}=\frac{1}{T}ln(U))$ ensures that the contribution is of order $1/\omega^0$. Similarly, $S_I^{(2)}(T)=-\frac{1}{2}\int_0^T dt_1\int_0^{t_1} dt_2\,[V_I(t_1),V_I(t_2)]$. After evaluating the nested time integrals and explicit division by $T$ yields a leading order correction term which represents ${O({\omega^{-1}})}$ process $H^{(1)}_{VV}$. This contains denominators of the form $(E_w-E_x+p_1\omega+p_2\omega)^{-1}$,$(E_w-E_x+p_1\omega)^{-1}$ and $(E_w-E_x+p_2\omega)^{-1}$. From this analysis, the allowed range of photon numbers $m$ can be inferred at a given perturbative order. As we see in $H^{(1)}_{VV}$ for sine drive $m$ can take values from $[-2,-1,0,1,2]$ as both $p_1,p_2=\pm 1$. More generally, the range for allowed $m$ values for any perturbation order $n$ will be from $-(n+1)$ to $n+1$. So, with this modified range we redefined our resonance diagnostic $\mathcal{R_{r}}=\frac{|\langle w|H^{(2k)}|x\rangle|}{min(|E_w^0-E_x^0\pm m\omega|)}$, where $m\in[-(2k+1),2k+1]$. We quantify the count of such resonance instances whenever $\mathcal{R_{r}}>1$ and also check the magnitude by computing the average value of all $\mathcal{R_{r}}$s. As shown in fig. \ref{res_count_all_mw}, the number of resonance counts is high and increases by orders of magnitude with perturbative order $2k$ for intermediate disorder strengths $\lambda=1,1.5$, indicating a resonance-dominated spectrum. However, we find in fig. \ref{res_avg_all_mw} and fig. \ref{res_rms_all_mw} that with increased $\lambda$ and $2k$, the average and rms $\mathcal{R_{r}}$ attains a maximum value, indicating that resonances are sparse but progressively stronger for such cases (In practice, such rare resonance condition are not achieved as for high $\lambda$s optimal truncation order is $2k\approx6$). In contrast, for intermediate $\lambda$, the number of resonant events increases substantially, implying that a larger fraction of effective Hamiltonian elements are affected by resonances (for $\lambda=1$, the average  and rms $\mathcal{R_{r}}$ attains a high value), which causes breakdown of agreement with exact numerics (specifically for $\lambda$=1 case).
%%%%%%%%%%%%%%%%%%%%%%%%%%%%%%%%%%%%%%%%%%%%%%%%%%%%%%%%%%%%%%%%%%%%%%%%%%%%%%%%%%%%%%%%%%%%%%%%%%%%%%%%%%%%%%%%%%%%%%%%%%%%%%%%%%%%

\subsubsection{Convergence analysis of VVPT using ratio test with different norms}
 
\begin{figure*}
\centering

\subfigure[]{
\includegraphics[width=0.24\linewidth]{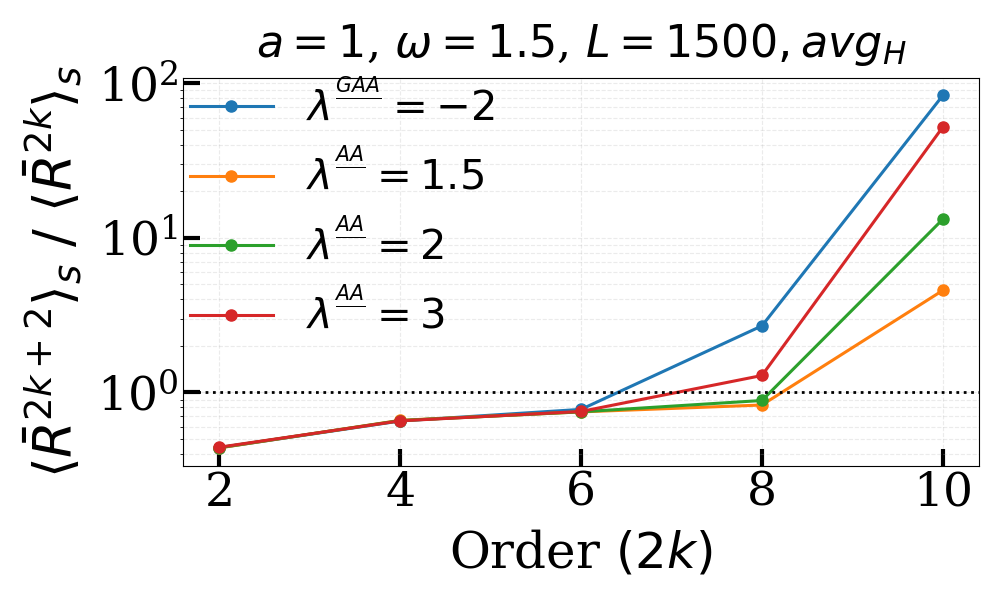}\label{avg_H_1500_a_1_om_1.5}}
\hspace{-0.38cm}
\subfigure[]{
\includegraphics[width=0.24\linewidth]{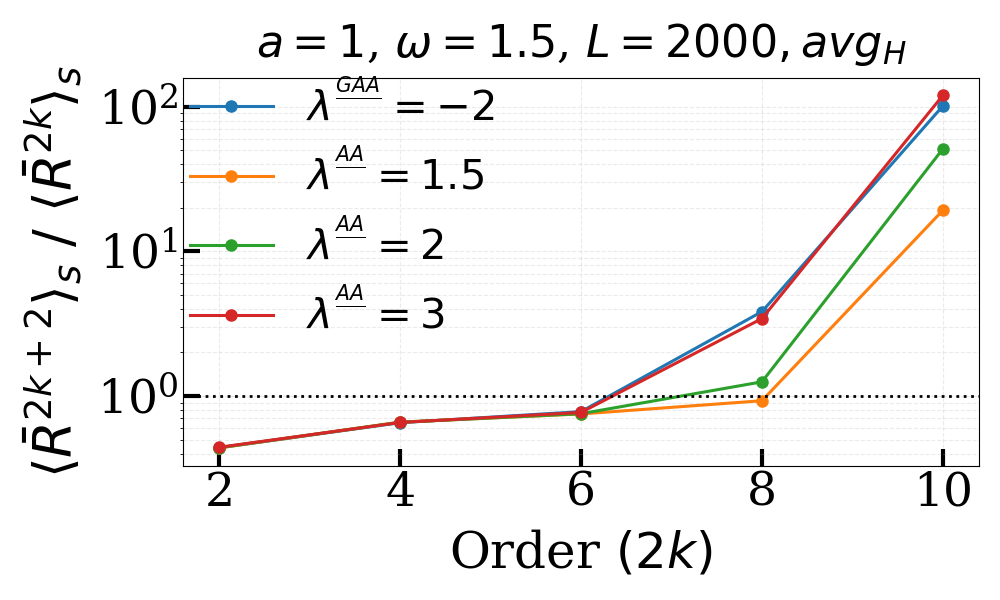}
\label{avg_H_2000_a_1_om_1.5}}
\subfigure[]{
\includegraphics[width=0.24\linewidth]{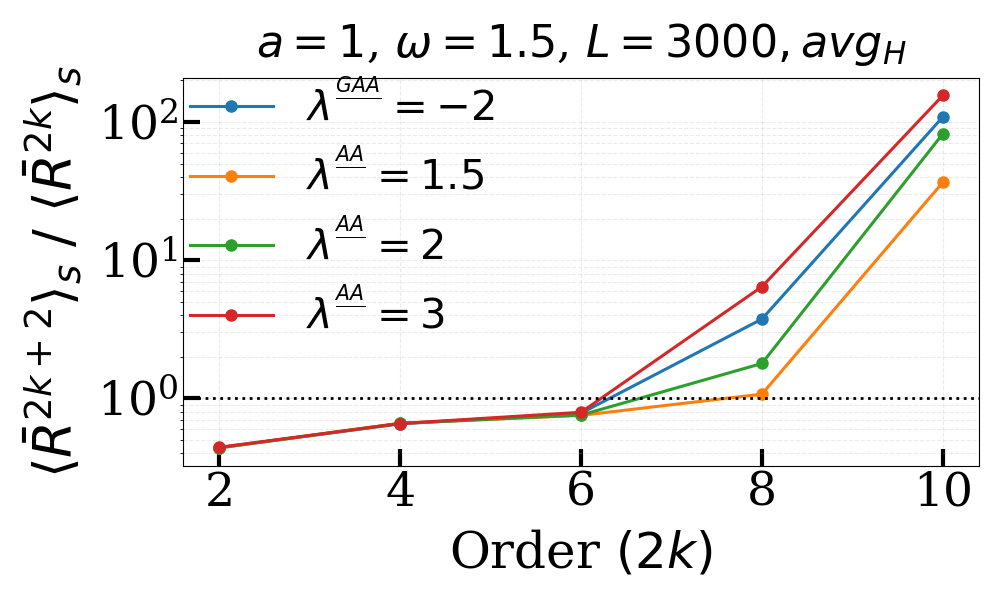}
\label{avg_H_3000_a_1_om_1.5}}
\hspace{-0.35cm}
\subfigure[]{
\includegraphics[width=0.24\linewidth]{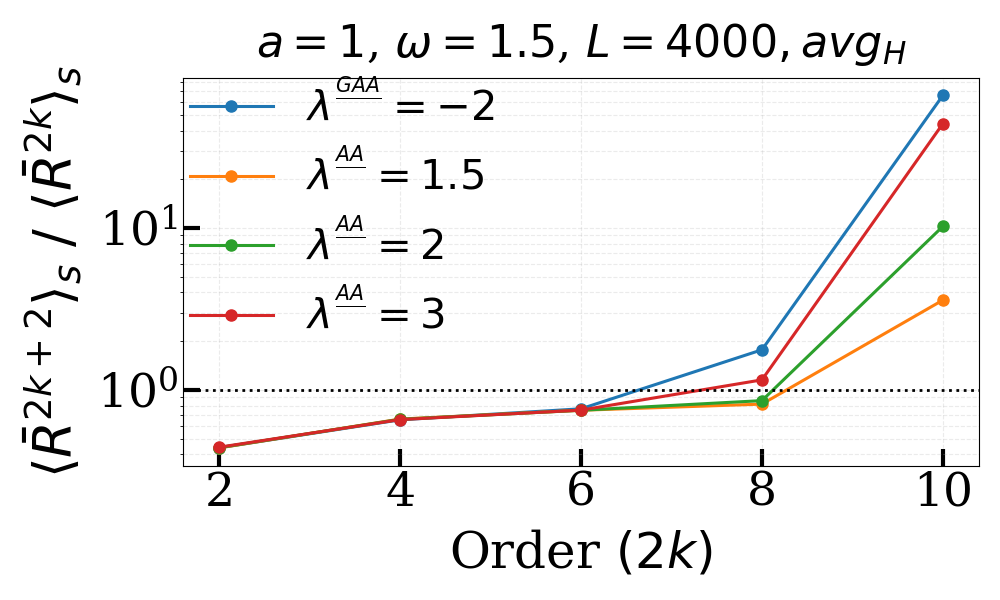}
\label{avg_H_4000_a_1_om_1.5}}\\ %new row
%%%%%%%%%%%%%%%%%%%%%%%%%%%%%%%%%%%%%%%%%%%%%%%%%%%%%%%
\subfigure[]{
\includegraphics[width=0.24\linewidth]{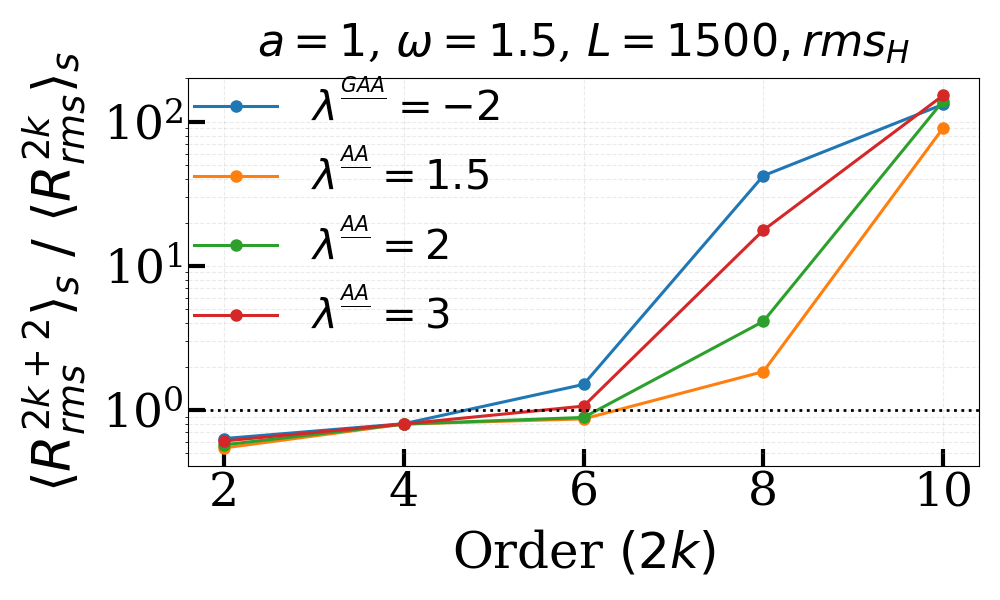}\label{rms_H_1500_a_1_om_1.5}}
\hspace{-0.38cm}
\subfigure[]{
\includegraphics[width=0.24\linewidth]{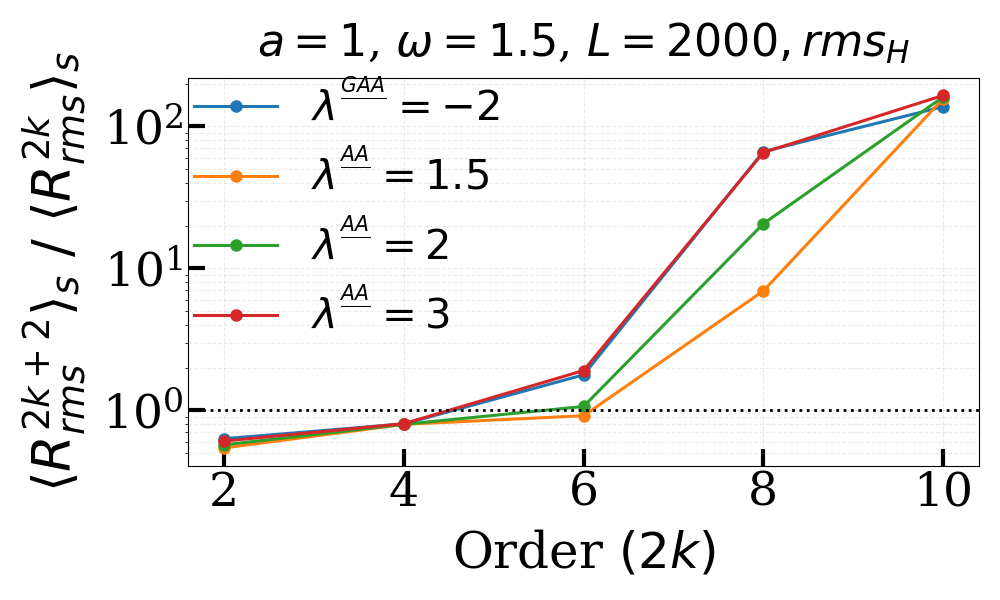}
\label{rms_H_2000_a_1_om_1.5}}
\subfigure[]{
\includegraphics[width=0.24\linewidth]{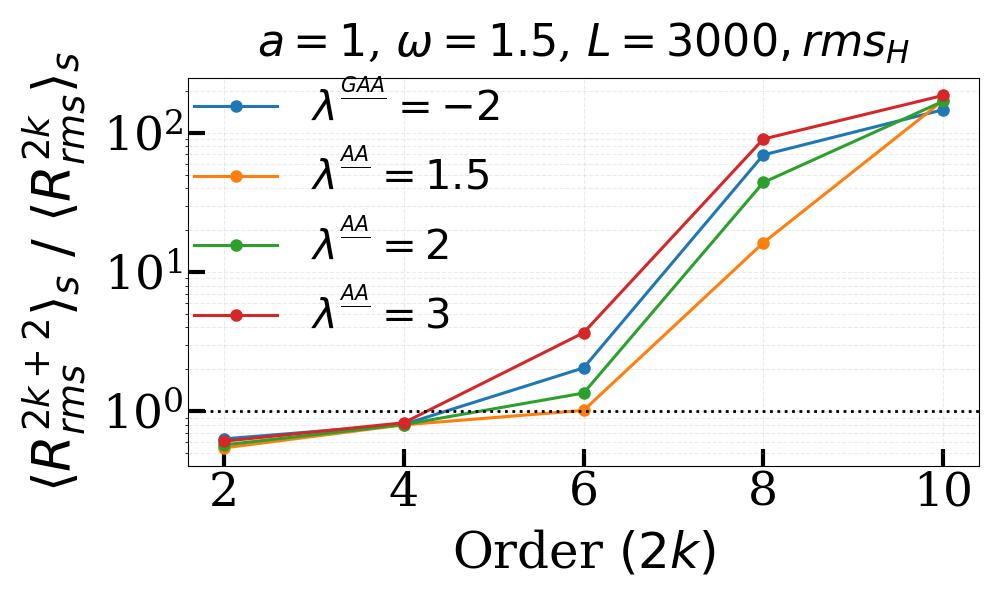}
\label{rms_H_3000_a_1_om_1.5}}
\hspace{-0.35cm}
\subfigure[]{
\includegraphics[width=0.24\linewidth]{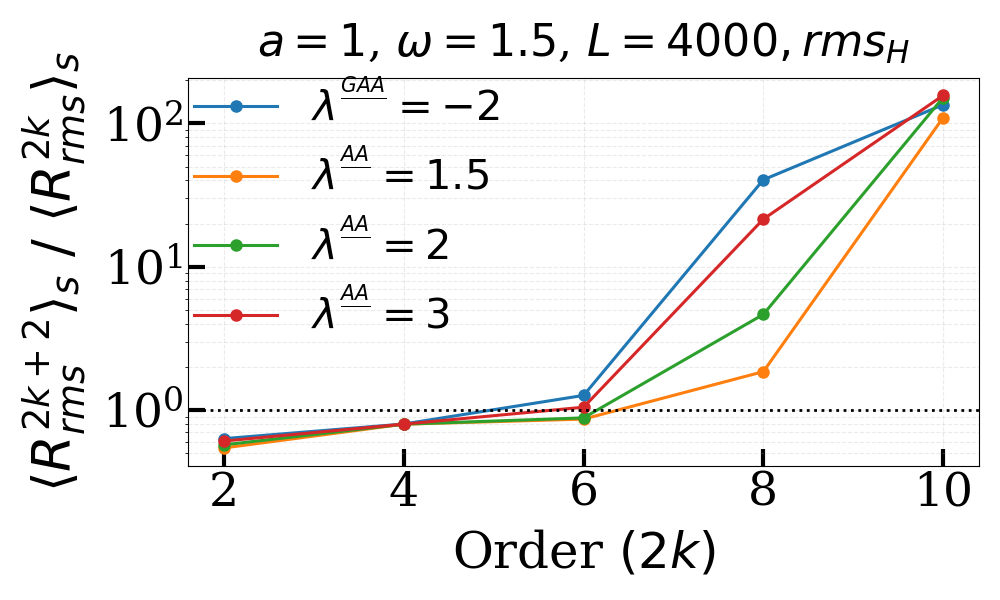}
\label{rms_H_4000_a_1_om_1.5}}\\ %new row
%%%%%%%%%%%%%%%%%%%%%%%%%%%%%%%%%%%%%%%%%%%%%%%%%%%%%%%
\subfigure[]{
\includegraphics[width=0.24\linewidth]{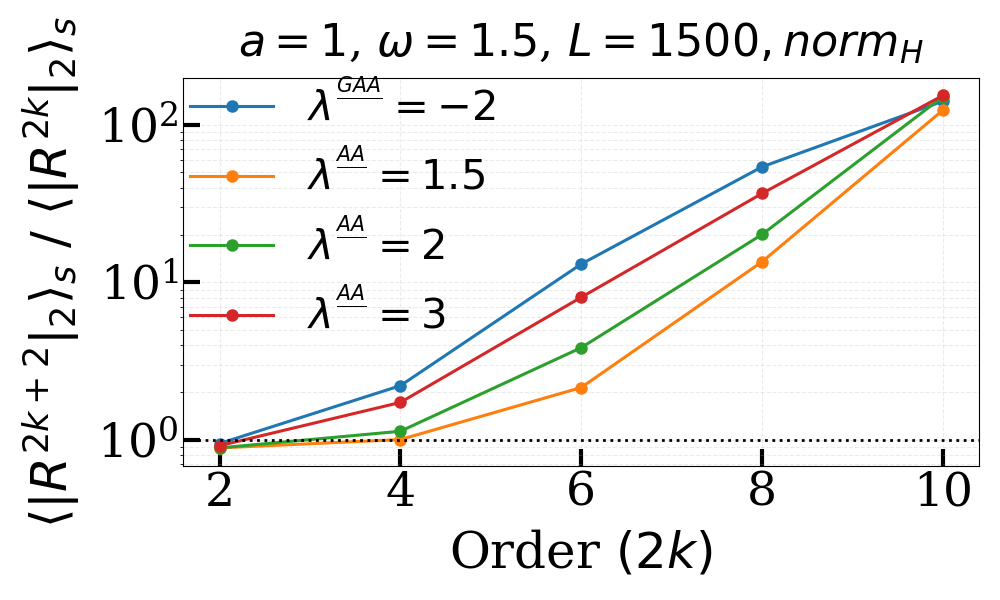}\label{norm_H_1500_a_1_om_1.5}}
\hspace{-0.38cm}
\subfigure[]{
\includegraphics[width=0.24\linewidth]{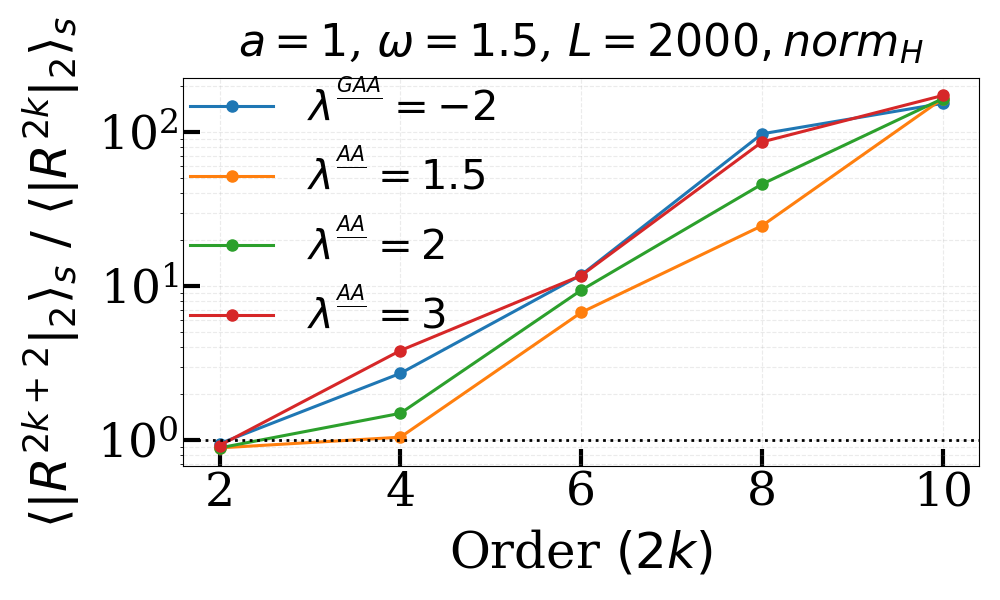}
\label{norm_H_2000_a_1_om_1.5}}
\subfigure[]{
\includegraphics[width=0.24\linewidth]{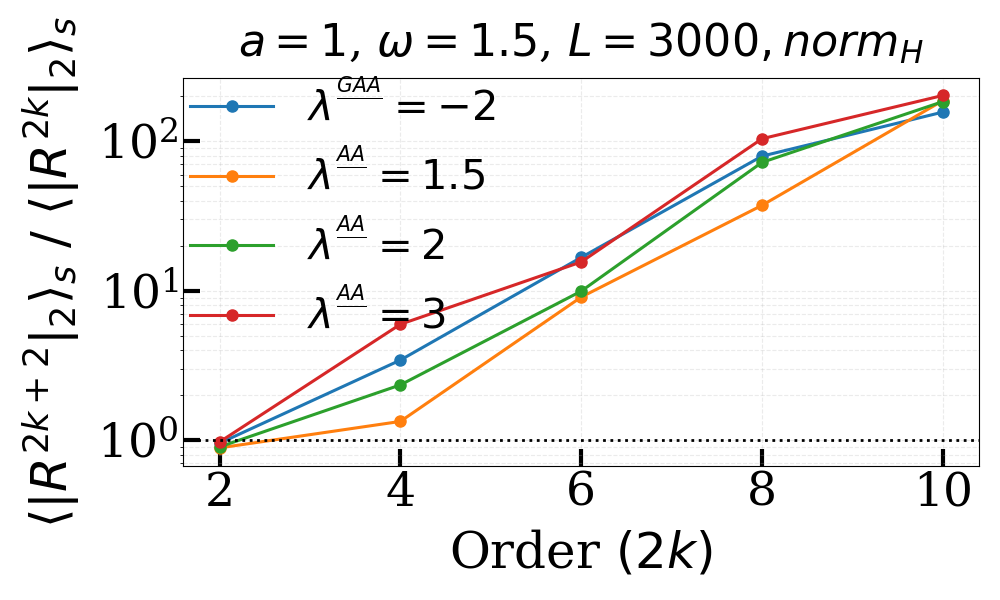}
\label{norm_H_3000_a_1_om_1.5}}
\hspace{-0.35cm}
\subfigure[]{
\includegraphics[width=0.24\linewidth]{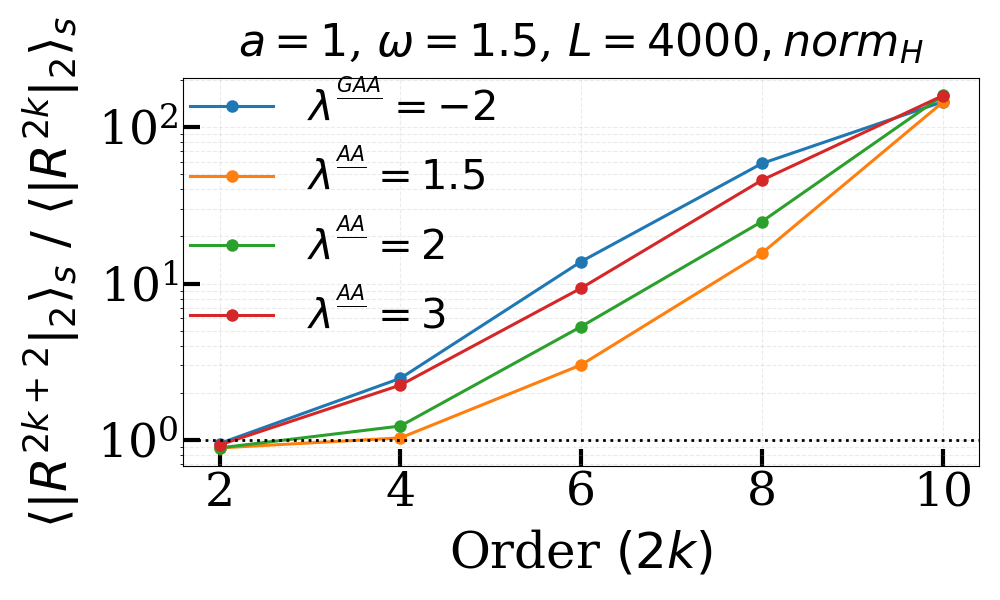}
\label{norm_H_4000_a_1_om_1.5}}\\  %new row
%%%%%%%%%%%%%%%%%%%%%%%%%%%%%%%%%%%%%%%%%%%%%%%%%%%%%%%
%%%%%%%%%%for omega=5.7, a=5%%%%%%%%%%%%%%%%%%%%%%%%%%%
%%%%%%%%%%%%%%%%%%%%%%%%%%%%%%%%%%%%%%%%%%%%%%%%%%%%%%%
\subfigure[]{
\includegraphics[width=0.24\linewidth]{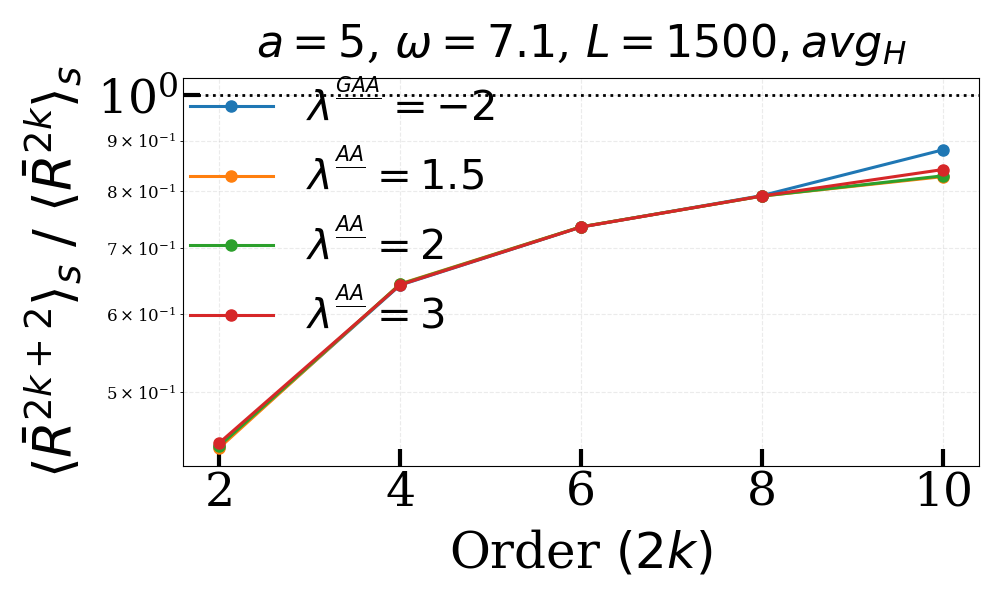}\label{avg_H_1500_a_5_om_7.1}}
\hspace{-0.38cm}
\subfigure[]{
\includegraphics[width=0.24\linewidth]{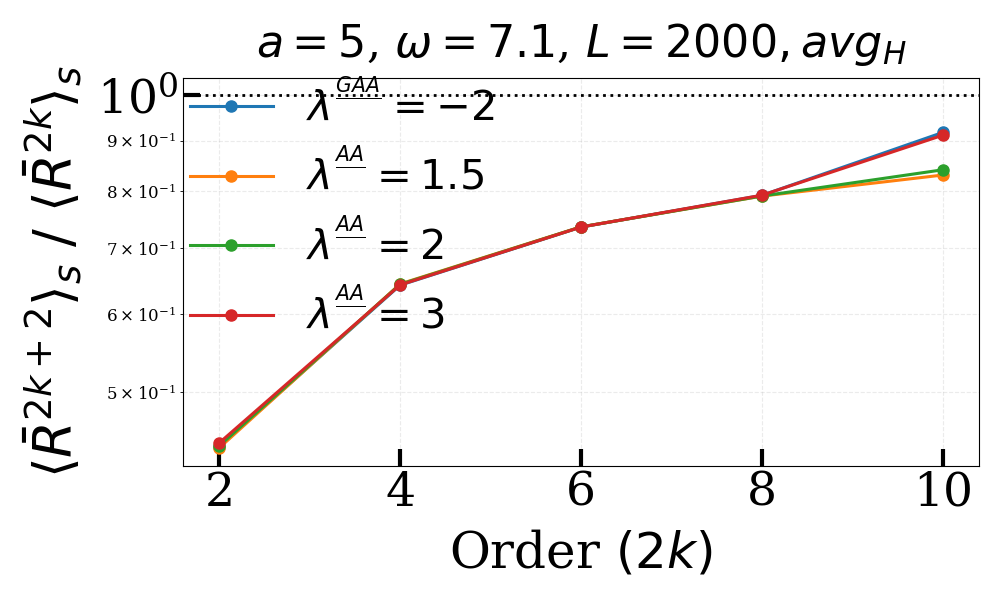}
\label{avg_H_2000_a_5_om_7.1}}
\subfigure[]{
\includegraphics[width=0.24\linewidth]{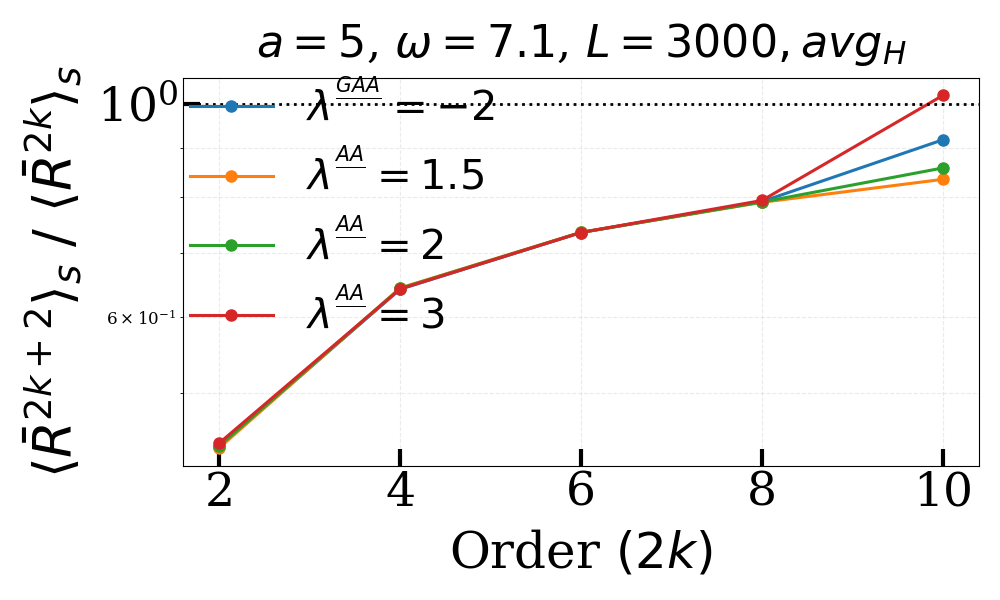}
\label{avg_H_3000_a_5_om_7.1}}
\hspace{-0.35cm}
\subfigure[]{
\includegraphics[width=0.24\linewidth]{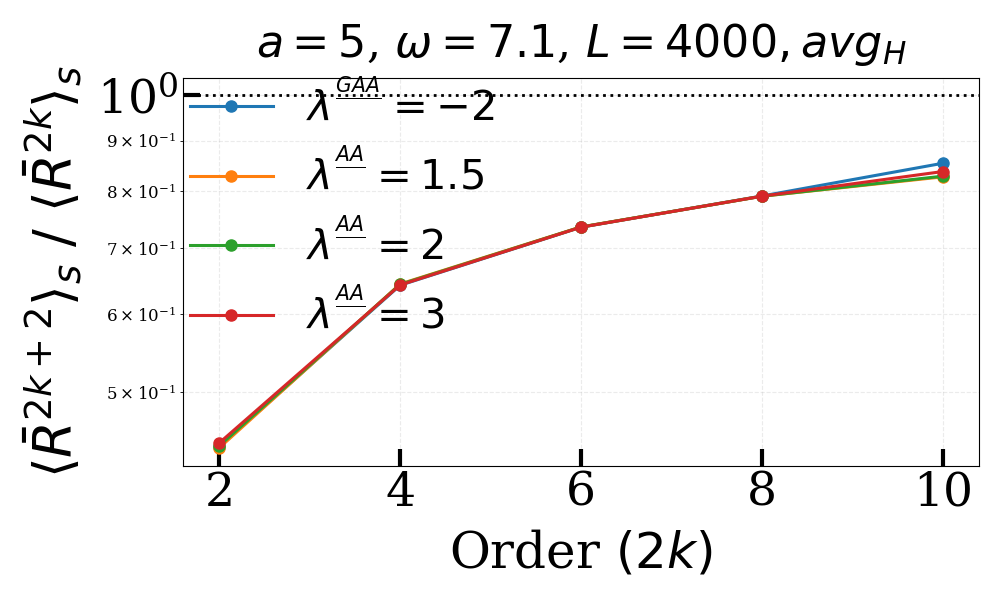}
\label{avg_H_4000_a_5_om_7.1}}\\ %new row
%%%%%%%%%%%%%%%%%%%%%%%%%%%%%%%%%%%%%%%%%%%%%%%%%%%%%%%
\subfigure[]{
\includegraphics[width=0.24\linewidth]{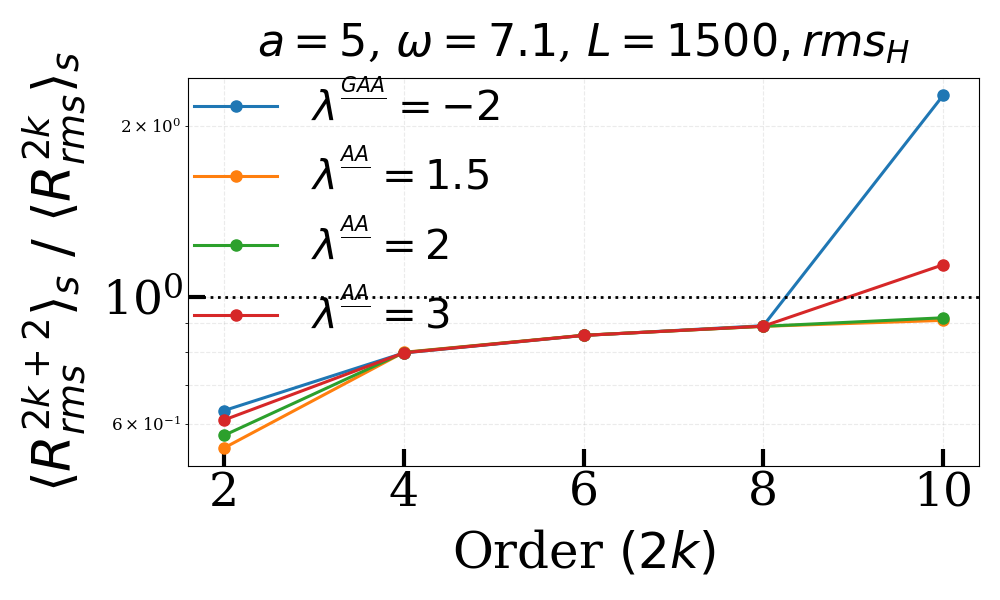}\label{rms_H_1500_a_5_om_7.1}}
\hspace{-0.38cm}
\subfigure[]{
\includegraphics[width=0.24\linewidth]{master/ratio_test_rms_N_2000_a_1_om_1.5.png}
\label{rms_H_2000_a_5_om_7.1}}
\subfigure[]{
\includegraphics[width=0.24\linewidth]{master/ratio_test_rms_N_3000_a_1_om_1.5.png}
\label{rms_H_3000_a_5_om_7.1}}
\hspace{-0.35cm}
\subfigure[]{
\includegraphics[width=0.24\linewidth]{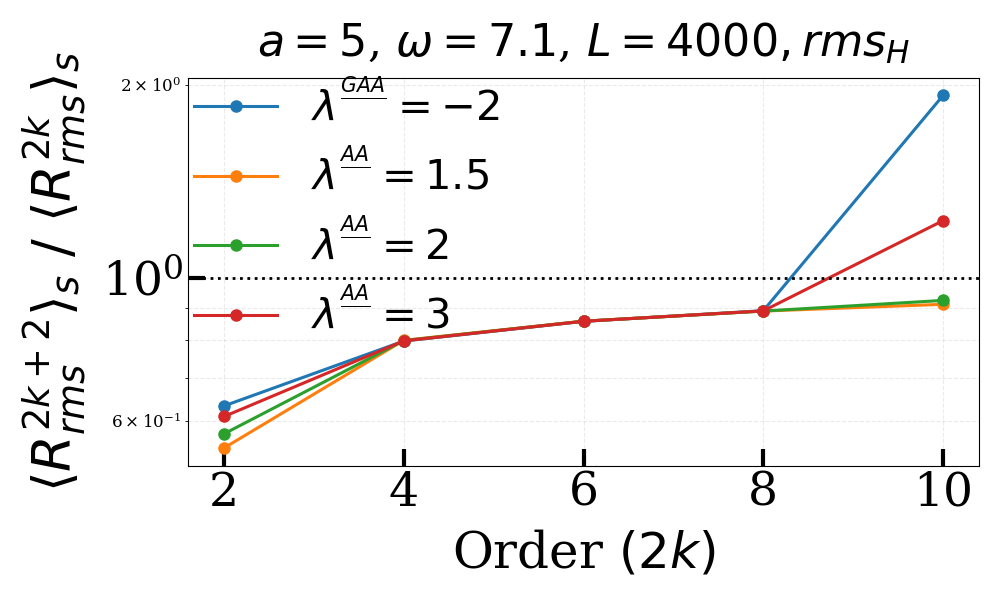}
\label{rms_H_4000_a_5_om_7.1}}\\ %new row
%%%%%%%%%%%%%%%%%%%%%%%%%%%%%%%%%%%%%%%%%%%%%%%%%%%%%%%
\subfigure[]{
\includegraphics[width=0.24\linewidth]{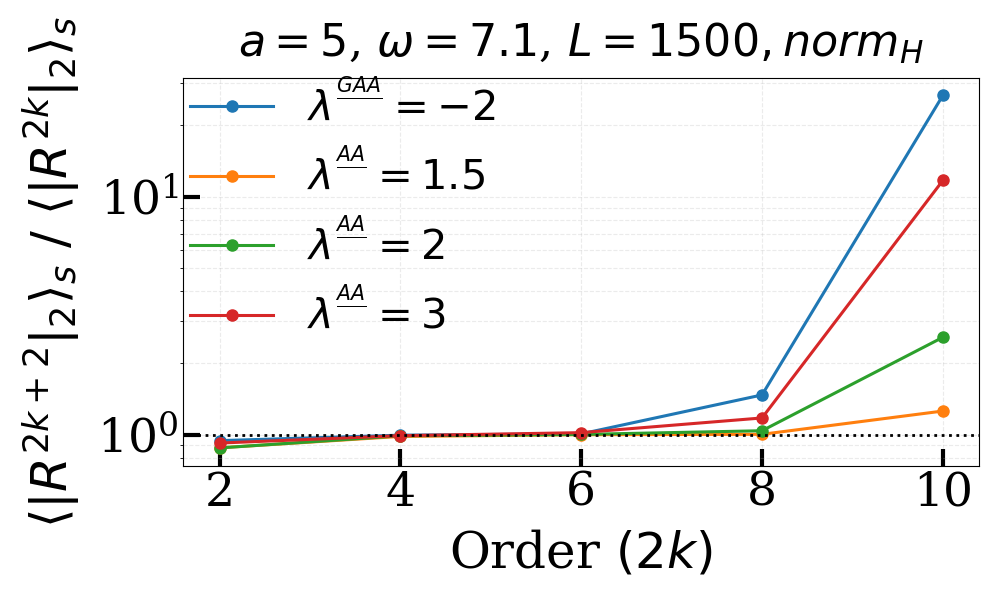}\label{norm_H_1500_a_5_om_7.1}}
\hspace{-0.38cm}
\subfigure[]{
\includegraphics[width=0.24\linewidth]{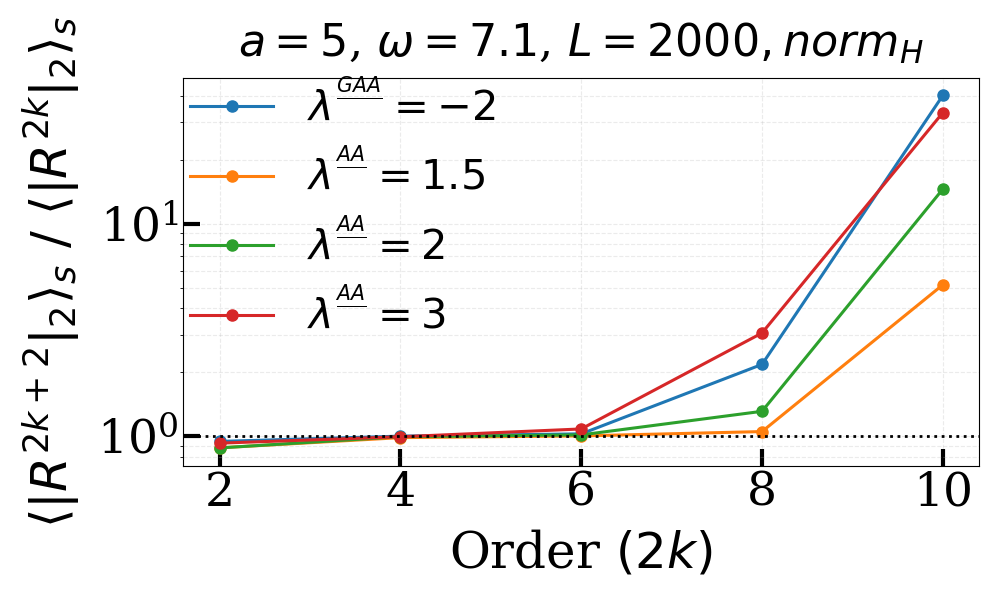}
\label{norm_H_2000_a_5_om_7.1}}
\subfigure[]{
\includegraphics[width=0.24\linewidth]{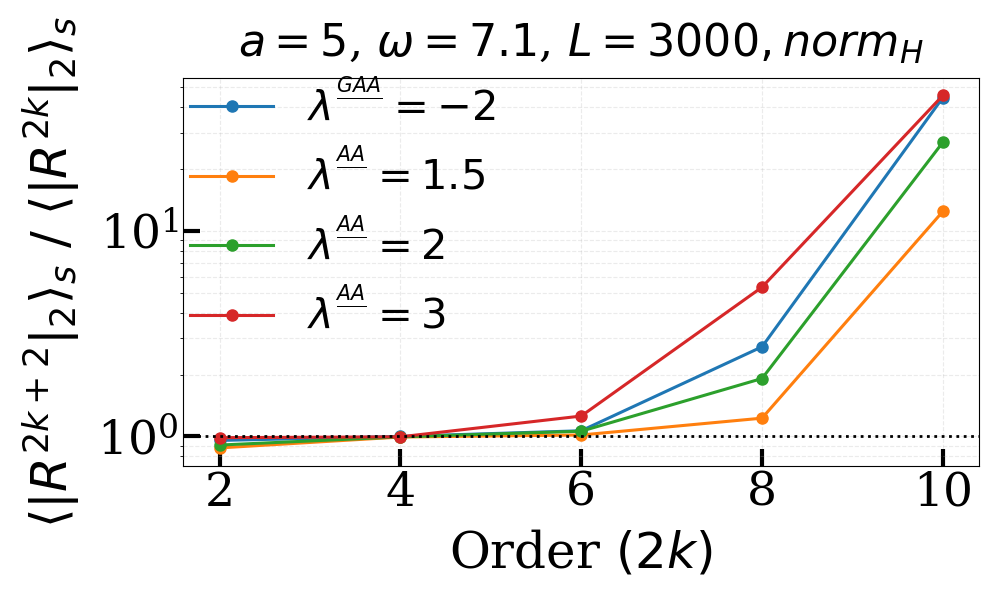}
\label{norm_H_3000_a_5_om_7.1}}
\hspace{-0.35cm}
\subfigure[]{
\includegraphics[width=0.24\linewidth]{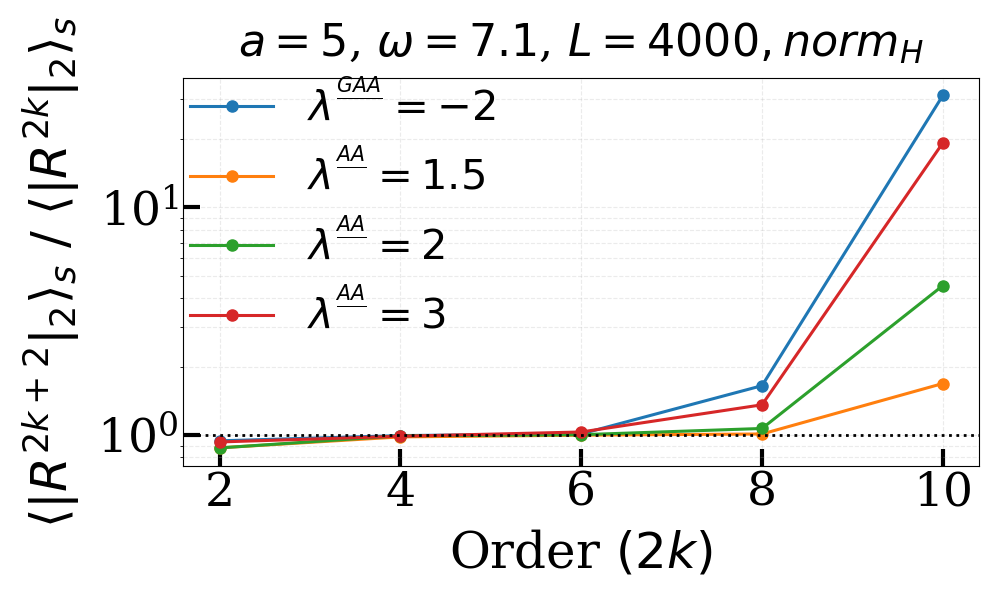}
\label{norm_H_4000_a_5_om_7.1}}\\ %new row
%%%%%%%%%%%%%%%%%%%%%%%%%%%%%%%%%%%%%%%%%%%%%%%%%%%%%%%
\captionsetup{justification=centerlast, width=\linewidth} 
\caption{{\bf{a)-d)}}Ratio test results for heuristic $n^{th}$-order VVPT using $avg_H$ for system sizes $L=1500,2000,3000$ and $4000$ at $(a,\omega)=(1,1.5)$ shown for AA potentials with $\lambda=1.5,2,3$ and for GAA model with $\lambda=-2,\beta=0.5$. The same parameters are used to compute $rms_H$ and $norm_H$ in {\bf{e)-h)}} and {\bf{i)-l)}} respectively. Panel {\bf{m)-p)}}, {\bf{q)-t)}} and {\bf{u)-x)}} show the corresponding analysis of $avg_H,rms _H$ and $norm_H$ for $(a,\omega)=(5,7.1)$. From this analysis of $avg_H$ across orders, we provide a criterion for selecting the optimal truncation order of the perturbation expansion.
}
\end{figure*}

\begin{figure*}
\centering

\subfigure[]{
\includegraphics[width=0.24\linewidth]{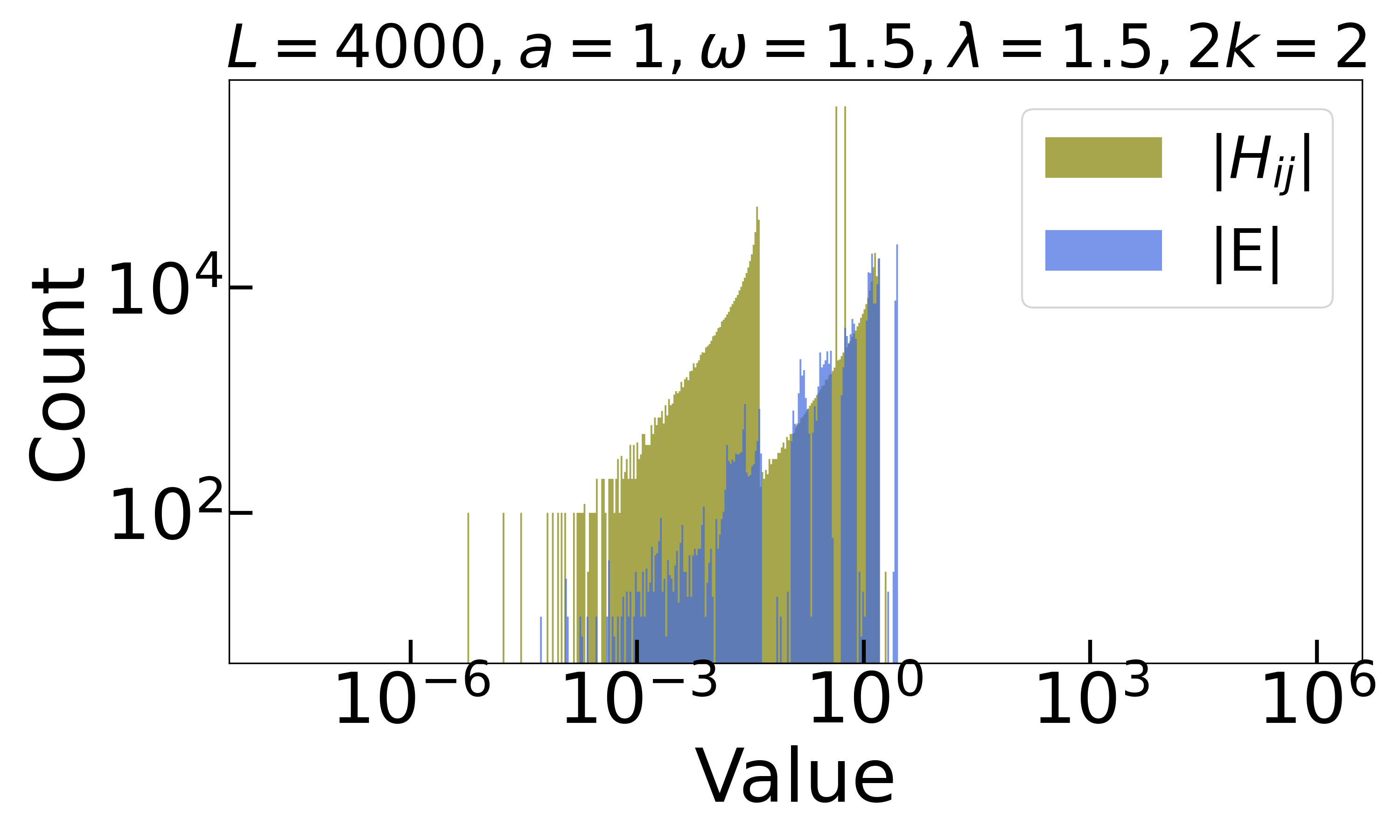}\label{O_2_lamb_1.5_N_4000_a_1_om_1.5_AA}}
\hspace{-0.35cm}
\subfigure[]{
\includegraphics[width=0.24\linewidth]{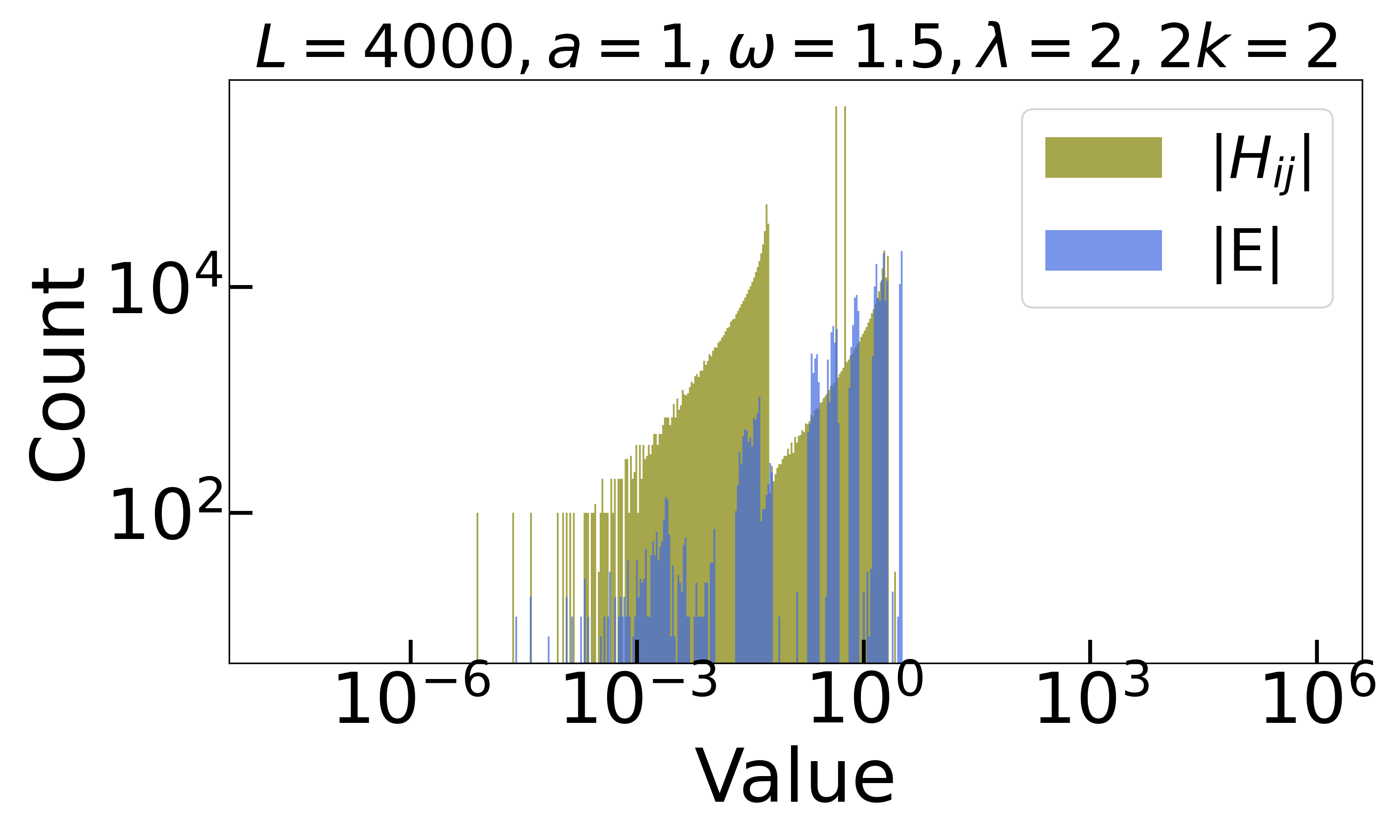}
\label{O_2_lamb_2_N_4000_a_1_om_1.5_AA}}
\subfigure[]{
\includegraphics[width=0.24\linewidth]{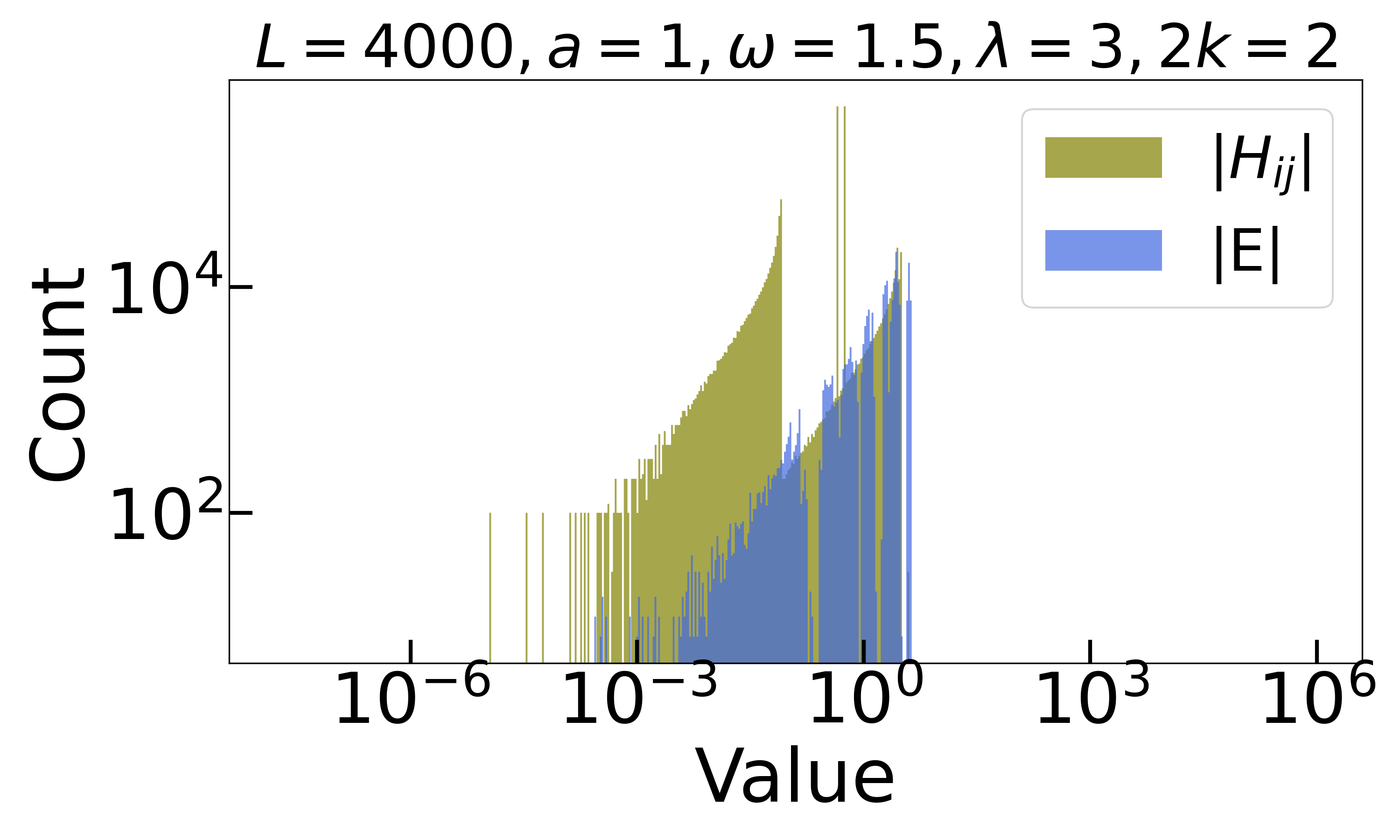}
\label{O_2_lamb_3_N_4000_a_1_om_1.5_AA}}
\hspace{-0.35cm}
\subfigure[]{
\includegraphics[width=0.24\linewidth]{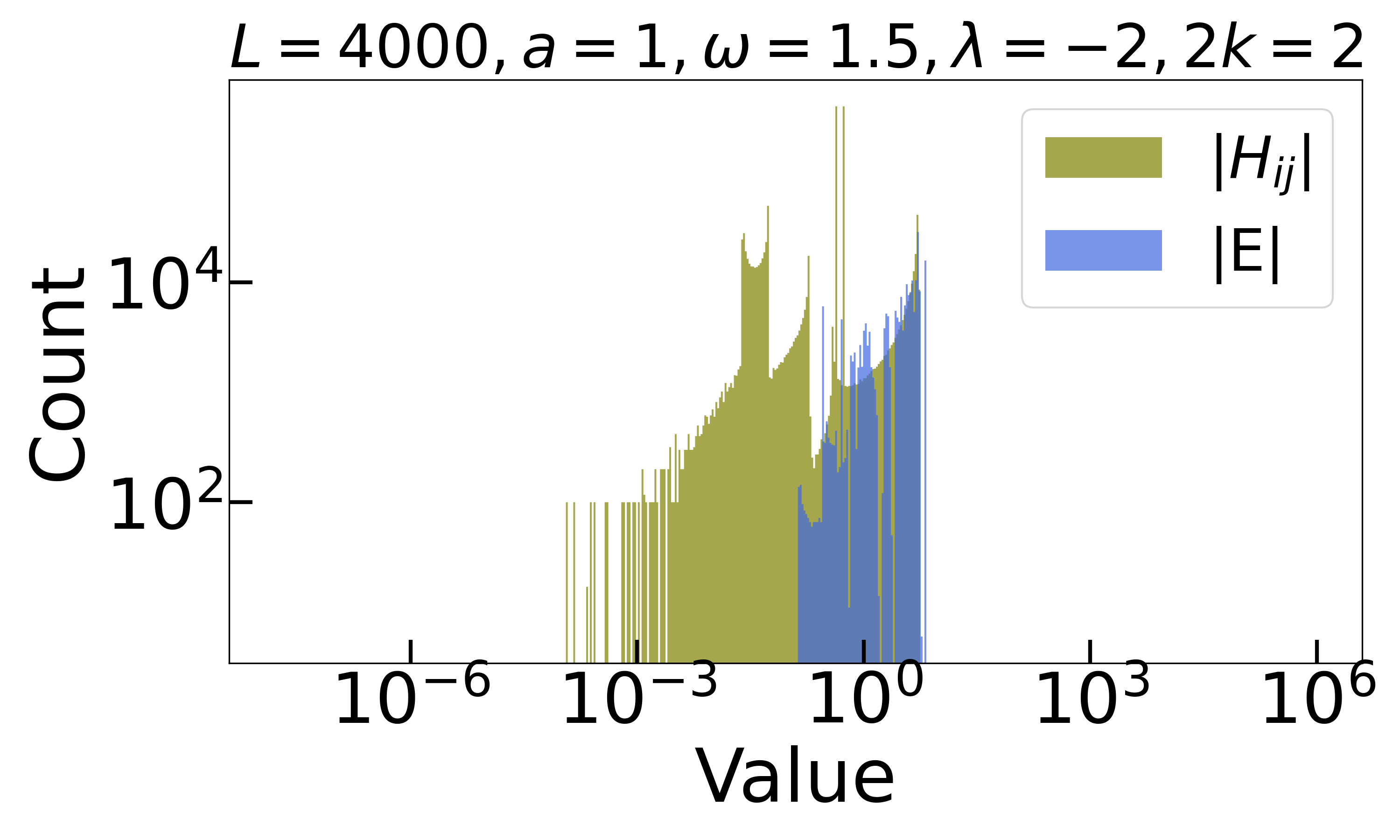}
\label{O_2_lamb_-2_N_4000_a_1_om_1.5_GAA}}\\ %new row
%%%%%%%%%%%%%%%%%%%%%%%%%%%%%%%%%%%%%%%%%%%%%%%%%%%%%%%
\subfigure[]{
\includegraphics[width=0.24\linewidth]{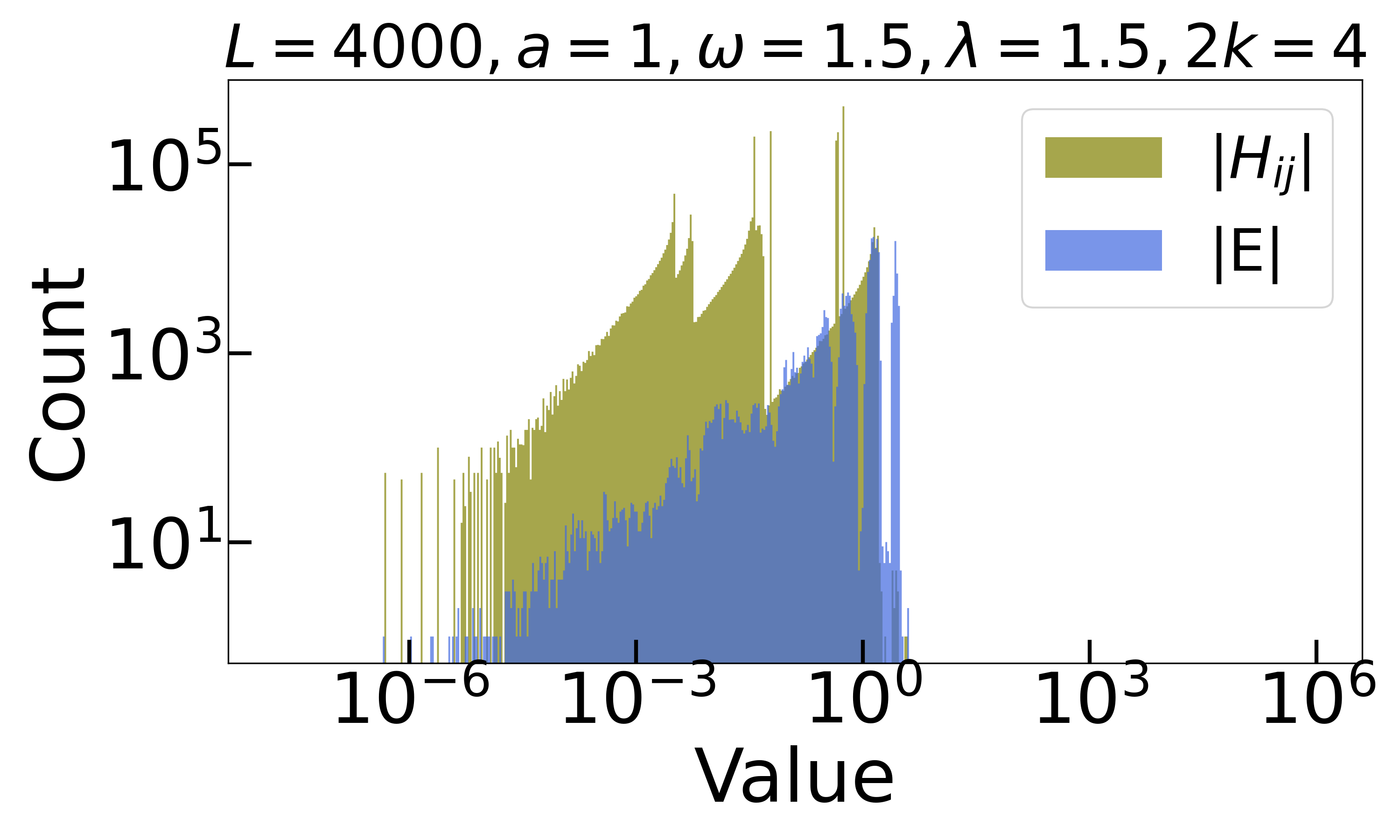}\label{O_4_lamb_1.5_N_4000_a_1_om_1.5_AA}}
\hspace{-0.35cm}
\subfigure[]{
\includegraphics[width=0.24\linewidth]{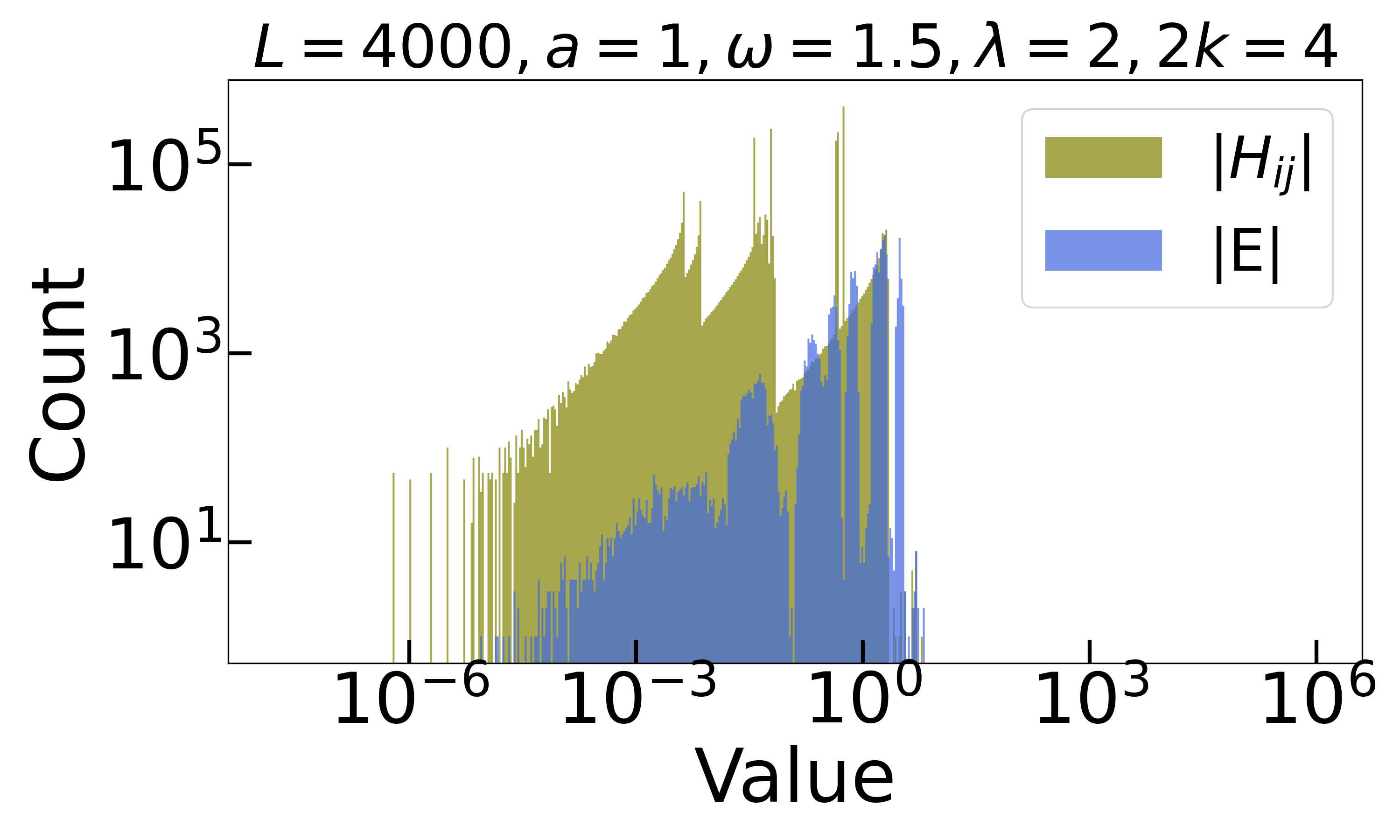}
\label{O_4_lamb_2_N_4000_a_1_om_1.5_AA}}
\subfigure[]{
\includegraphics[width=0.24\linewidth]{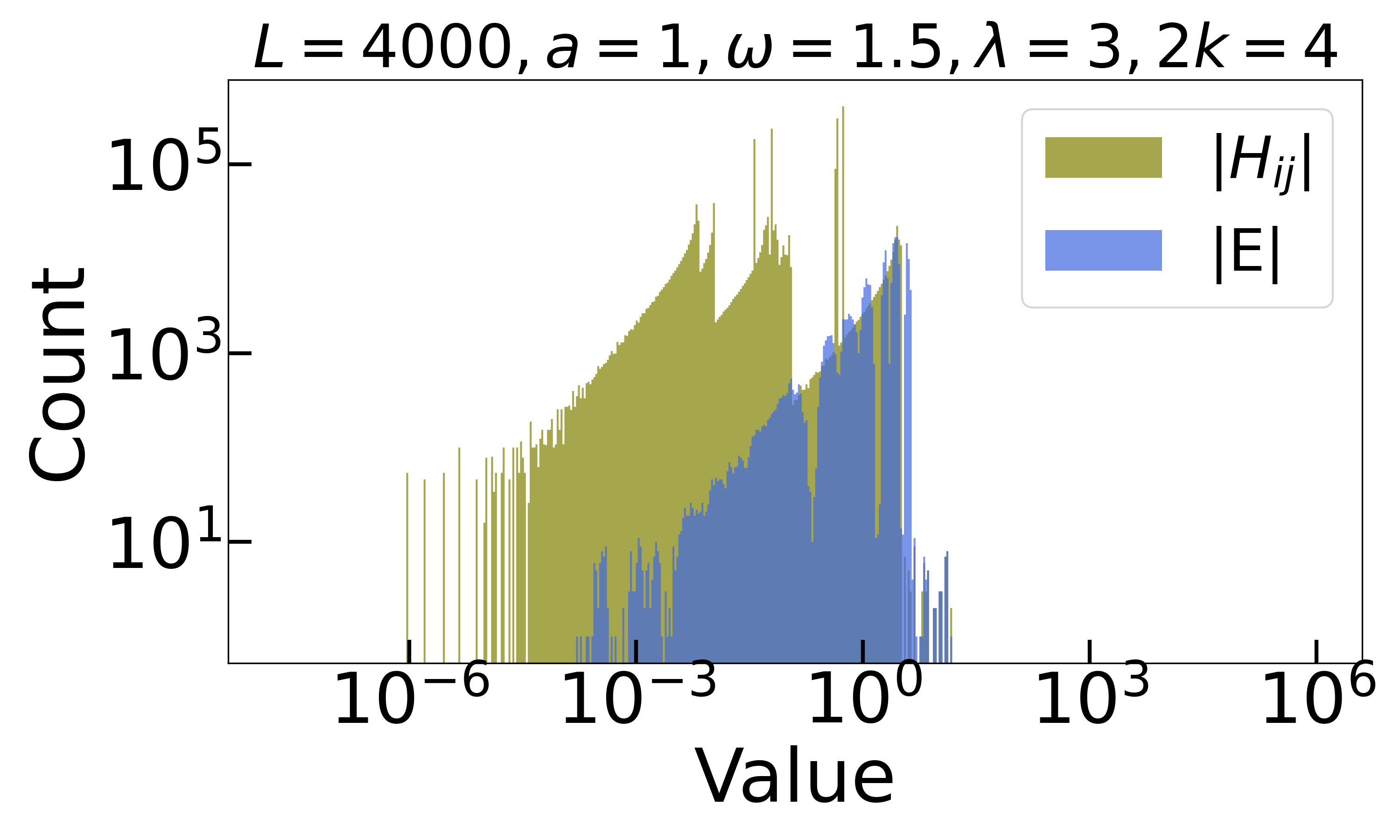}
\label{O_4_lamb_3_N_4000_a_1_om_1.5_AA}}
\hspace{-0.35cm}
\subfigure[]{
\includegraphics[width=0.24\linewidth]{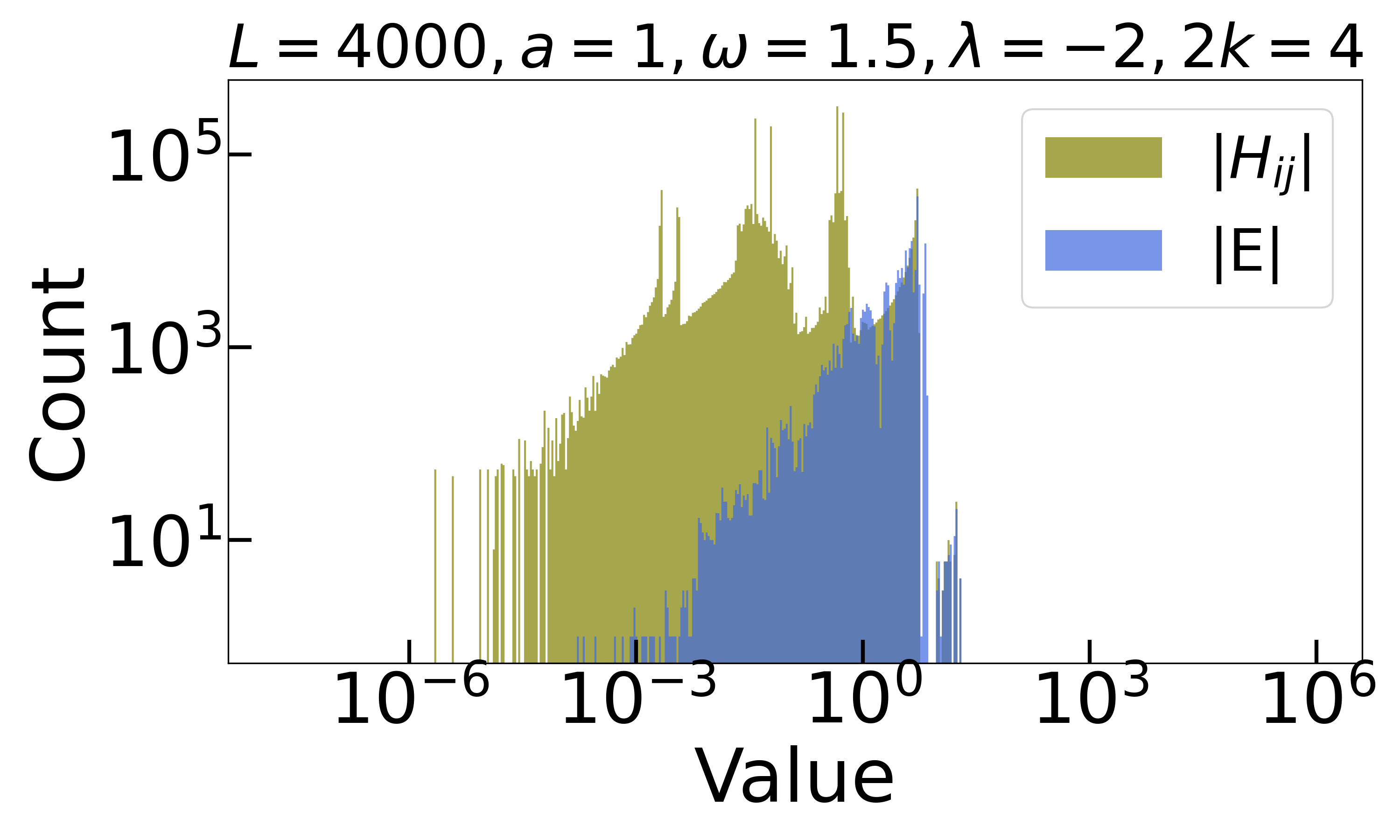}
\label{O_4_lamb_-2_N_4000_a_1_om_1.5_GAA}}\\ %new row
%%%%%%%%%%%%%%%%%%%%%%%%%%%%%%%%%%%%%%%%%%%%%%%%%%%%%%%
\subfigure[]{
\includegraphics[width=0.24\linewidth]{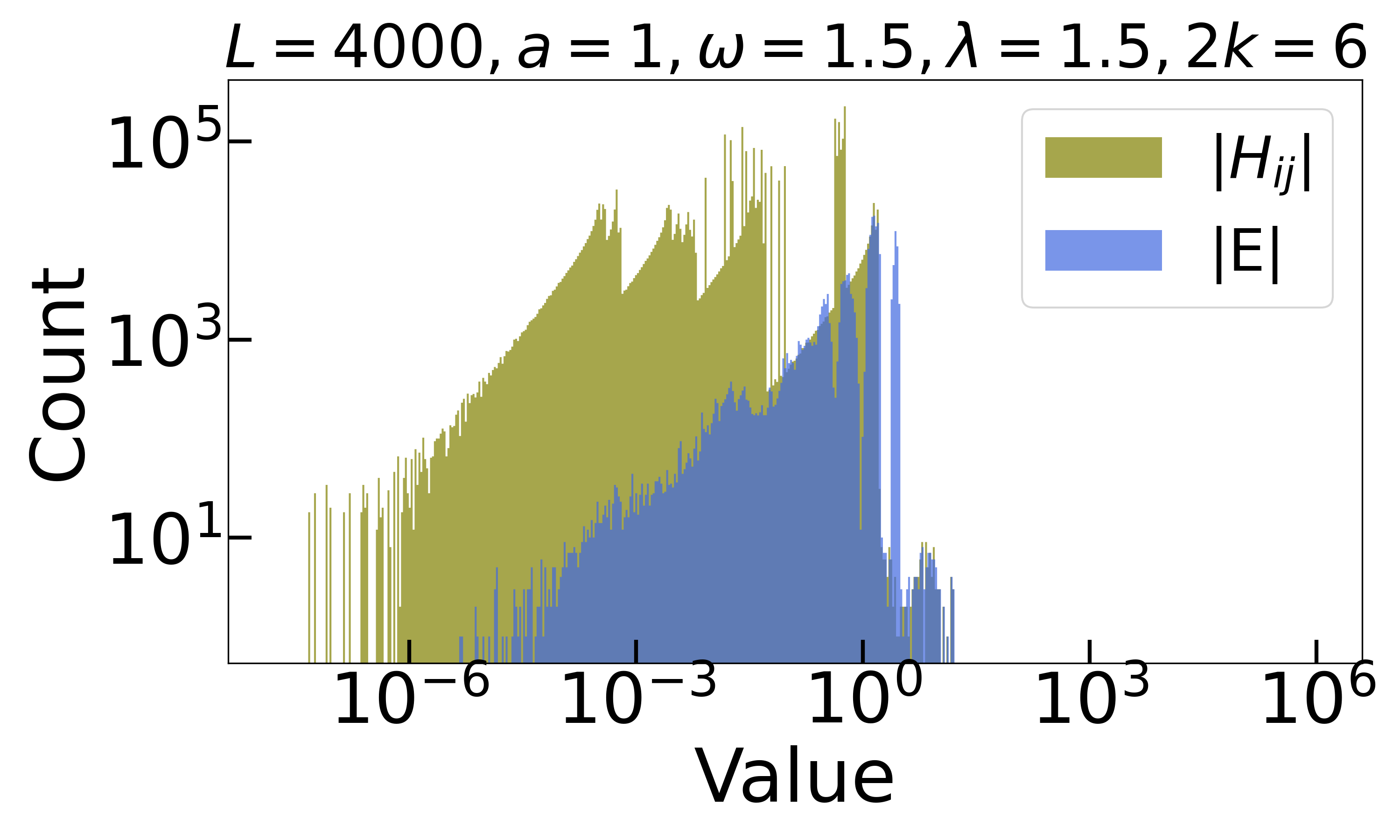}\label{O_6_lamb_1.5_N_4000_a_1_om_1.5_AA}}
\hspace{-0.35cm}
\subfigure[]{
\includegraphics[width=0.24\linewidth]{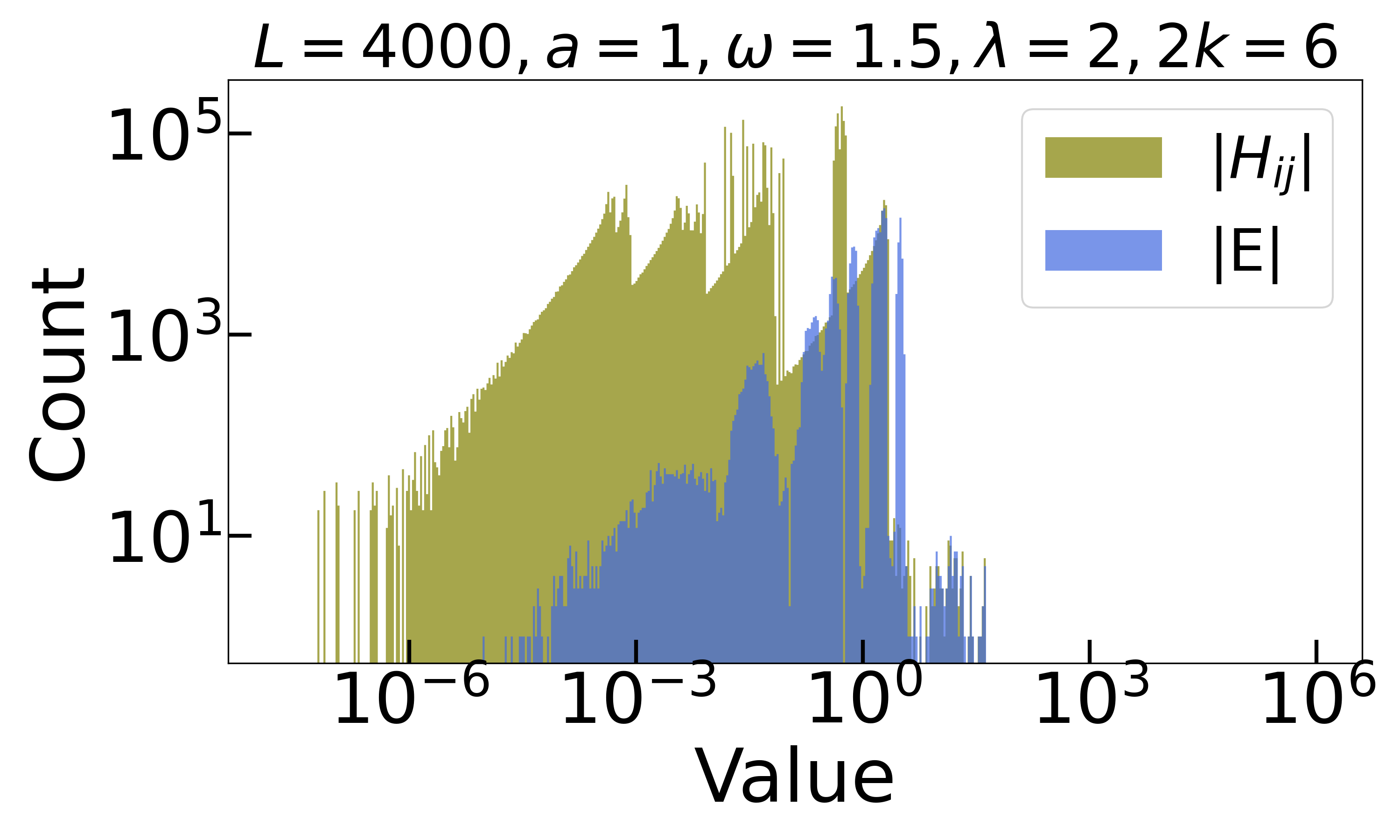}
\label{O_6_lamb_2_N_4000_a_1_om_1.5_AA}}
\subfigure[]{
\includegraphics[width=0.24\linewidth]{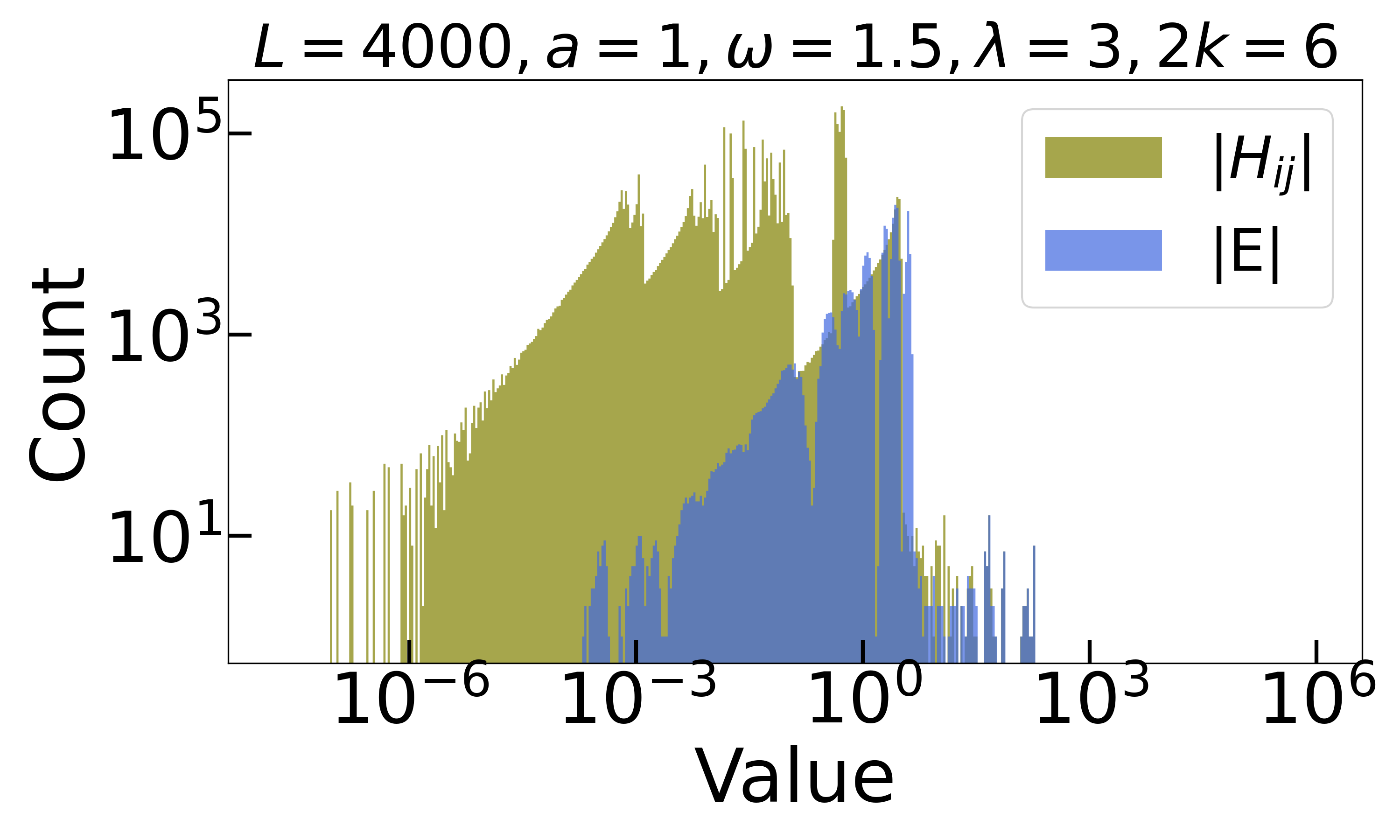}
\label{O_6_lamb_3_N_4000_a_1_om_1.5_AA}}
\hspace{-0.35cm}
\subfigure[]{
\includegraphics[width=0.24\linewidth]{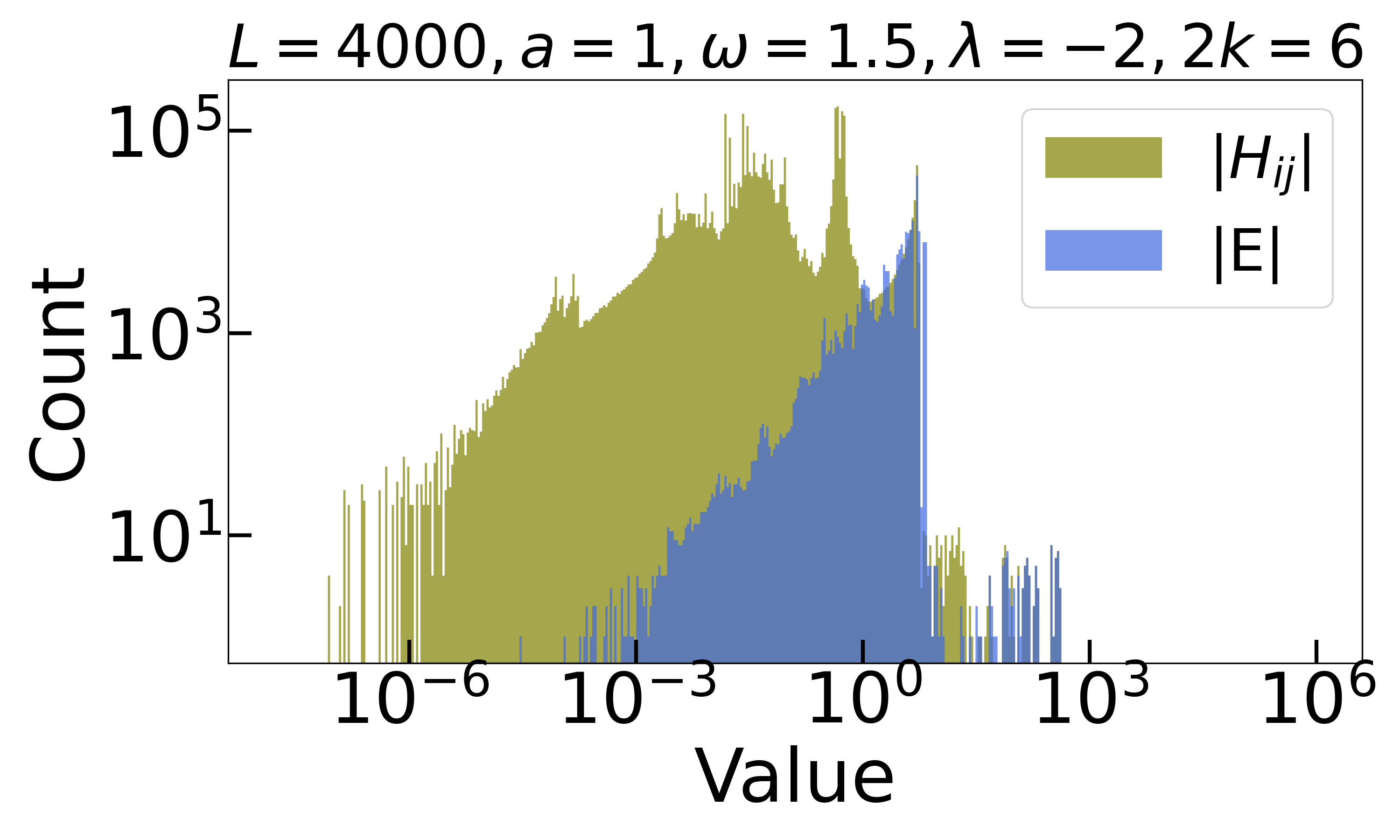}
\label{O_6_lamb_-2_N_4000_a_1_om_1.5_GAA}}\\ %new row
%%%%%%%%%%%%%%%%%%%%%%%%%%%%%%%%%%%%%%%%%%%%%%%%%%%%%%%
\subfigure[]{
\includegraphics[width=0.24\linewidth]{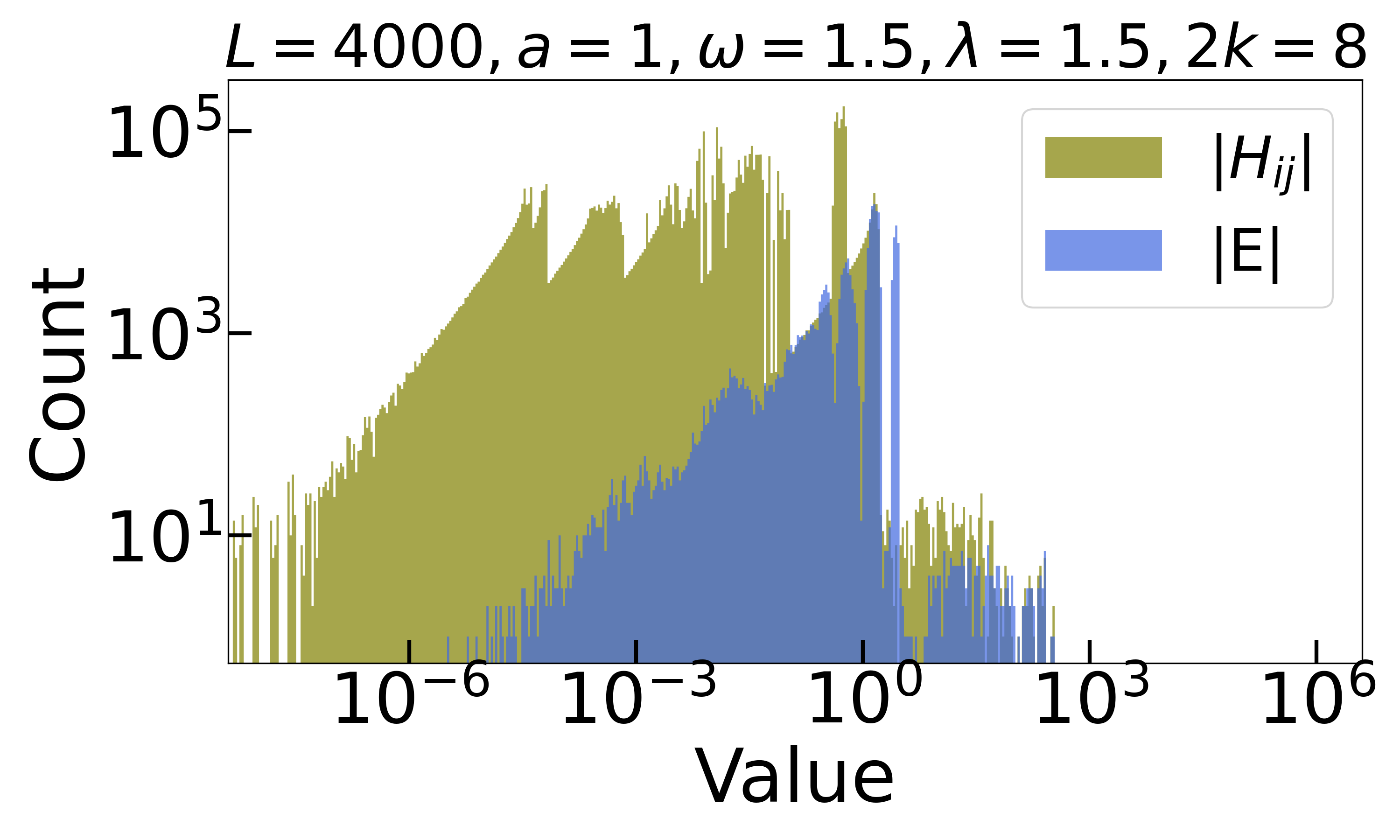}\label{O_8_lamb_1.5_N_4000_a_1_om_1.5_AA}}
\hspace{-0.35cm}
\subfigure[]{
\includegraphics[width=0.24\linewidth]{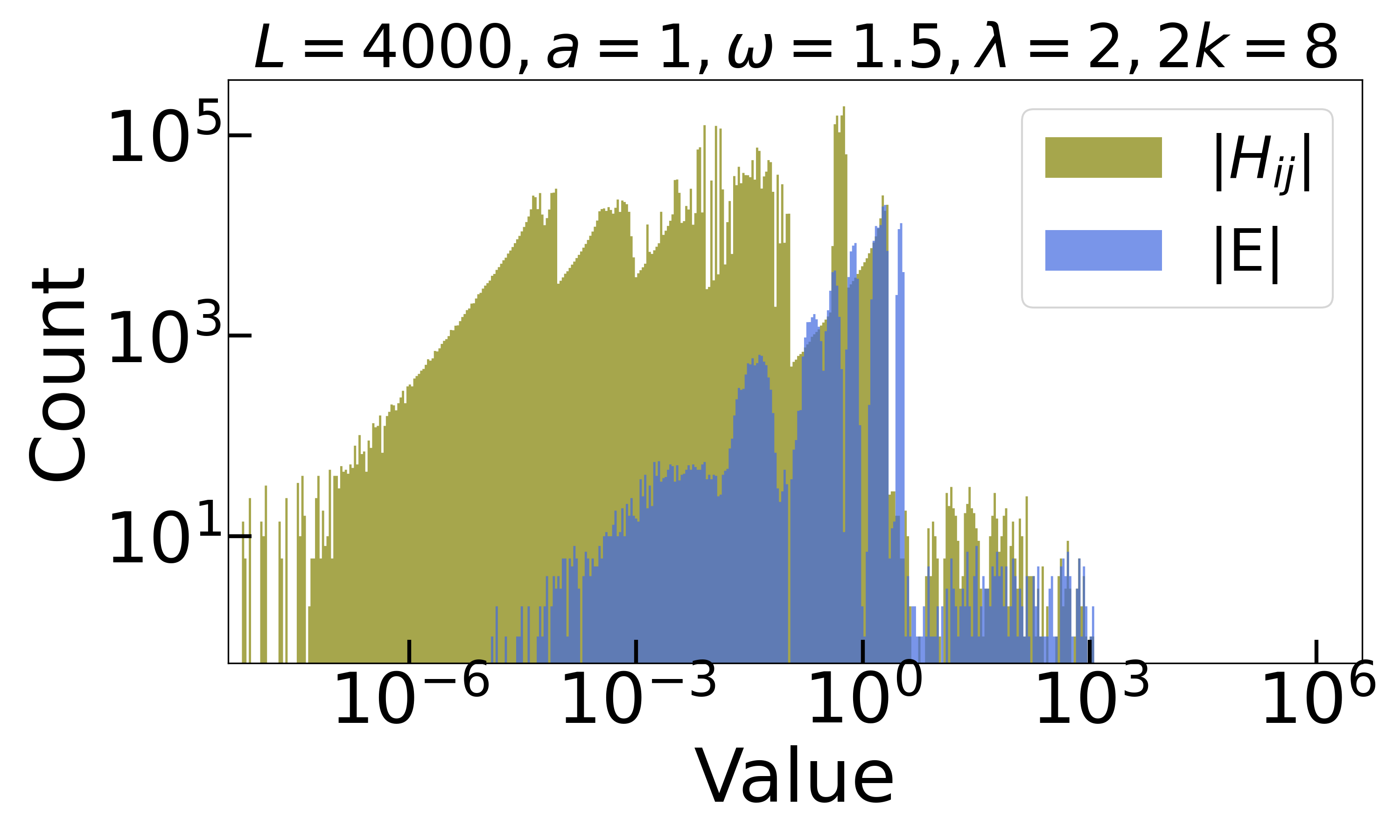}
\label{O_8_lamb_2_N_4000_a_1_om_1.5_AA}}
\subfigure[]{
\includegraphics[width=0.24\linewidth]{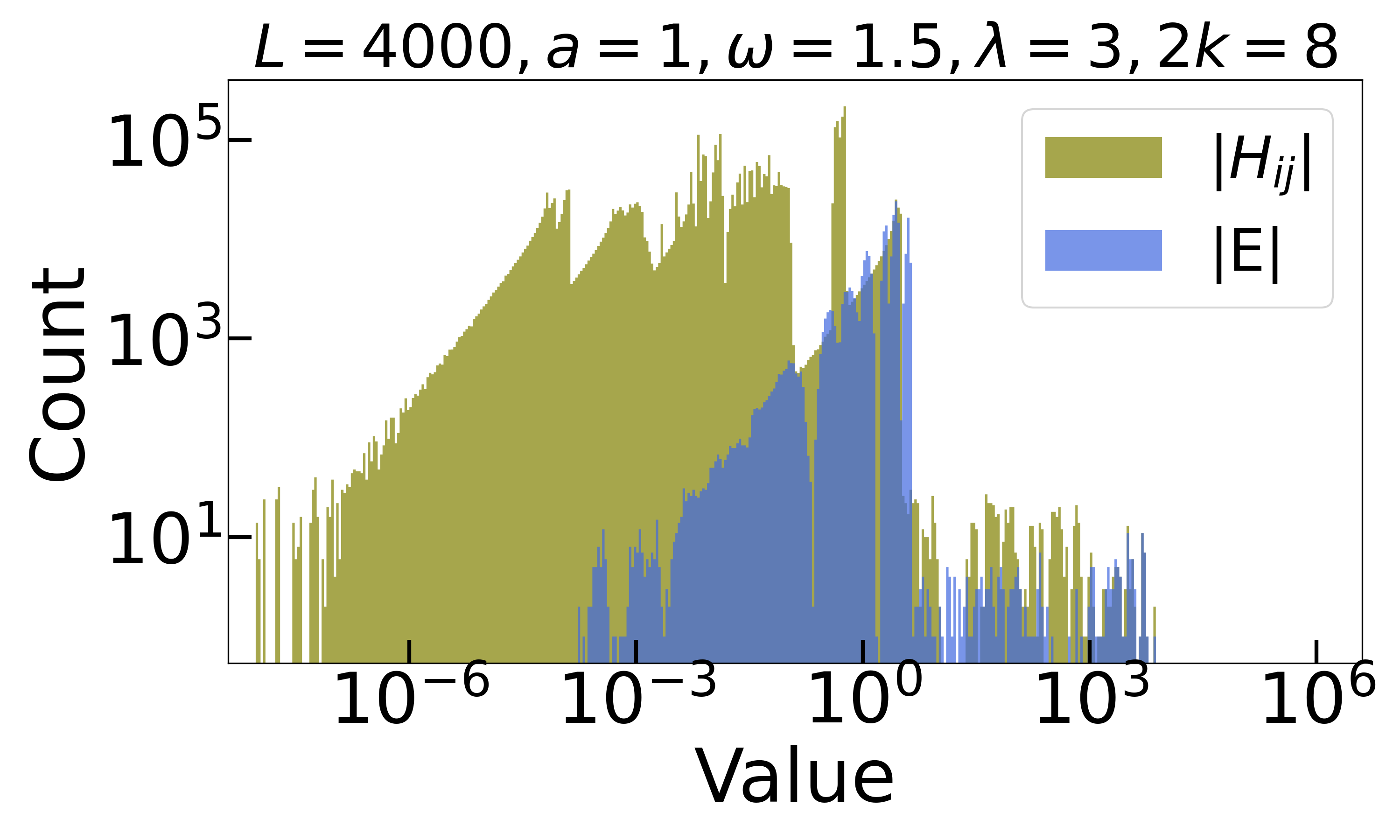}
\label{O_8_lamb_3_N_4000_a_1_om_1.5_AA}}
\hspace{-0.35cm}
\subfigure[]{
\includegraphics[width=0.24\linewidth]{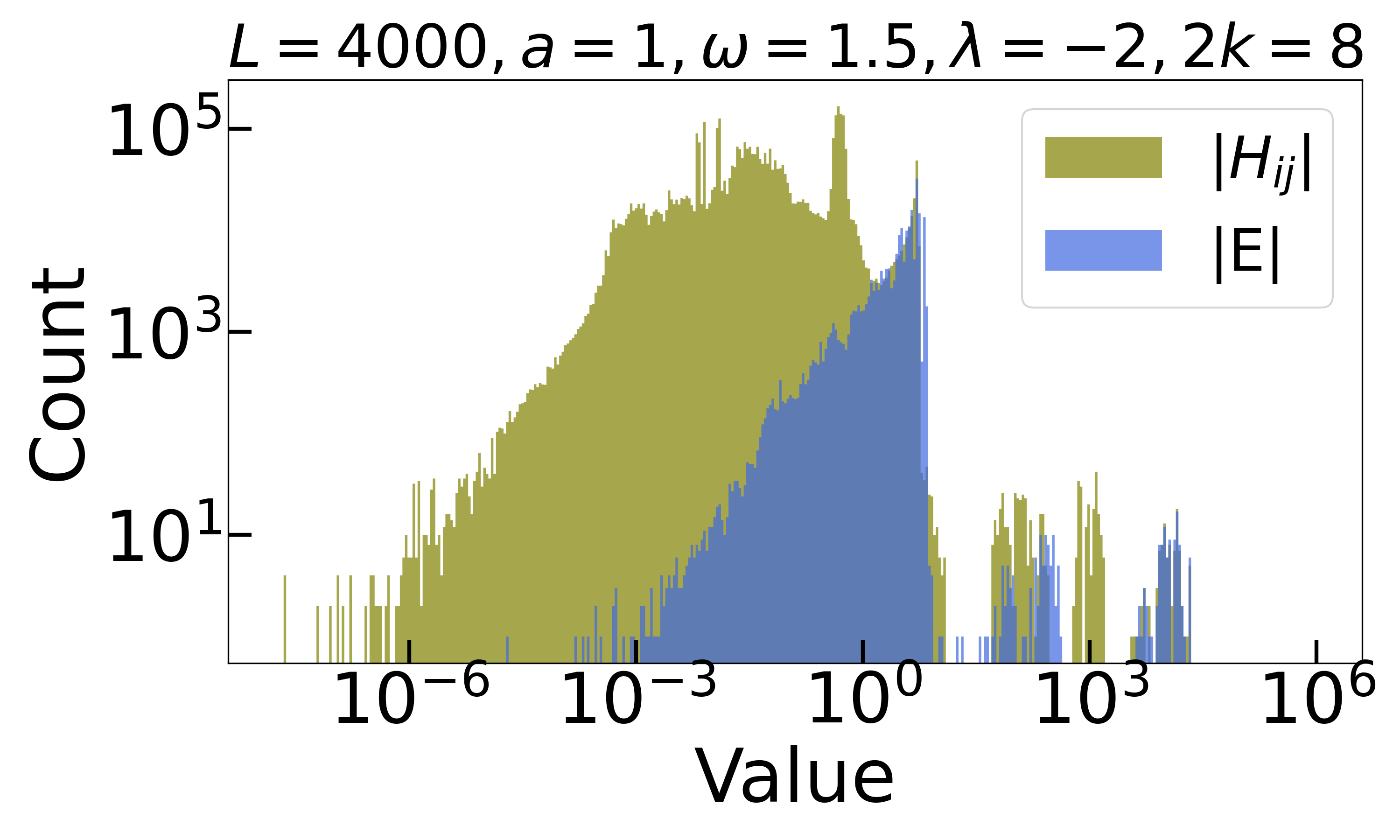}
\label{O_8_lamb_-2_N_4000_a_1_om_1.5_GAA}}\\ %new row
%%%%%%%%%%%%%%%%%%%%%%%%%%%%%%%%%%%%%%%%%%%%%%%%%%%%%%%
\subfigure[]{
\includegraphics[width=0.24\linewidth]{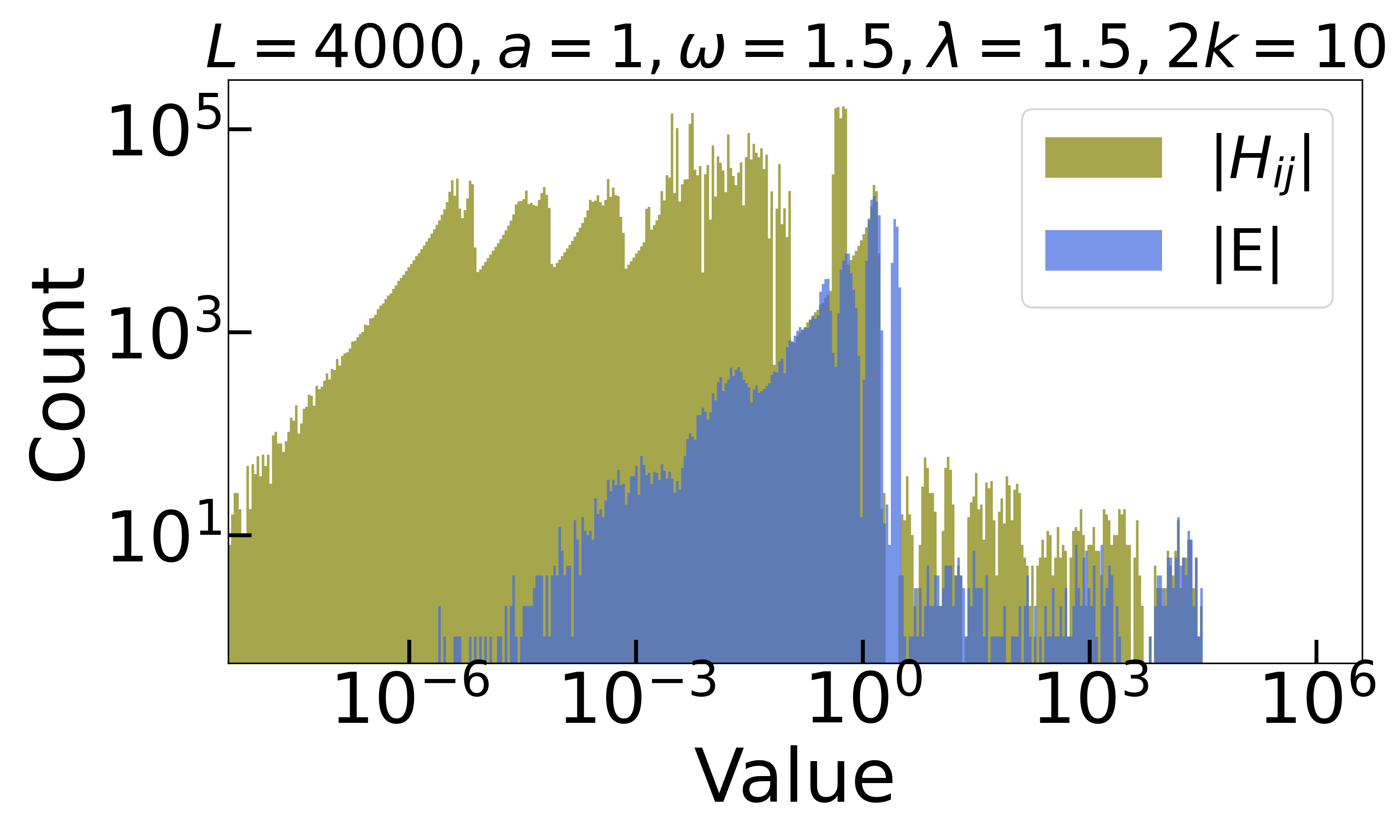}\label{O_10_lamb_1.5_N_4000_a_1_om_1.5_AA}}
\hspace{-0.35cm}
\subfigure[]{
\includegraphics[width=0.24\linewidth]{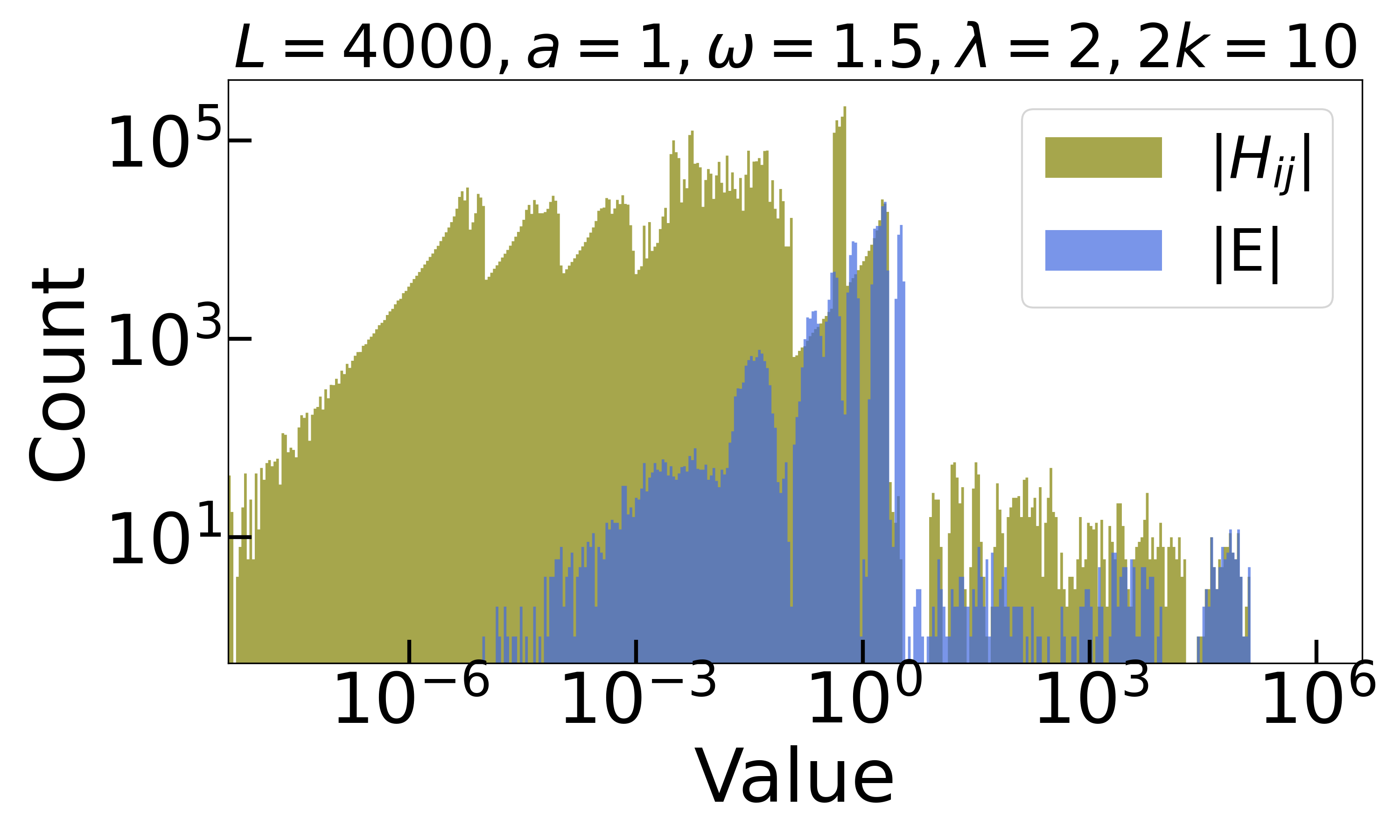}
\label{O_10_lamb_2_N_4000_a_1_om_1.5_AA}}
\subfigure[]{
\includegraphics[width=0.24\linewidth]{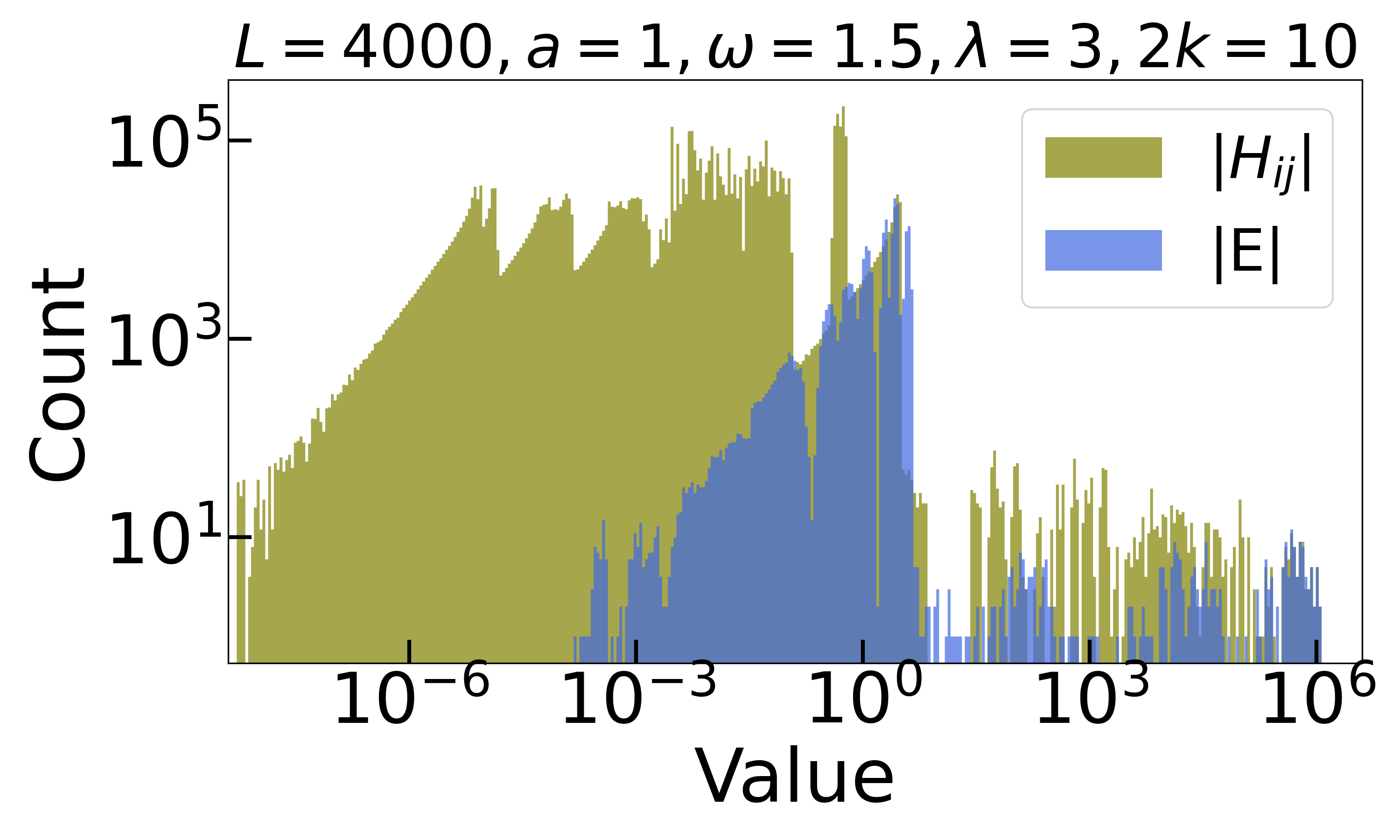}
\label{O_10_lamb_3_N_4000_a_1_om_1.5_AA}}
\hspace{-0.35cm}
\subfigure[]{
\includegraphics[width=0.24\linewidth]{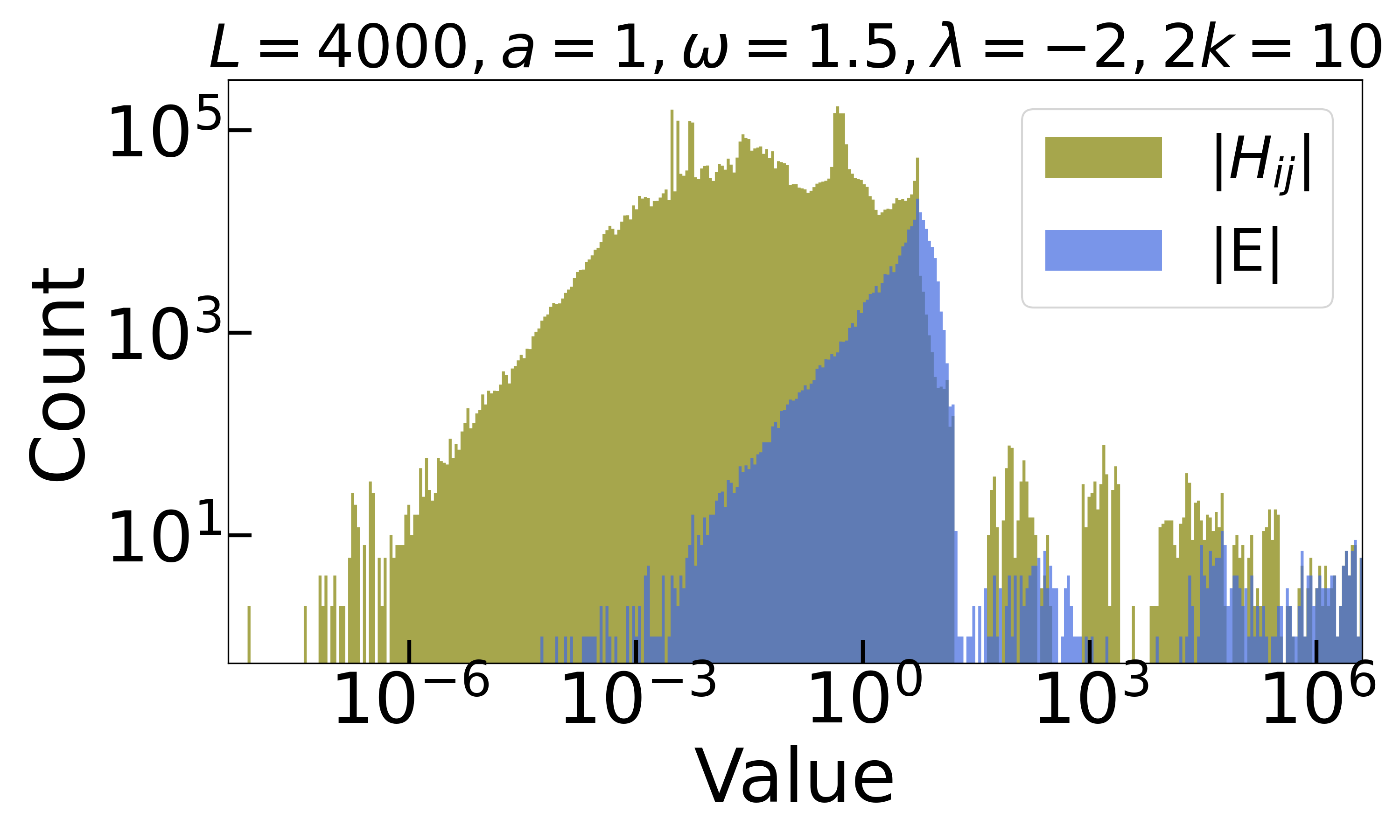}
\label{O_10_lamb_-2_N_4000_a_1_om_1.5_GAA}}\\ %new row
%%%%%%%%%%%%%%%%%%%%%%%%%%%%%%%%%%%%%%%%%%%%%%%%%%%%%%%
%%%%%%%%%%%%%%%%%%%%%%%%%%%%%%%%%%%%%%%%%%%%%%%%%%%%%%%
%%%%%%%%%%%%%%%%%%%%%%%%%%%%%%%%%%%%%%%%%%%%%%%%%%%%%%%
\captionsetup{justification=centerlast, width=\linewidth} 
\caption{{\bf{a)-d)}} Represents matrix element distributions $|H_{i,j}|$ and eigenvalue distributions $|E|$ for the AA model with $\lambda=1.5,2,3$ and the GAA model with $\lambda=-2,\beta=0.5$, evaluated at $(a,\omega)=(1,1.5)$ and lower order $2k=2$. Panels {\bf{e)-h)}}, {\bf{i)-l)}}, {\bf{m)-p)}} and {\bf{q)-t)}} show the corresponding distributions after including higher order corrections $2k=4,6,8$ and $10$ respectively to the second order effective Hamiltonian. As $2k$ increases, the $H_{i,j}$ distribution develops a pronounced  HET and LET. It is evident the $|E|$ distribution is significantly more sensitive to HET, exhibiting a monotonic growth in its tail consistent with the increase in large $|H_{i,j}|$ elements of $H_{eff}$.}
\end{figure*}

\begin{figure*}
\centering

\subfigure[]{
\includegraphics[width=0.24\linewidth]{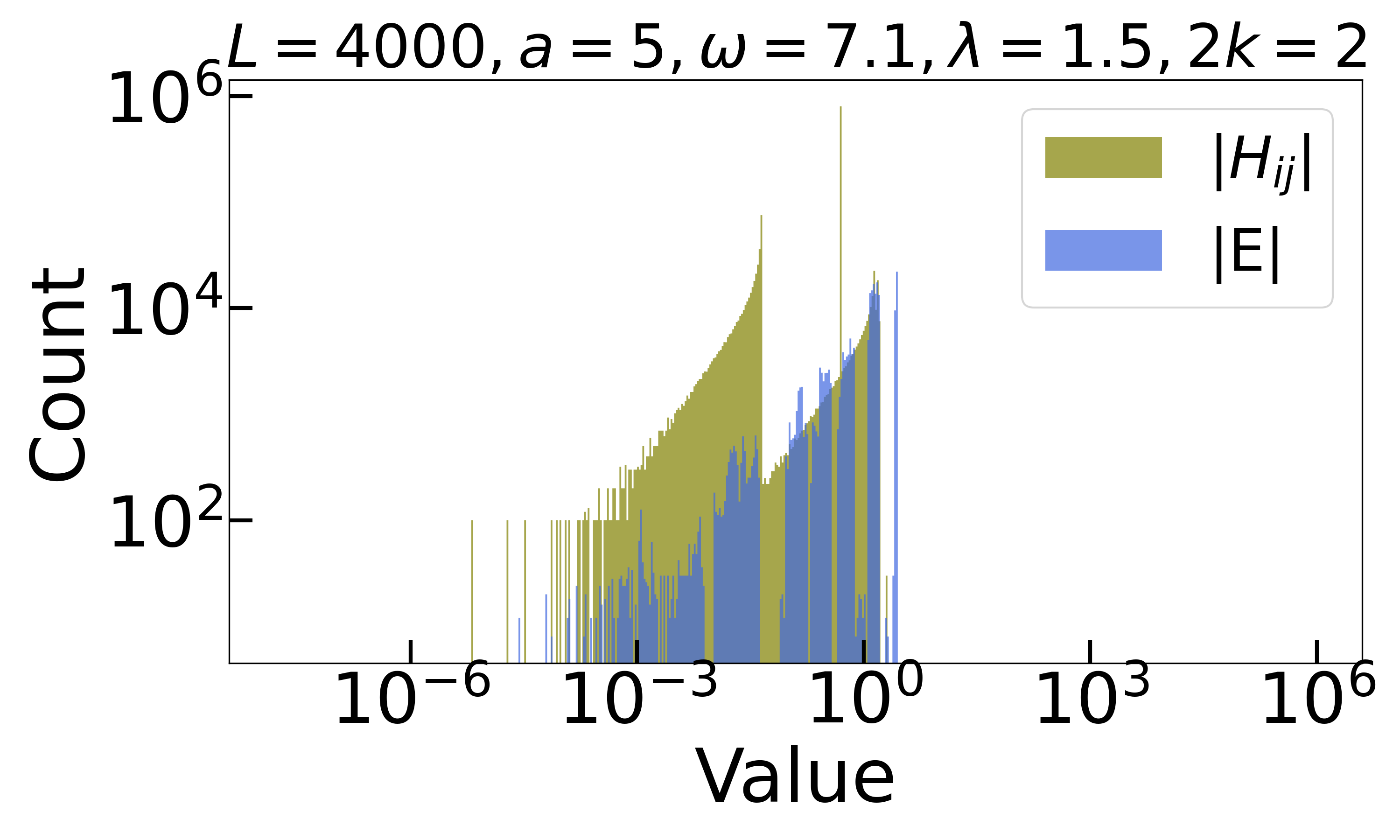}\label{O_2_lamb_1.5_N_4000_a_5_om_7.1_AA}}
\hspace{-0.35cm}
\subfigure[]{
\includegraphics[width=0.24\linewidth]{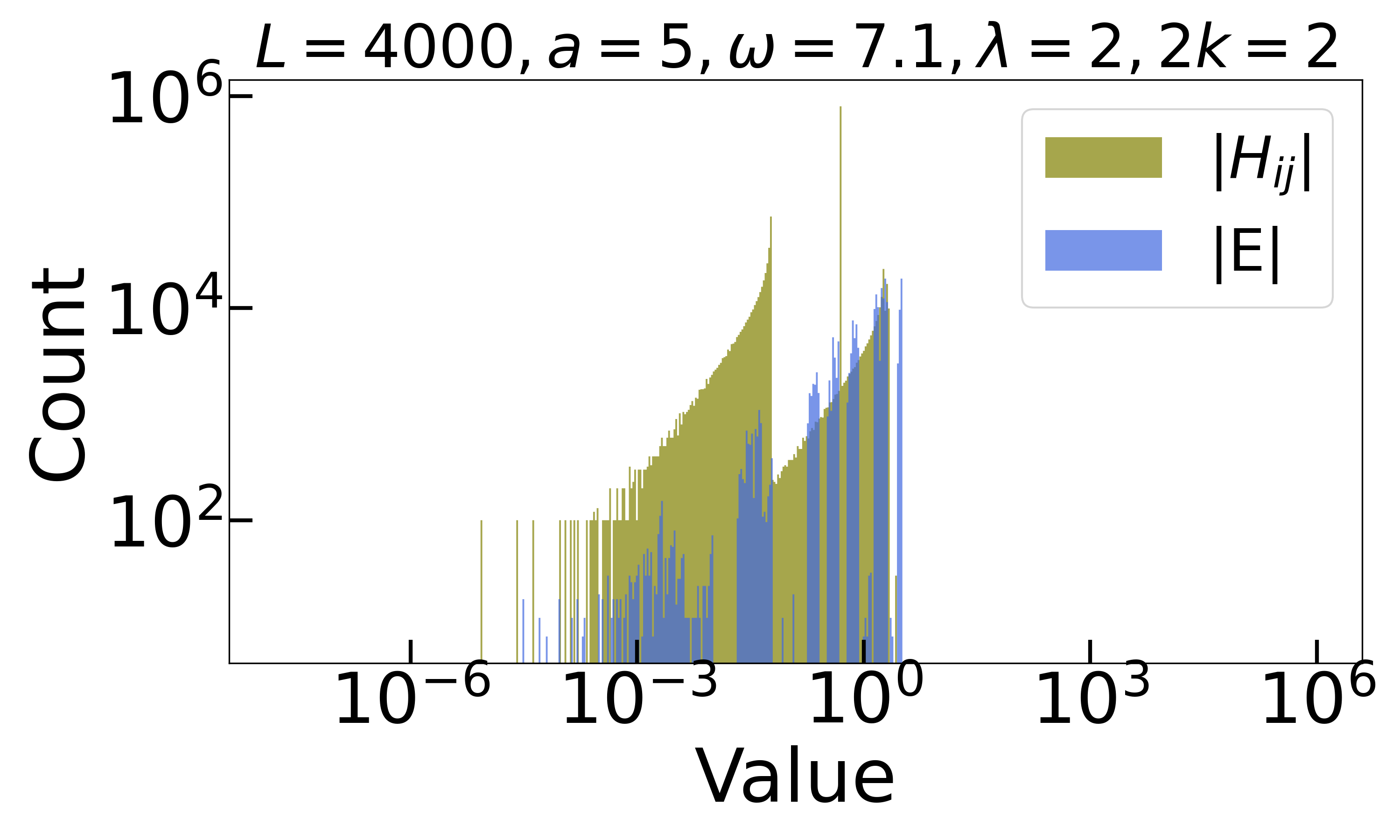}
\label{O_2_lamb_2_N_4000_a_5_om_7.1_AA}}
\subfigure[]{
\includegraphics[width=0.24\linewidth]{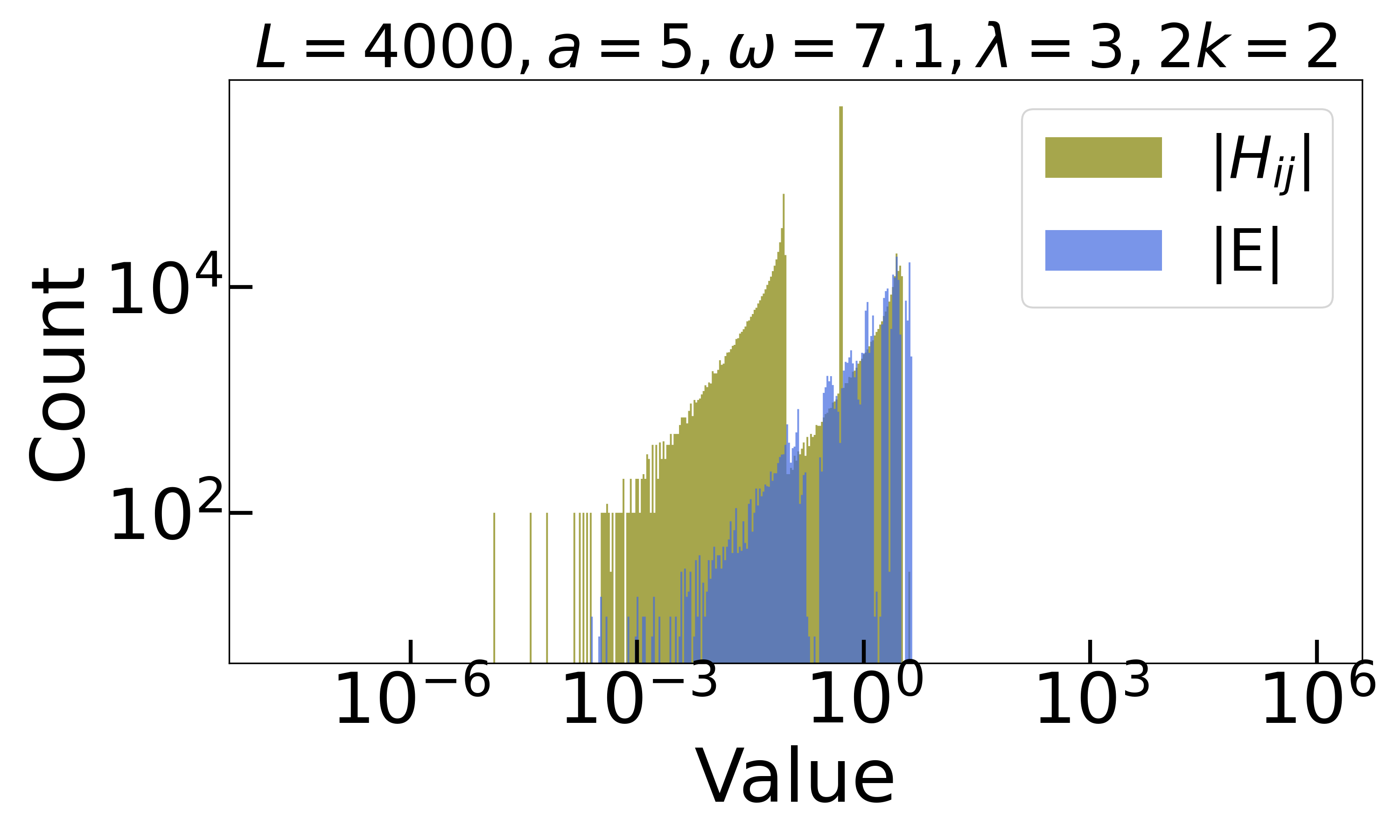}
\label{O_2_lamb_3_N_4000_a_5_om_7.1_AA}}
\hspace{-0.35cm}
\subfigure[]{
\includegraphics[width=0.24\linewidth]{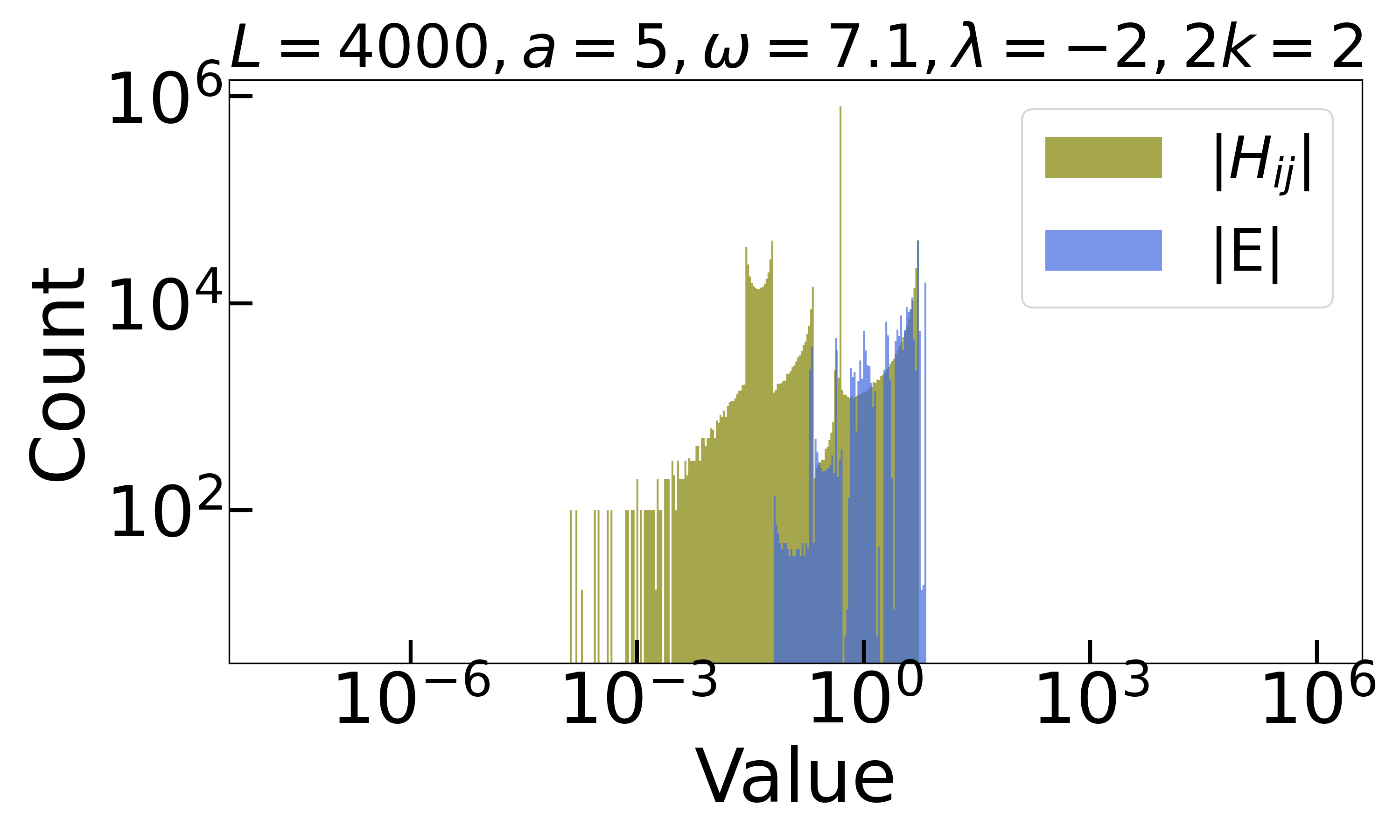}
\label{O_2_lamb_-2_N_4000_a_5_om_7.1_GAA}}\\ %new row
%%%%%%%%%%%%%%%%%%%%%%%%%%%%%%%%%%%%%%%%%%%%%%%%%%%%%%%
\subfigure[]{
\includegraphics[width=0.24\linewidth]{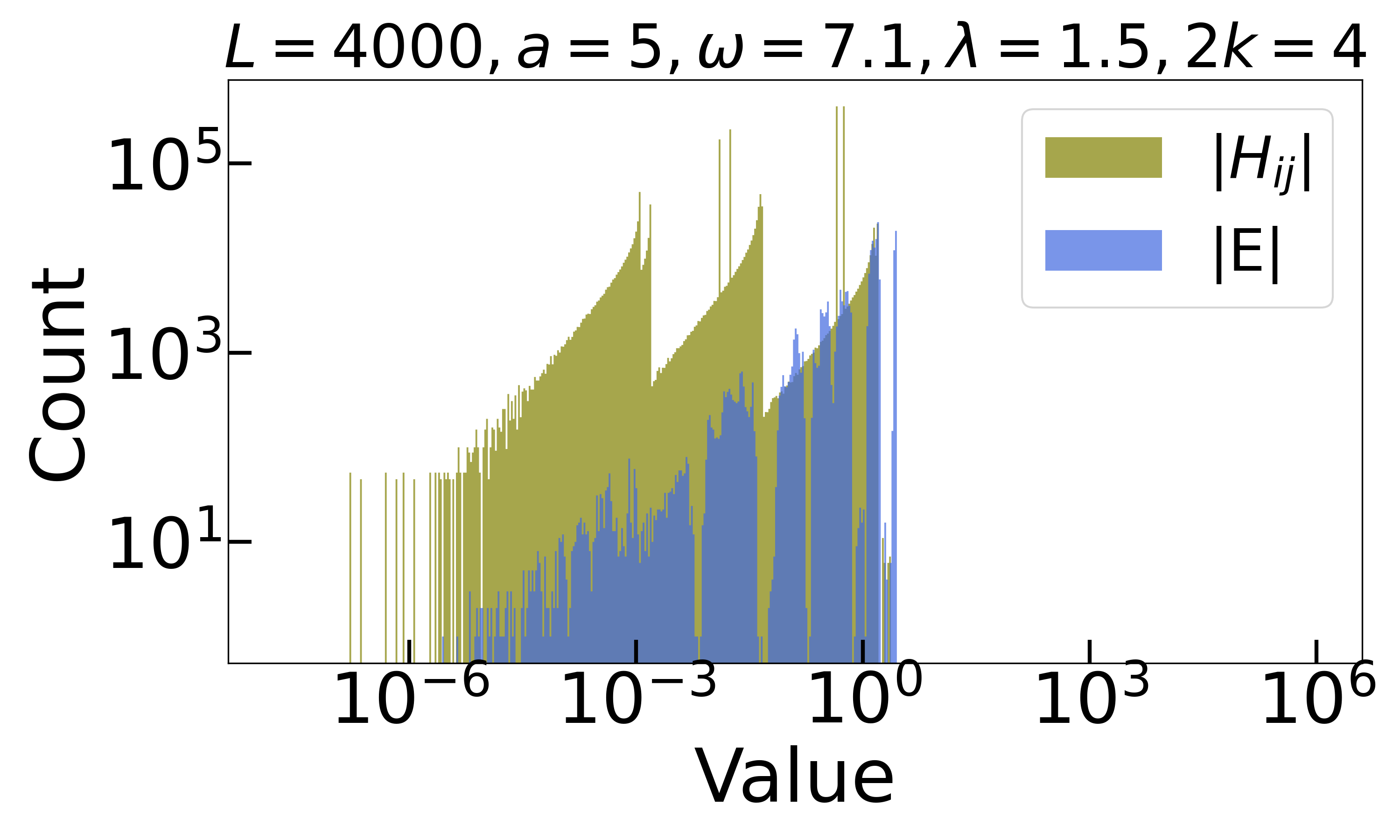}\label{O_4_lamb_1.5_N_4000_a_5_om_7.1_AA}}
\hspace{-0.35cm}
\subfigure[]{
\includegraphics[width=0.24\linewidth]{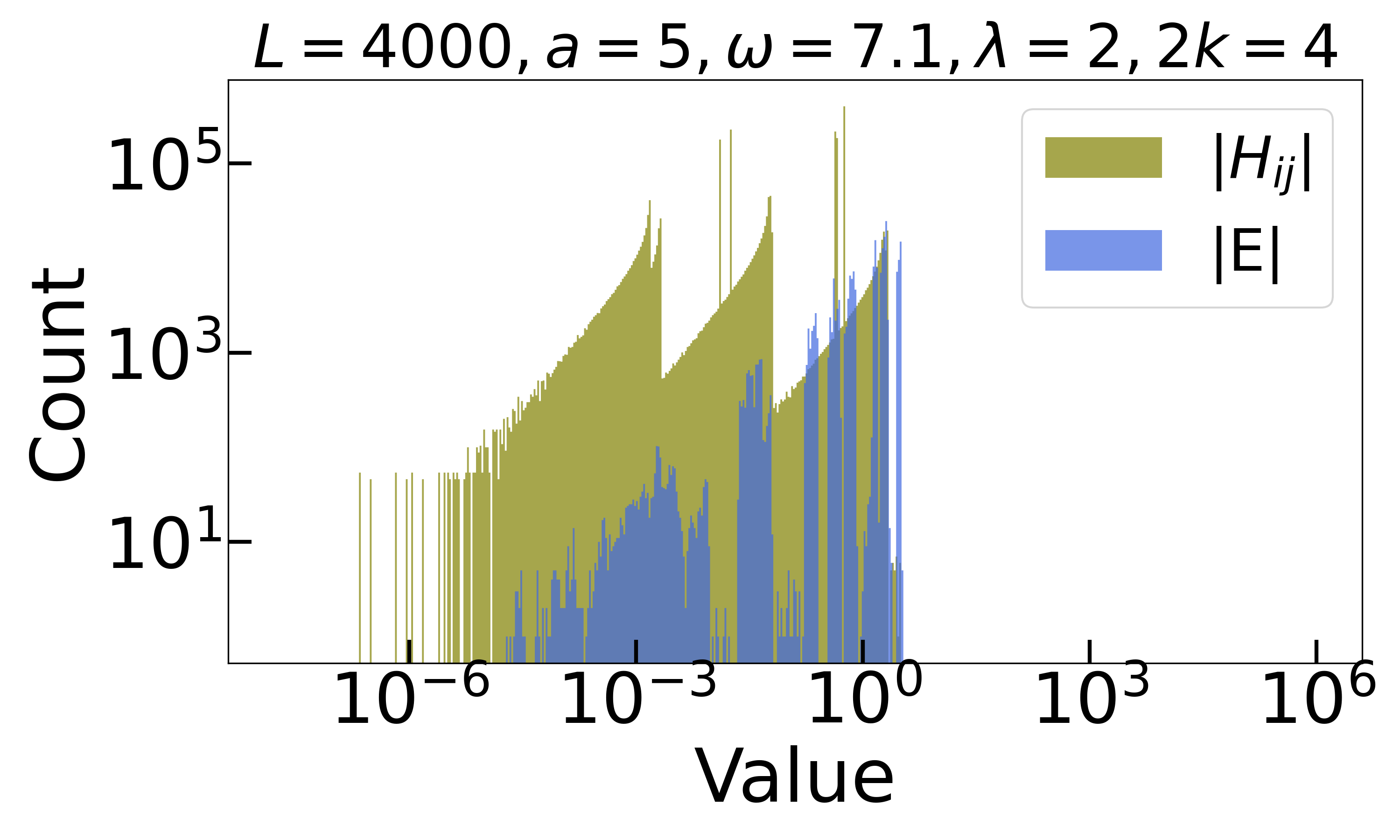}
\label{O_4_lamb_2_N_4000_a_5_om_7.1_AA}}
\subfigure[]{
\includegraphics[width=0.24\linewidth]{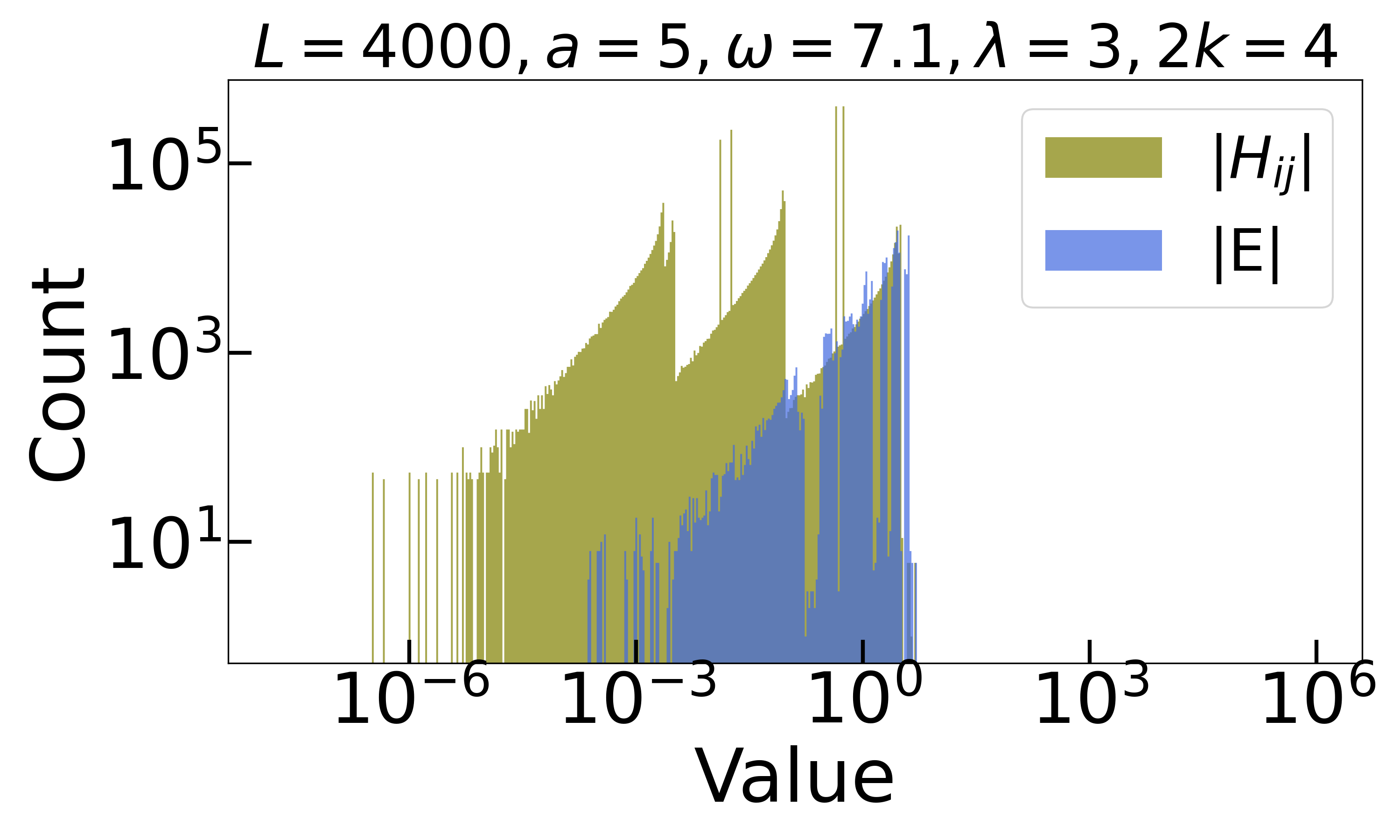}
\label{O_4_lamb_3_N_4000_a_5_om_7.1_AA}}
\hspace{-0.35cm}
\subfigure[]{
\includegraphics[width=0.24\linewidth]{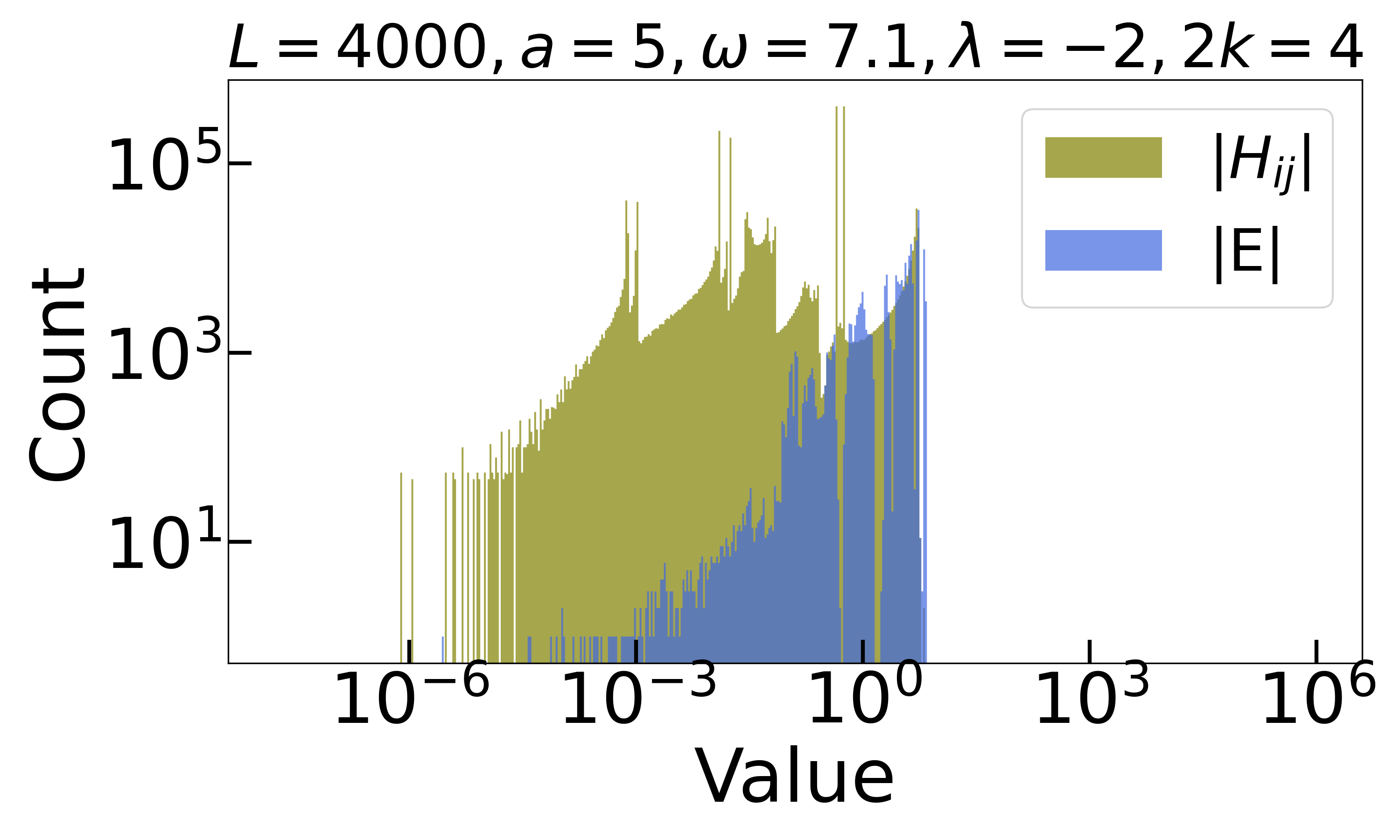}
\label{O_4_lamb_-2_N_4000_a_5_om_7.1_GAA}}\\ %new row
%%%%%%%%%%%%%%%%%%%%%%%%%%%%%%%%%%%%%%%%%%%%%%%%%%%%%%%
\subfigure[]{
\includegraphics[width=0.24\linewidth]{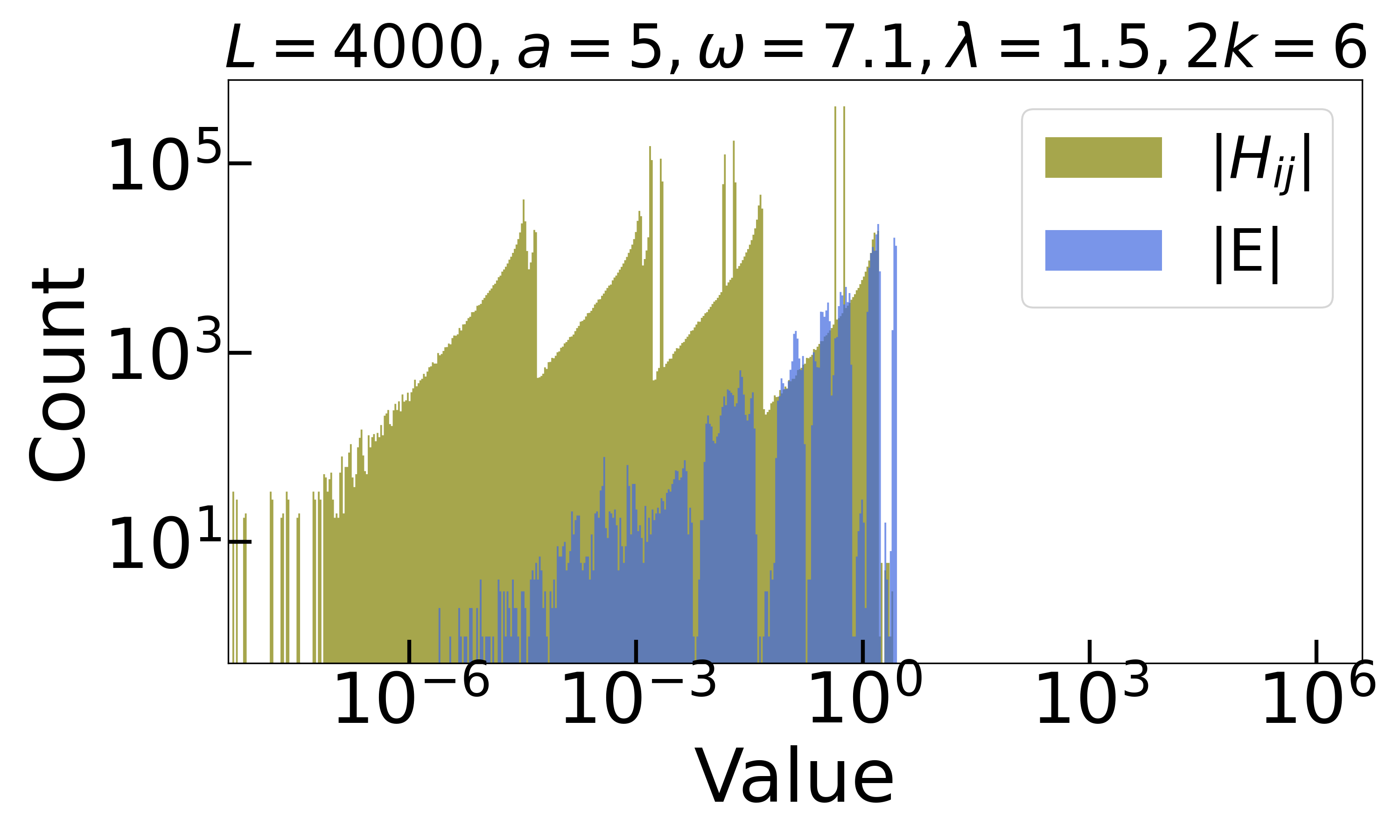}\label{O_6_lamb_1.5_N_4000_a_5_om_7.1_AA}}
\hspace{-0.35cm}
\subfigure[]{
\includegraphics[width=0.24\linewidth]{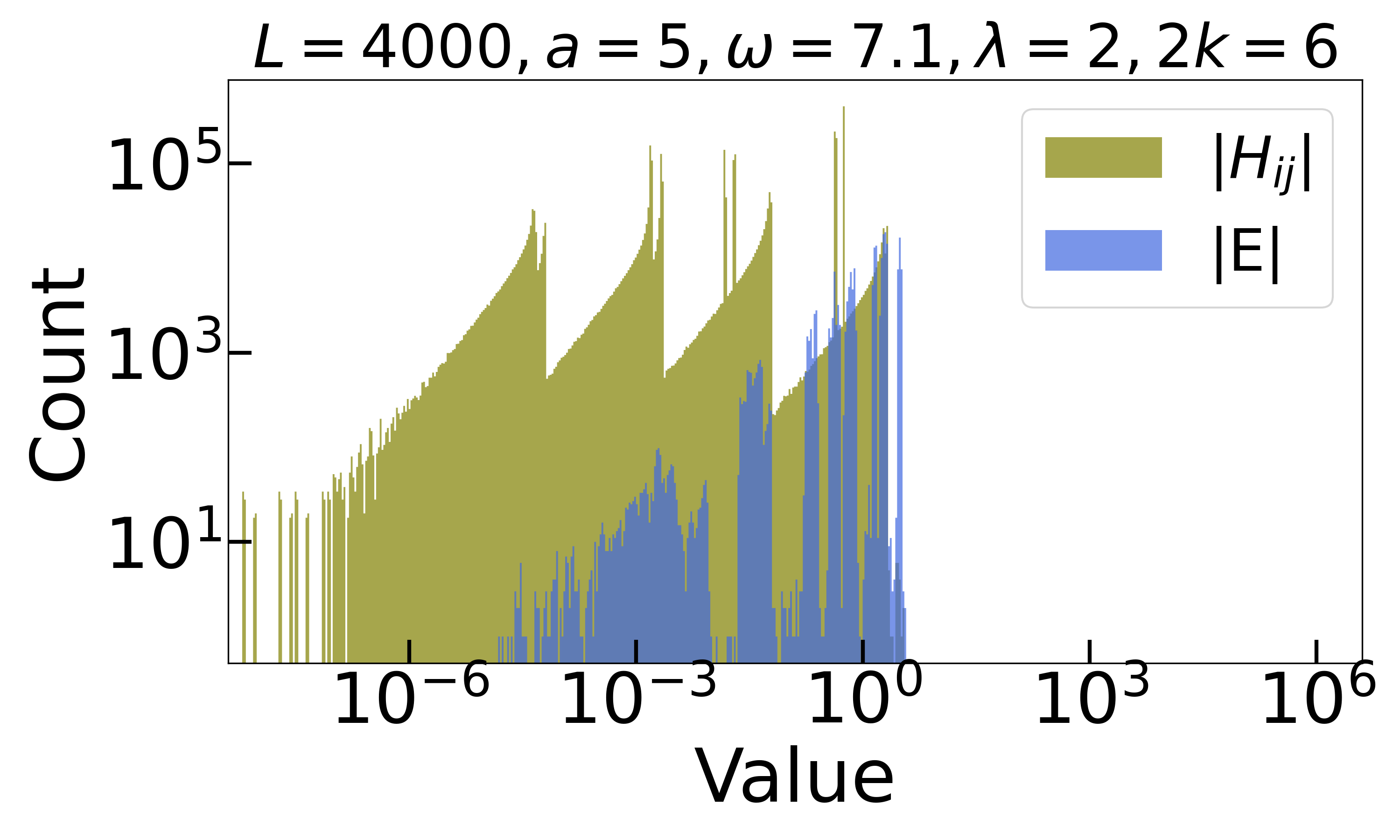}
\label{O_6_lamb_2_N_4000_a_5_om_7.1_AA}}
\subfigure[]{
\includegraphics[width=0.24\linewidth]{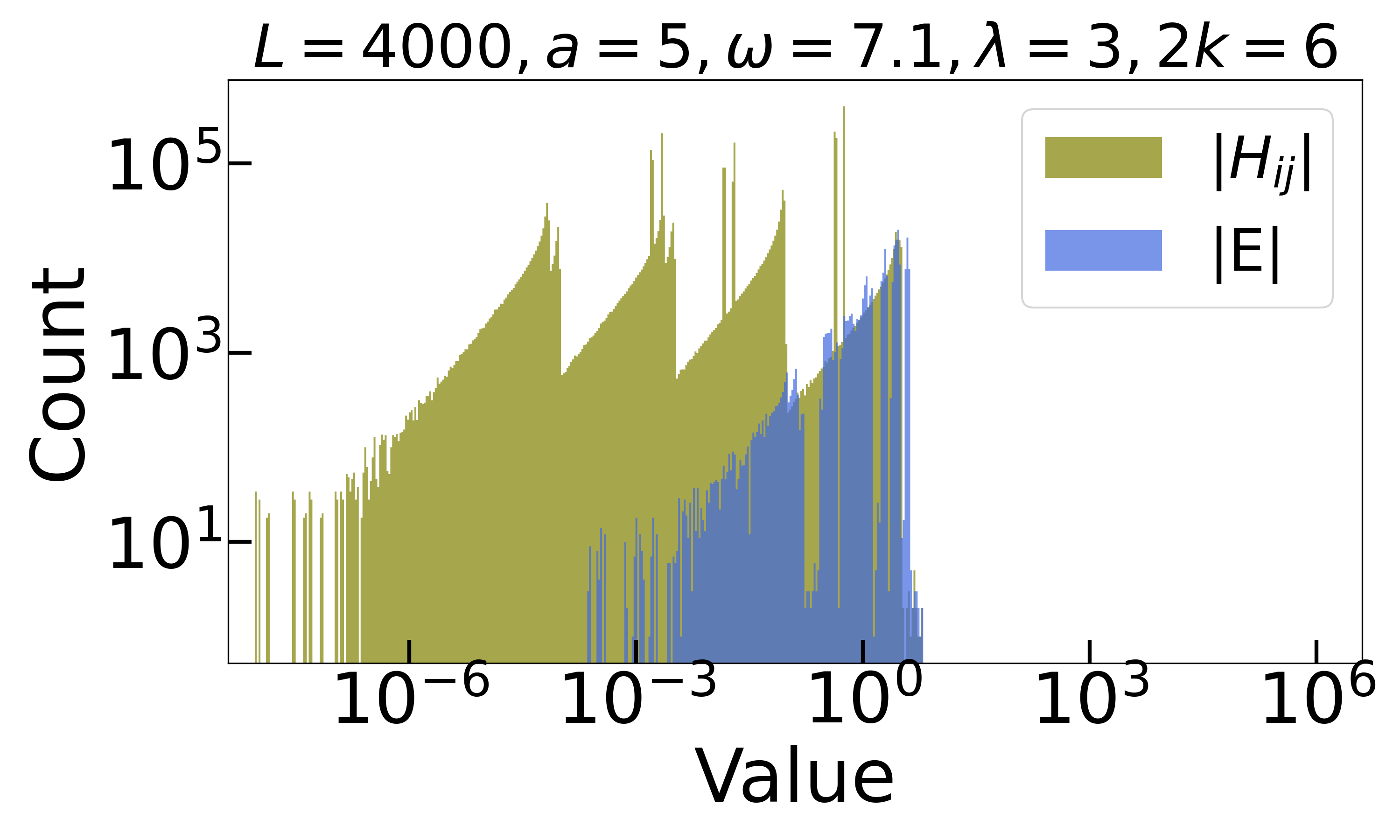}
\label{O_6_lamb_3_N_4000_a_5_om_7.1_AA}}
\hspace{-0.35cm}
\subfigure[]{
\includegraphics[width=0.24\linewidth]{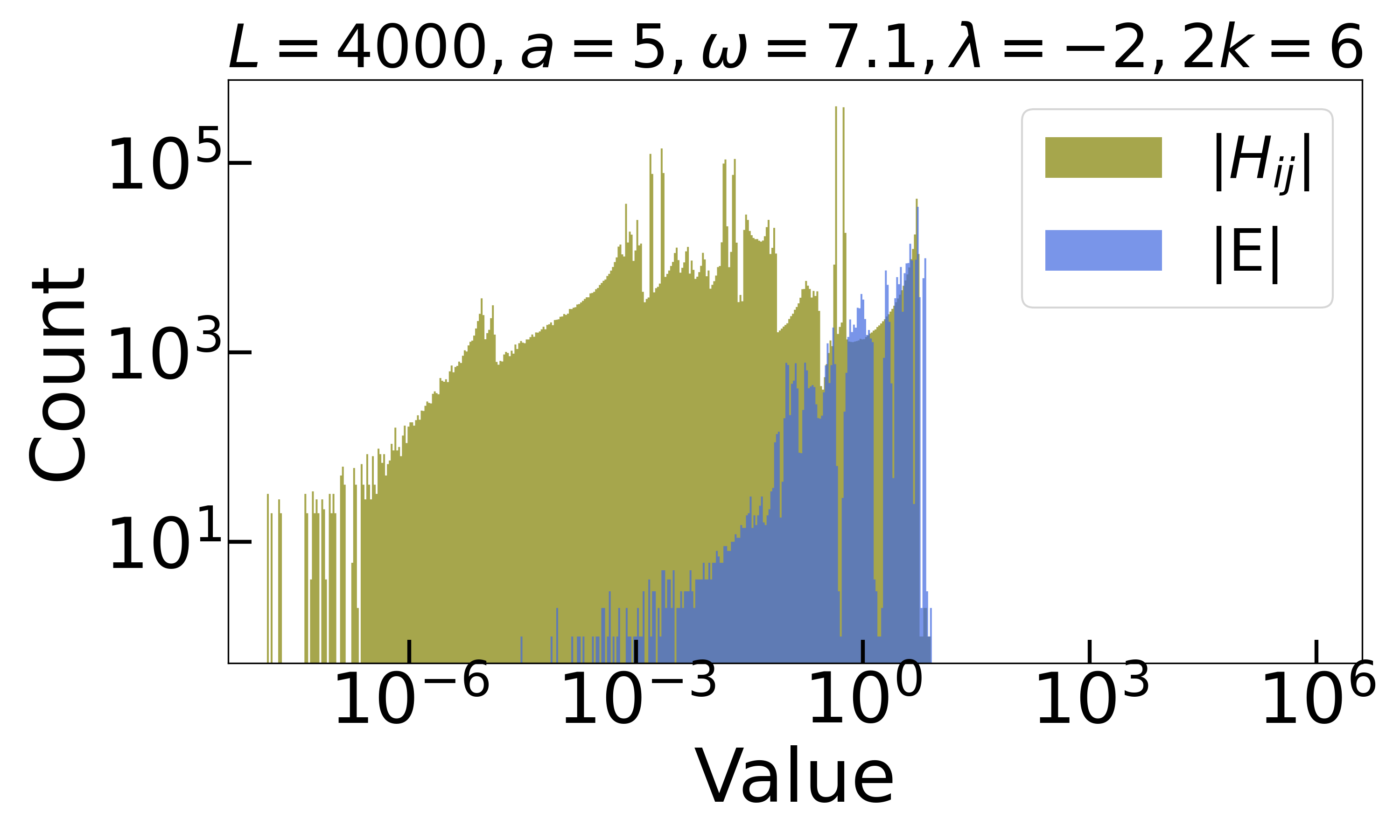}
\label{O_6_lamb_-2_N_4000_a_5_om_7.1_GAA}}\\ %new row
%%%%%%%%%%%%%%%%%%%%%%%%%%%%%%%%%%%%%%%%%%%%%%%%%%%%%%%
\subfigure[]{
\includegraphics[width=0.24\linewidth]{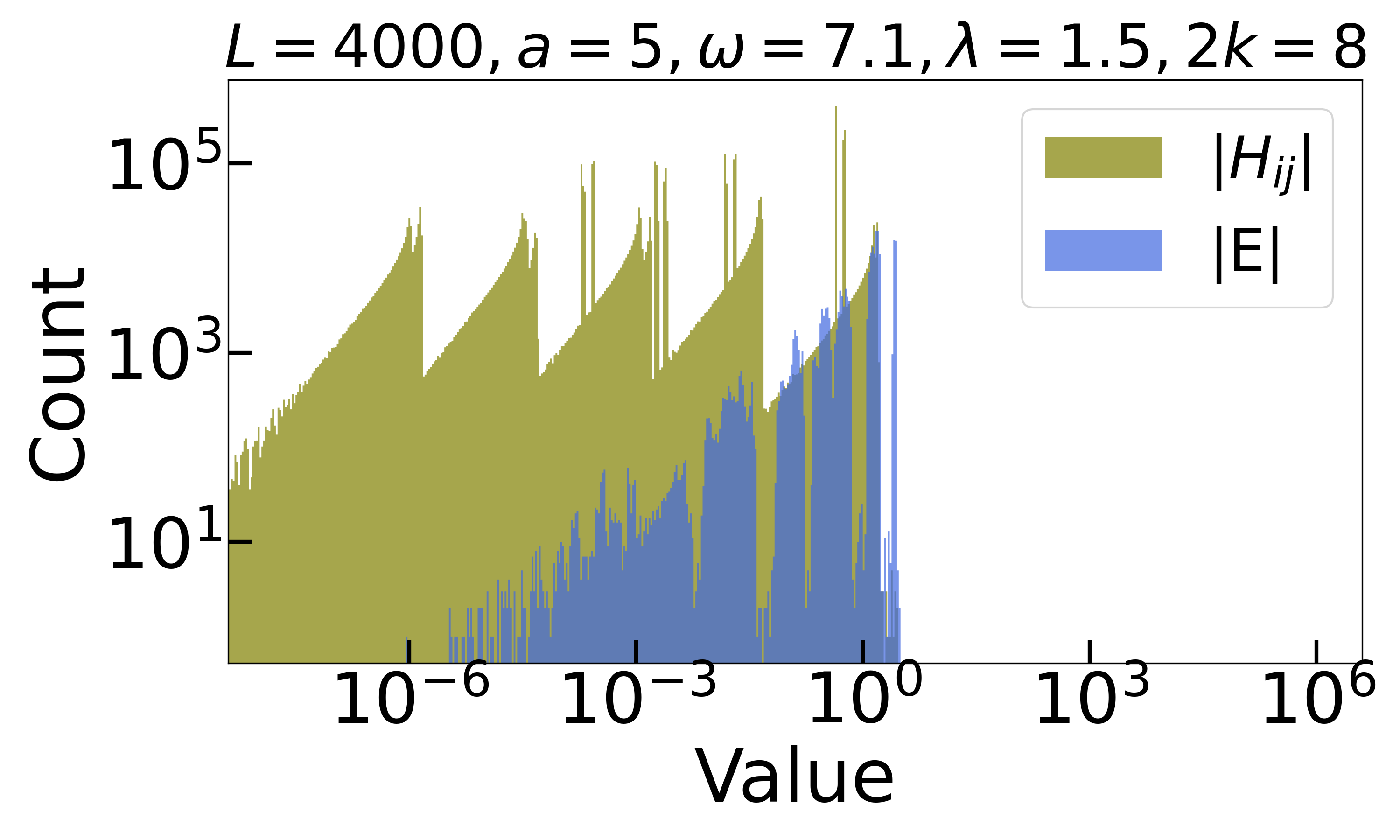}\label{O_8_lamb_1.5_N_4000_a_5_om_7.1_AA}}
\hspace{-0.35cm}
\subfigure[]{
\includegraphics[width=0.24\linewidth]{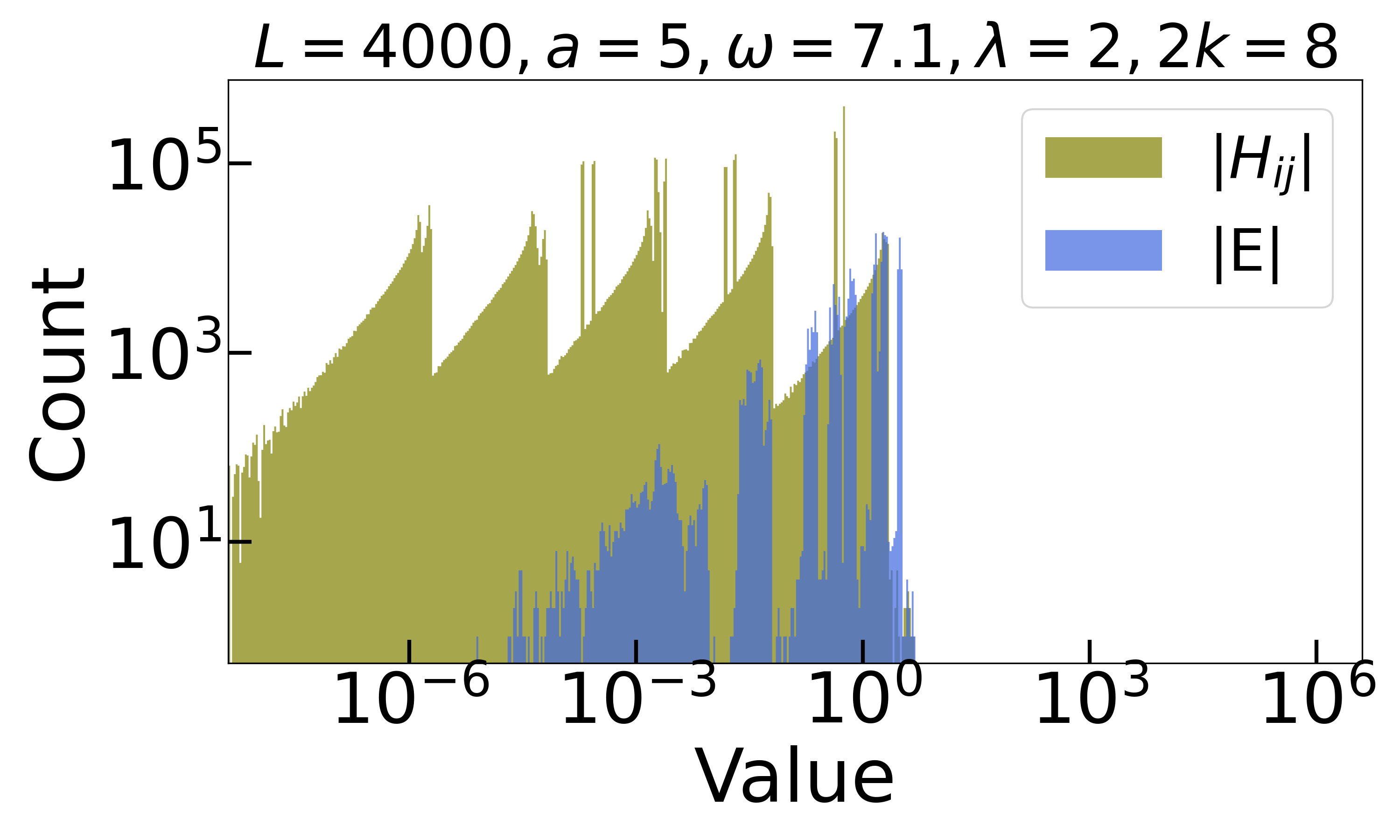}
\label{O_8_lamb_2_N_4000_a_5_om_7.1_AA}}
\subfigure[]{
\includegraphics[width=0.24\linewidth]{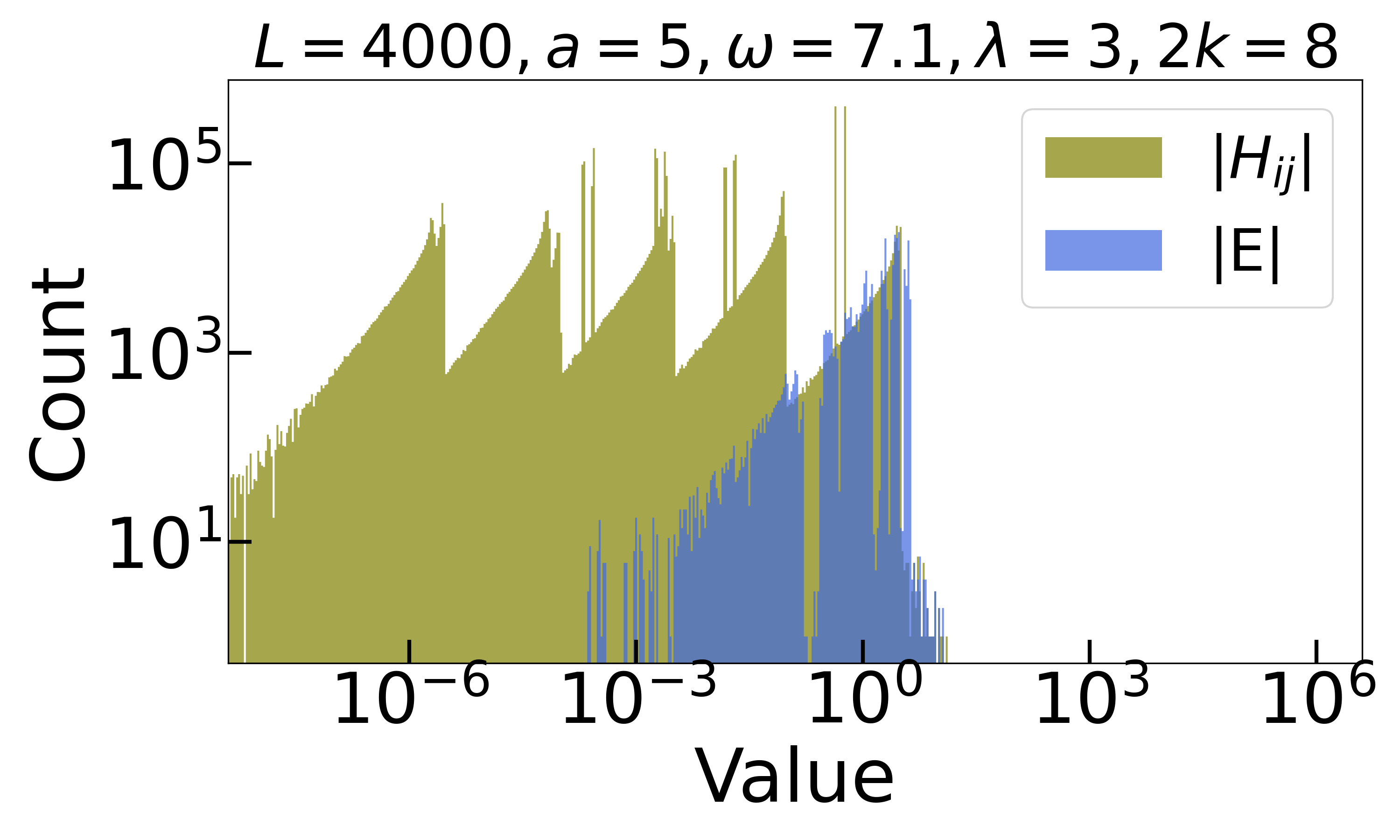}
\label{O_8_lamb_3_N_4000_a_5_om_7.1_AA}}
\hspace{-0.35cm}
\subfigure[]{
\includegraphics[width=0.24\linewidth]{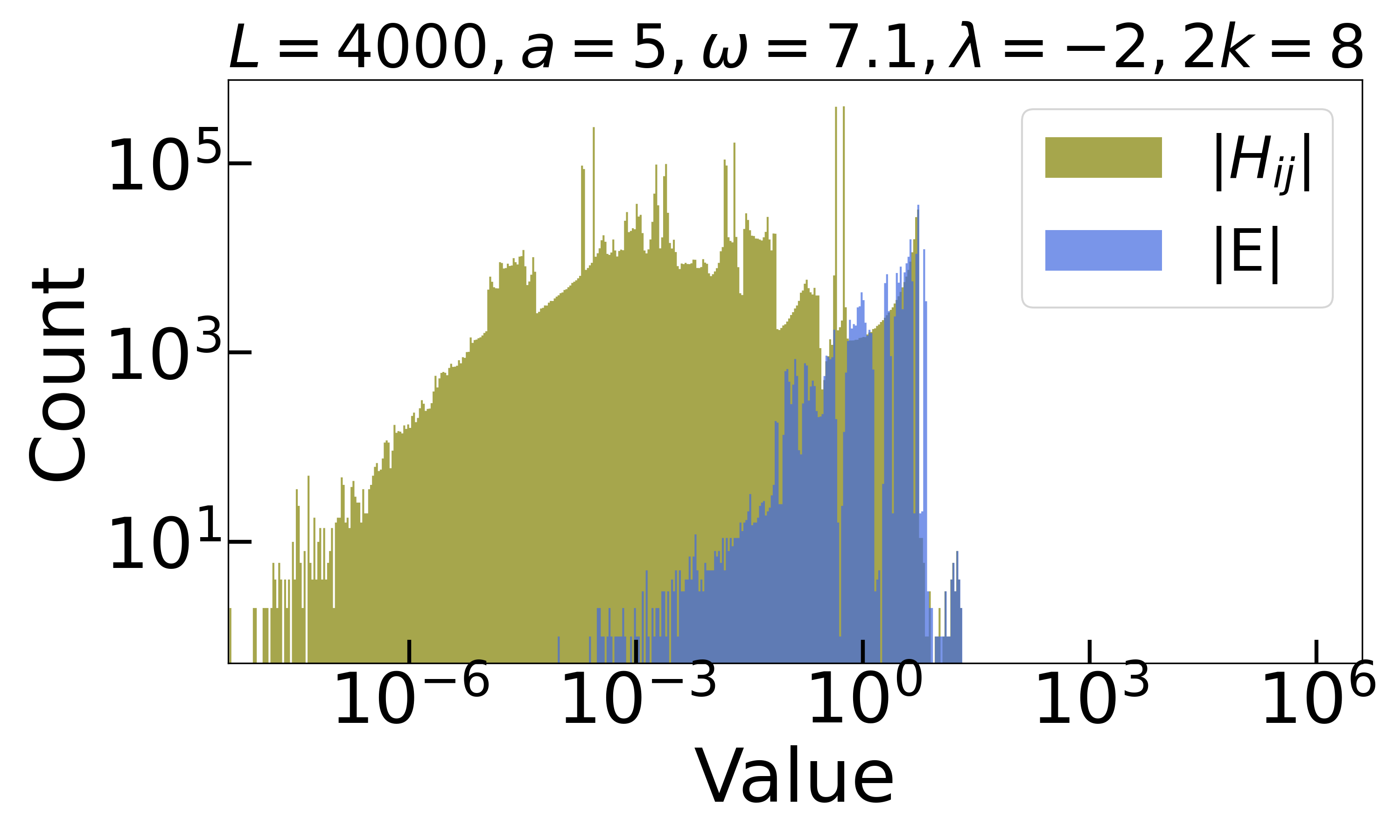}
\label{O_8_lamb_-2_N_4000_a_5_om_7.1_GAA}}\\ %new row
%%%%%%%%%%%%%%%%%%%%%%%%%%%%%%%%%%%%%%%%%%%%%%%%%%%%%%%
\subfigure[]{
\includegraphics[width=0.24\linewidth]{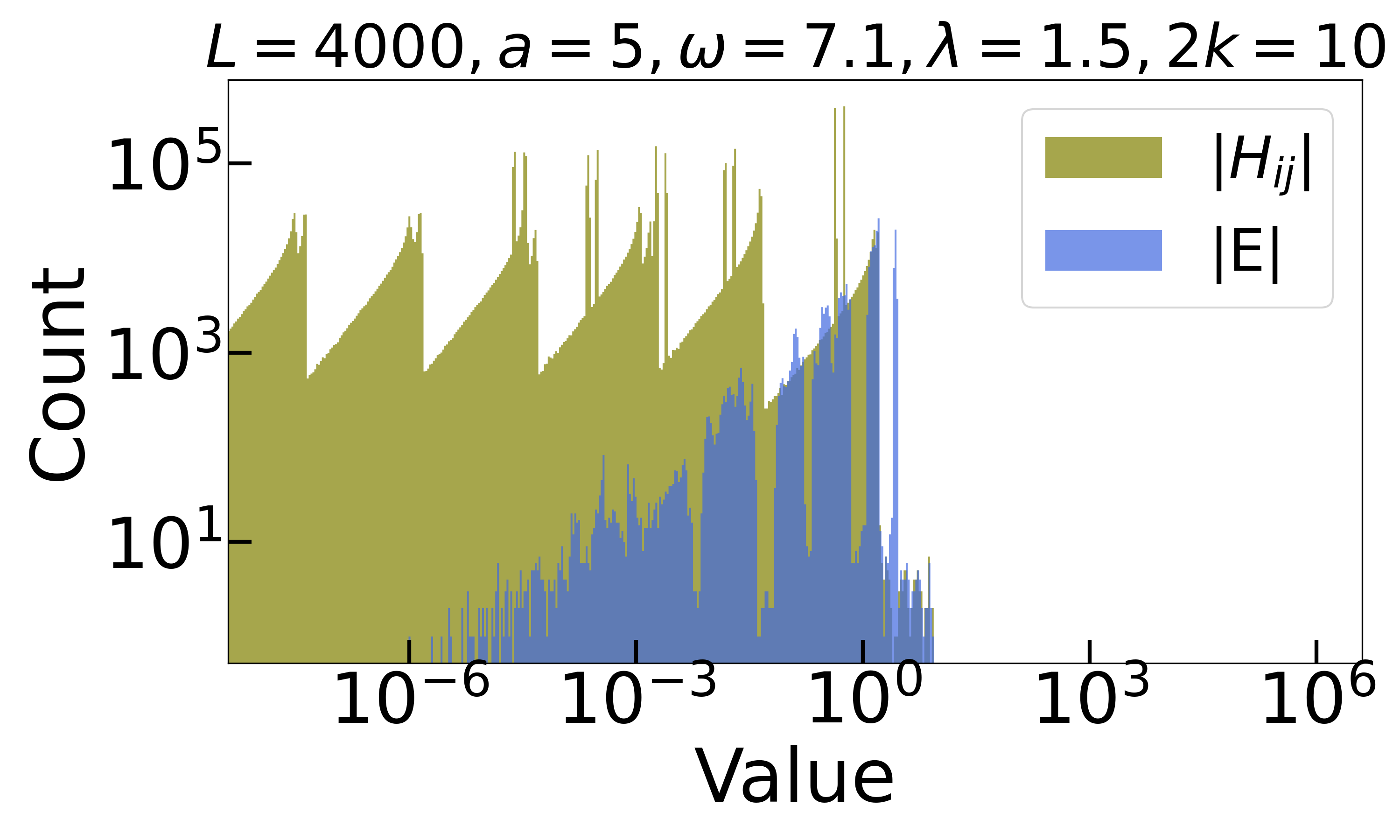}\label{O_10_lamb_1.5_N_4000_a_5_om_7.1_AA}}
\hspace{-0.35cm}
\subfigure[]{
\includegraphics[width=0.24\linewidth]{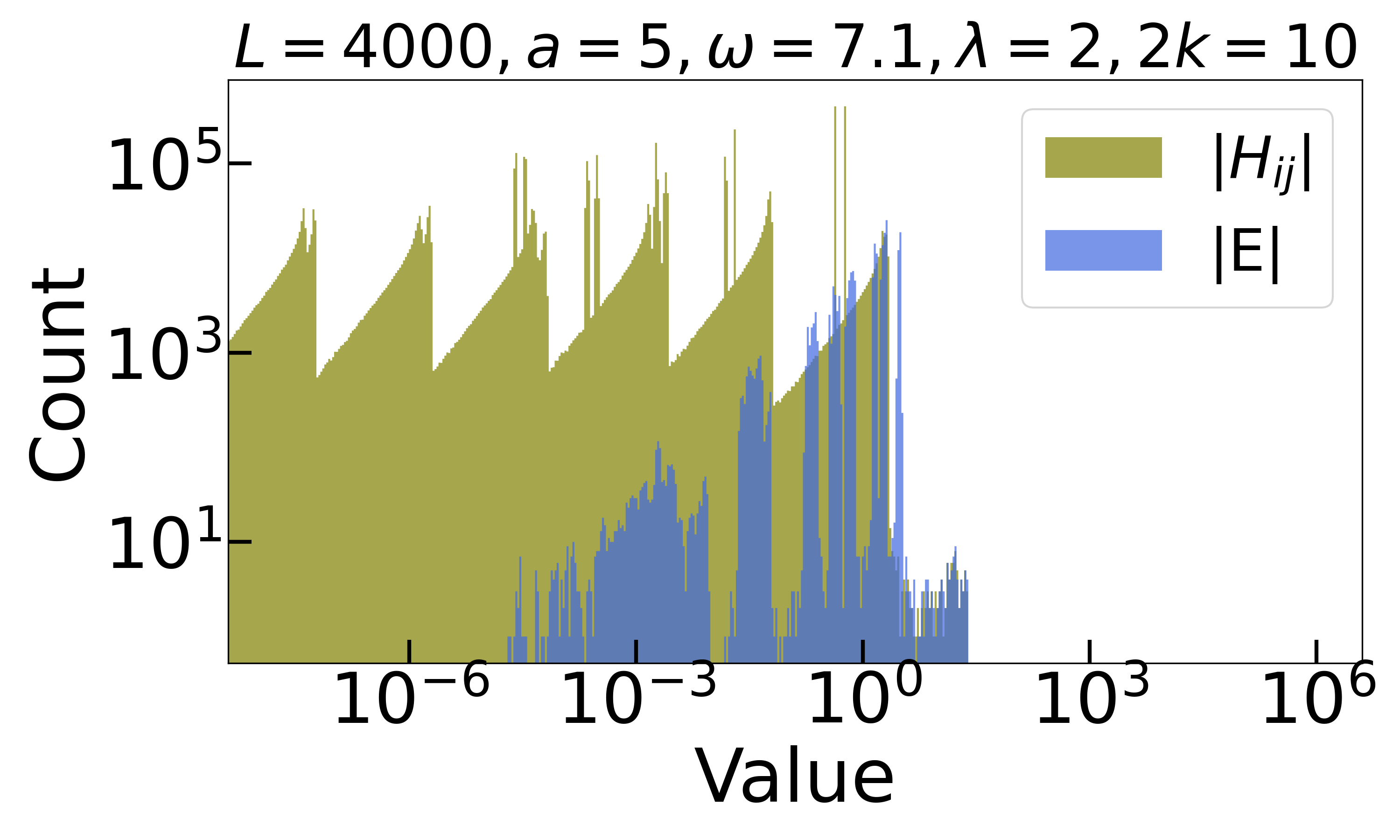}
\label{O_10_lamb_2_N_4000_a_5_om_7.1_AA}}
\subfigure[]{
\includegraphics[width=0.24\linewidth]{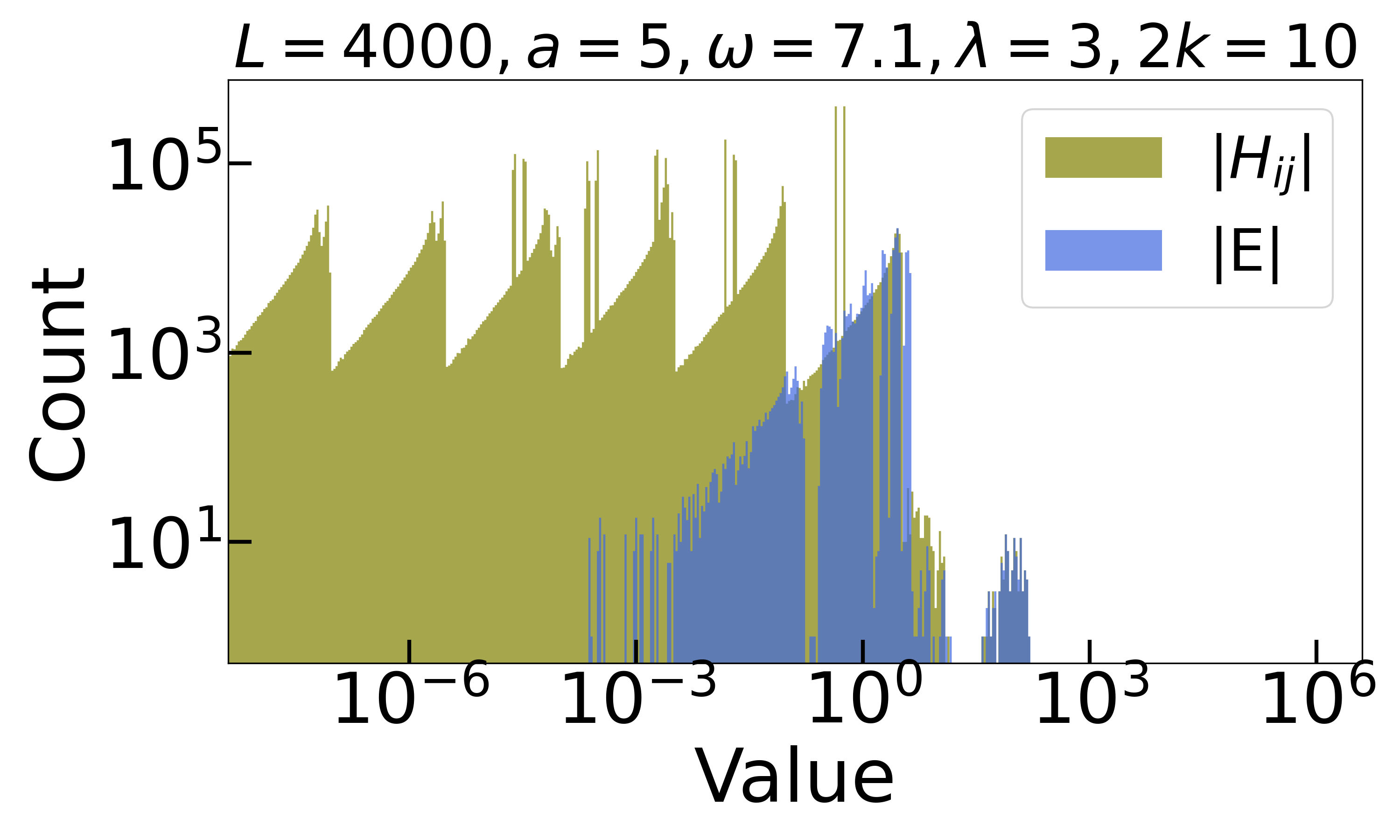}
\label{O_10_lamb_3_N_4000_a_5_om_7.1_AA}}
\hspace{-0.35cm}
\subfigure[]{
\includegraphics[width=0.24\linewidth]{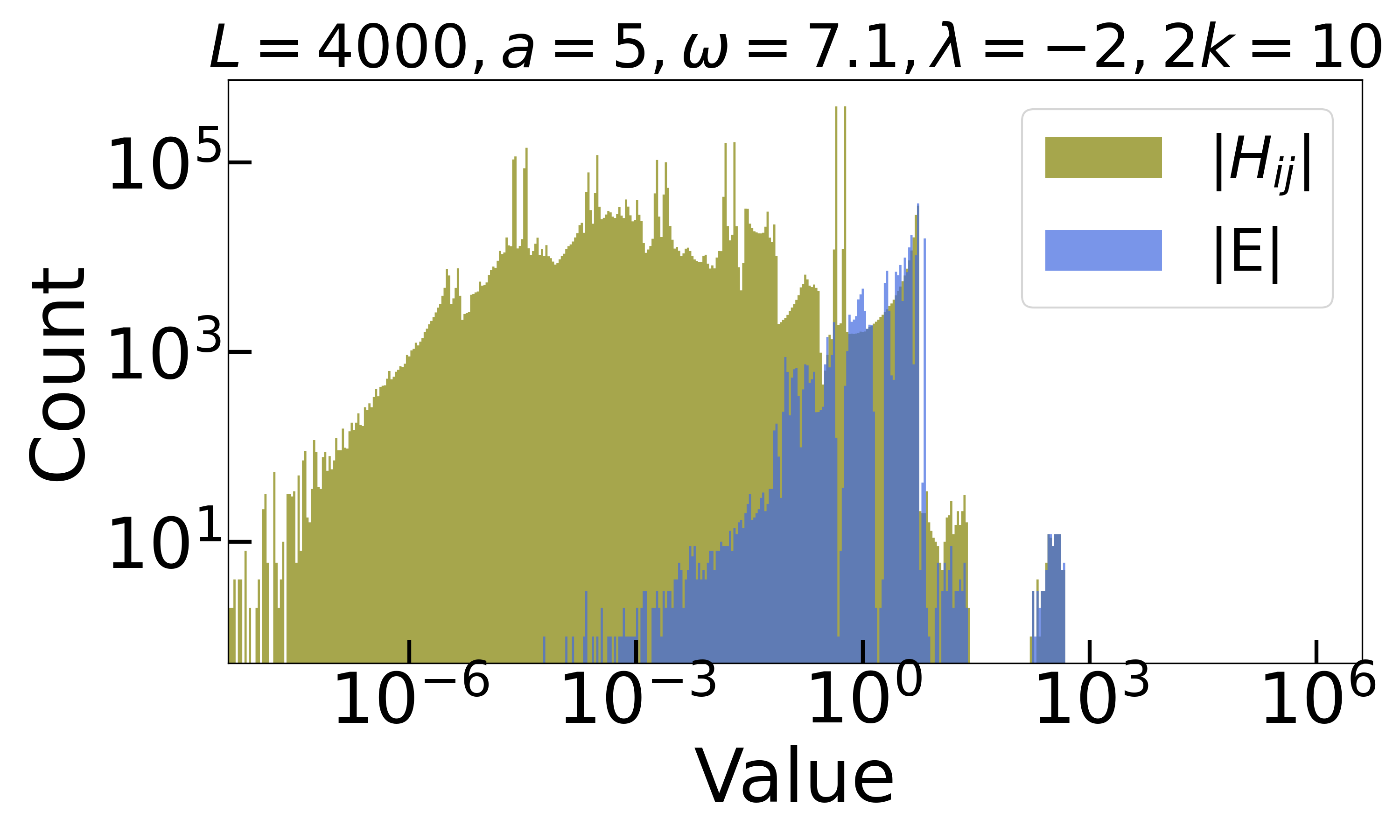}
\label{O_10_lamb_-2_N_4000_a_5_om_7.1_GAA}}\\ %new row
%%%%%%%%%%%%%%%%%%%%%%%%%%%%%%%%%%%%%%%%%%%%%%%%%%%%%%%
%%%%%%%%%%%%%%%%%%%%%%%%%%%%%%%%%%%%%%%%%%%%%%%%%%%%%%%
%%%%%%%%%%%%%%%%%%%%%%%%%%%%%%%%%%%%%%%%%%%%%%%%%%%%%%%
\captionsetup{justification=centerlast, width=\linewidth} 
\caption{{\bf{a)-d)}} Represents matrix element distributions $|H_{i,j}|$ and eigenvalue distributions $|E|$ for the AA model with $\lambda=1.5,2,3$ and the GAA model with $\lambda=-2,\beta=0.5$, evaluated at $(a,\omega)=(5,7.1)$ and lower order $2k=2$. Panels {\bf{e)-h)}}, {\bf{i)-l)}}, {\bf{m)-p)}} and {\bf{q)-t)}} show the corresponding distributions after including higher order corrections $2k=4,6,8$ and $10$ respectively to the second order effective Hamiltonian. As $2k$ increases, the $H_{i,j}$ distribution develops a pronounced  HET and LET. It is evident the $|E|$ distribution is significantly more sensitive to HET, exhibiting a monotonic growth in its tail consistent with the increase in large $|H_{i,j}|$ elements of $H_{eff}$.
}
\end{figure*}

We introduce three kinds of ratio test protocols to determine the  optimal truncation order of the VVPT. In each case, the ratio test is either defined by the matrix element (denoted by $H_{i,j}$) or eigenvalue of $H_{eff}$ at each $\mathcal{O}(2k)$ , which essentially represents the underlying $\bar{R}^{2k}_{i,j}$. We define, 
\begin{enumerate}
    \item $avg_H=\bar{R}^{2k+2}/\bar{R}^{2k}$, where $\bar{R}^{2k}=\langle|\bar{R}^{2k}_{i,j}|\rangle=\frac{\displaystyle \sum_{i,j:\, H_{ij}\neq 0} |H_{ij}|}{N_{nz}}$.
    \item $rms_H=R_{rms}^{2k+2}/R_{rms}^{2k}$, where $ R_{rms}^{2k}=\Bigg[\sqrt{\frac{1}{N_{\text{nz}}}\displaystyle \sum_{i,j\in \text{nonzero}} (R_{i,j})^2}\Bigg]_{2k}=\Bigg[\sqrt{\frac{1}{N_{\text{nz}}}\displaystyle \sum_{i,j\in \text{nonzero}} (H_{i,j})^2}\Bigg]_{2k}$.
    \item $norm_H=|R^{2k+2}|_2/|R^{2k}|_2$, where $|R^{2k}|_2=\Big[||H||_2\Big]_{2k}$, i.e. the spectral norm.
\end{enumerate}
Here, $N_{nz}$ denotes the number of nonzero matrix elements (amplitude strength $\geq10^{-8}$) of the effective Hamiltonian (H) at a given order $2k$.
We found three ratio tests capture distinct aspects of effective H. The quantity $avg_H$ reflects overall growth of matrix elements $H_{i,j}$ when higher order correction terms are included. In contrast, $rms_H$ amplifies the presence of strong couplings both in hoppings and onsite terms. While spectral norm ($norm_H$) represents the largest eigenvalue associated with the strongest {$i-j$} couplings. In this letter, we primarily rely on $avg_H$, which enables us to incorporate maximum permissible perturbative order, avoiding divergences of the effective H. This criterion also ensures the resulting truncated expansion remains in closest agreement with exact numerics, beyond which the perturbative series fails to provide meaningful results. 
In figures \ref{avg_H_1500_a_1_om_1.5}-\ref{avg_H_4000_a_1_om_1.5}, \ref{rms_H_1500_a_1_om_1.5}-\ref{rms_H_4000_a_1_om_1.5} and \ref{norm_H_1500_a_1_om_1.5}-\ref{norm_H_4000_a_1_om_1.5}, we illustrated the $avg_H,rms_H$ and $norm_H$ results respectively, for various $L$ given in the main text at $a=1,\omega=1.5$. These plots demonstrate how $avg_H$ allows us to access higher order perturbation expansion, as $\bar{R}^{2k+2}/\bar{R}^{2k}<1$ for increasingly large $2k$ compared to $rms_H,norm_H$. Here $\langle\ldots\rangle_s$ represents the sample average over different random number sequences. The same analysis is shown for a non-universal $a=5,\omega=7.1$ in figures \ref{avg_H_1500_a_5_om_7.1}-\ref{norm_H_4000_a_5_om_7.1} and the corresponding $|H_{i,j}|$ and $|E|$ eigenvalue distributions are given in plots \ref{O_2_lamb_1.5_N_4000_a_1_om_1.5_AA}-\ref{O_10_lamb_-2_N_4000_a_1_om_1.5_GAA} and \ref{O_2_lamb_1.5_N_4000_a_5_om_7.1_AA}-\ref{O_10_lamb_-2_N_4000_a_5_om_7.1_GAA} for $(a,\omega)=(1,1.5)$ and $(5,7.1)$, respectively. 
The $|H_{i,j}|$ distributions clearly show that inclusion of higher order perturbative terms generates both a high-end tail (HET) with $|H_{i,j}|>10^0$ and a low-end tail (LET) with weights $<10^{-3}$. The plots show that $|E|$ has greater sensitivity towards strong couplings as the eigenvalue spectrum develops a tail at HET, which grows monotonically with $|H_{i,j}|$s of higher order. This behavior also explains why  $a=5,\omega=7.1$ remains well behaved across all ratio tests, unlike $a=1,\omega=1.5$; as the former develops a broader LET compared to its HET, effectively suppressing the emergence of large matrix elements and promoting divergence at higher order. The reason for such behavior is that the latter is closer to the radius of convergence compared to the former one, which in turn promotes divergence at early orders of perturbation theory.

%=========================================
%\newpage
%==========================================================
%==========================================================
%=============== ratio test plots =========================
%==========================================================
%==========================================================

%###############################################################
%================================================================
%=============== distribution plots =============================
%%%%%%%%%%%%%%%% for omega=1.5, a=1   %%%%%%%%%%%%%%%%%%%%%%%%%%%
%================================================================
%================================================================

%###############################################################
%================================================================
%=============== distribution plots =============================
%%%%%%%%%%%%%%%% for omega=7.1, a=5   %%%%%%%%%%%%%%%%%%%%%%%%%%%
%================================================================
%================================================================

 %XXXXXXXXXXXXXXXXXXXXXXXXXXXXXXXXXXXXXXXXXXXXXXXXXXXXXXXXXXXXXX   
%xxxxxxxxxxxxxxxxxxxxxxxxxxxxxxxxxxxxxxxxxxxxxxxxxxxxxxx
%=======================================================
% \typeout{get arXiv to do 4 passes: Label(s) may have changed. Rerun}

\end{document}